\documentclass[aps,prd,twocolumn,showpacs,superscriptaddress,nofootinbib,floatfix,letterpaper]{revtex4}

\newif\ifnotoverleaf
\notoverleaftrue

\pdfoutput=1
\usepackage[T1]{fontenc}
\usepackage{setspace}
\usepackage{amsmath}
\usepackage{graphicx}
\usepackage{placeins}
\usepackage{units}
\usepackage{xspace}
\usepackage{tabulary}
\usepackage[caption=false]{subfig}
\usepackage{multirow}
\usepackage{color}

\newcommand{\blue}[1]{\textcolor{blue}{#1}}
\newcommand{\red}[1]{\textcolor{red}{#1}}

\renewcommand{\blue}[1]{{#1}}
\renewcommand{\red}[1]{{#1}}

\newcommand{\eerec}{\ensuremath{E_{e}^{\mathrm{reco}}}}
\newcommand{\eetrue}{\ensuremath{E_{e}^{\mathrm{true}}}}

\newcommand{\therec}{\ensuremath{\theta_{e}^{\mathrm{reco}}}}
\usepackage{hyperref}

\newcommand{\sizecheck}{0} 
\newcommand{\PRDsupp}{1}   
\ifnum\PRDsupp=0
  
\else
  
\fi

\newif\ifpdf
\ifx\pdfoutput\undefined
   \pdffalse
\else
   \pdfoutput=1
   \pdftrue
\fi
\ifpdf
   \usepackage{graphicx}
   \usepackage{epstopdf}
\else
   \usepackage{graphicx}
\fi

\begin{document}

\title{Neutrino-electron elastic scattering for flux determination at the DUNE oscillation experiment} 

\author{Chris M.\ Marshall}\affiliation{Lawrence Berkeley National Laboratory, Berkeley, California, USA}
\author{Kevin S. McFarland}\affiliation{University of Rochester, Rochester, New York, USA}
\author{Callum Wilkinson}\affiliation{University of Bern, Bern, Switzerland}

\date{\today}

\pacs{13.15.+g,13.66-a}
\begin{abstract}
We study the feasibility of using neutrino-electron elastic scattering to measure the neutrino flux in the DUNE neutrino oscillation experiment.  The neutrino-electron scattering cross section is precisely known, and the kinematics of the reaction allow determination of the incoming neutrino energy by precise measurement of the energy and angle of the recoiling electron.  For \blue{several possible near detectors, we perform an analysis of their ability to measure neutrino flux in the presence of backgrounds and uncertainties.}  \blue{With realistic assumptions about detector masses,} we find that a liquid argon detector, even with limitations due to angular resolution, is able to perform better than less dense detectors with more precise event-by-event neutrino energy measurements. We find that the absolute flux normalization uncertainty can be reduced from $\sim$8\% to $\sim$2\%, and the uncertainty on the flux shape can be reduced by $\sim$20-30\%.
\end{abstract}

\ifnum\sizecheck=0  
\maketitle
\fi

\section{Introduction}
\label{sec:motivation}

The Deep Underground Neutrino Experiment (DUNE) is designed to measure CP violation in neutrino oscillations by making precise measurements of the neutrino flavor oscillations $\nu_\mu\not\to\nu_\mu$ and $\nu_\mu\to\nu_e$, and their antineutrino analogues, as a function of the neutrino energy, $E_{\nu}$~\cite{duneCDRphys}.  DUNE uses a wideband neutrino beam peaked at 2.5 GeV, and with 92\% of the muon-neutrino flux in the energy range $0.5$--$5$ GeV~\cite{duneCDRlbnf}. The DUNE far detector (FD) will measure neutrino-argon interactions, and infer the neutrino energy from the observed final state particles~\cite{duneCDRdet}. In addition to oscillation parameters, these measurements are sensitive to several inputs, each of which has significant, ${\cal O}$(10\%), uncertainties: the neutrino-argon interaction cross sections, the relationship between the true and inferred $E_{\nu}$, and the neutrino flux. To achieve its physics goals, in particular the measurement of CP violation, the DUNE near detector (ND) must constrain the uncertainties on the predicted event spectra to the level of $\sim$2-3\%~\cite{duneCDRphys}.

Neutrino cross-section uncertainties are energy dependent, and affect both the rate of interactions and the energy reconstruction.  Near detector measurements of neutrino-argon charged-current interactions are extremely valuable, but typically constrain a product of flux and cross sections. The near and far detectors see different neutrino fluxes, primarily because of oscillations, which limits the ability to extrapolate these ND constraints to the FD. The flux as a function of $E_{\nu}$ is {\em a priori} poorly predicted, primarily because of hadron production uncertainties as described in Section~\ref{sec:lbnf_beam}. This makes it difficult for the near detector to simultaneously measure the flux of neutrinos and to study the mechanisms by which neutrinos interact.

A helpful way to break this degeneracy is to separately measure the flux of neutrinos as a function of energy at the near detector. This can be done by selecting a sample of events for which the cross section as a function of energy is known. For example, at $E_{\nu} \gg 1$~GeV, the neutrino interaction cross section for events with low energy transfer, $\nu$, is roughly constant with neutrino energy.  This way to measure flux, referred to as the ``low-$\nu$'' technique, has been used to study total and deep-inelastic neutrino cross sections as a function of neutrino energies~\cite{Quintas:1992yv,Yang:2000ju,Tzanov:2005kr,Adamson:2009ju,DeVan:2016rkm,Ren:2017xov}. However, the DUNE first and second oscillation maxima occur at $\sim$2.5 GeV and $\sim$0.8 GeV, respectively. At these energies, it is difficult to select a sample with $\nu/E_{\nu}$ sufficiently small, and the low-$\nu$ technique breaks down. In this energy range, neutrino-electron elastic scattering is the only process with a known cross section.

In this paper, we demonstrate the feasibility of measuring the neutrino flux as a function of energy in the DUNE near detector using neutrino-electron elastic scattering. The LBNF beamline simulation is used to produce a flux of neutrinos at the near detector location, including the effect of the beam dispersion, as described in Section~\ref{sec:lbnf_beam}. We use Geant4 simulations to study the expected electron angular resolution in the liquid argon time projection chamber (LAr TPC) of the DUNE ND in Section~\ref{sec:dune-nd}. Similar studies are performed for a high-resolution gaseous detector, and a plastic scintillator detector. \blue{A detector with perfect electron reconstruction and background rejection is also considered as a limiting case}. Section~\ref{sec:study-framework} describes the details of the flux fits, and presents the results. \blue{In Section~\ref{sec:IMD}, we discuss the potential for using inverse muon decay to further constrain the high energy $\nu_{\mu}$ flux. Finally, in Section~\ref{sec:conclusions}, we present our conclusions.}

\section{Neutrino electron scattering}
\label{sec:nue_scat_intro}

Neutrino-electron elastic scattering, $\nu e^-\to\nu e^-$, is precisely predicted by the electroweak theory because it is a $2\to 2$ process that involves only weak interactions of fundamental leptons.  
In the limits that the neutrino energy $E_\nu$ is much greater than the electron mass $m_e$ and far below the energies required for resonant $W$ boson production, $E_\nu\ll \frac{M_W^2}{2m_e}$, the $\nu e^{-} \rightarrow \nu e^{-}$ cross section for neutrinos or antineutrinos is given at tree level by
\begin{equation}
\frac{d\sigma(\nu e^-\to\nu e^-)}{dy} = \frac{G^2_{\mathrm{F}} s}{\pi}\left[
  C_{\mathrm{LL}}^2+C_{\mathrm{LR}}^2(1-y)^2\right] .
\label{eqn:tree-xsec}
\end{equation}
Here, $G_{\mathrm{F}}$ is the Fermi weak coupling constant, $s$ is the Mandelstam invariant representing the square of the total energy in the center-of-mass frame, and $y\equiv T_e/E_\nu$ where $T_e$ is the electron kinetic energy.  The couplings $C_{\mathrm{LL}}$ and $C_{\mathrm{LR}}$ are different for neutrinos and antineutrinos and depend on flavor.  For $\nu_\mu$ and $\nu_\tau$, $C_{\mathrm{LL}}=-\frac{1}{2}+\sin^2\theta_W$ and $C_{\mathrm{LR}}=\sin^2\theta_W$, where $\theta_W$ is the Weinberg angle, and in the corresponding antineutrino couplings, the values for $C_{\mathrm{LL}}$ and $C_{\mathrm{LR}}$ are interchanged.  For $\nu_e$ ($\bar\nu_e$), the value of one of the couplings, $C_{\mathrm{LL}}$ ($C_{\mathrm{LR}}$), is $\frac{1}{2}+\sin^2\theta_W$ because of interfering contributions from neutral-current interaction that is present for all flavors and from a charged-current interaction that is present only for electron neutrinos.  Electroweak radiative corrections to the process are few-percent corrections and are discussed in detail in Appendix~\ref{sec:radiative}. 

The theoretical uncertainty of the neutrino-electron elastic scattering cross section from uncertainties in the parameters and radiative corrections is small~\cite{Erler:2013xha}.  \blue{Recent work~\cite{Tomalak:2019} has shown that the limiting uncertainty comes from hadronic loops in radiative corrections which results in a few permille uncertainty.}  Therefore a measurement of the reaction can be used to measure neutrino flux at this precision.  At the $\sim {\cal O}(1)$~GeV neutrino energies of DUNE, this cross section is approximately $10^{-4}$ of the total charged-current $\nu_\mu$ cross section; therefore the number of events is small and backgrounds may be substantial.  ~\red{However, for realistic near detector sizes, the event sample is expected to be sufficiently large in the DUNE beam to allow for statistical precision on a neutrino-electron elastic scattering sample to be ${\cal O}(1\%)$~\cite{Tian:2015dba}.}

The angle of the final state electron with respect to the neutrino, $\theta_e$, is
\begin{equation}
1-\cos\theta_e = \frac{m_e(1-y)}{E_e},
\label{eqn:angle}
\end{equation} 
\blue{where $E_{e}$ is the energy of the final state electron.} Therefore at neutrino energies $\sim$1~GeV, such as for DUNE, where $m_e\ll E_\nu$, the
final state electron is very forward.  A measurement of the angle and electron energy determines $y$, and thus also the neutrino energy.

Another neutrino-electron scattering process with a well-known cross section is inverse muon decay (IMD), $\nu_\mu e^-\to\nu_e\mu^-$.  This process has a threshold energy of $E_{\mathrm{min}}=\frac{m_{\mu}^{2}-m_e^2}{2m_{e}} \approx 11$ GeV, and
a total cross section given at tree level by~\cite{Bardin:1986dk}
\begin{equation}
\sigma=\frac{(s-m_\mu^2)^2 G_{\mathrm{F}}^2}{s\pi}+{\cal O}\left( \frac{m_e^2G_{\mathrm{F}}}{s}\right) .
\label{eqn:IMD}
\end{equation}
\noindent The spectrum of muons emitted for a fixed neutrino energy in the lab frame, $E_\nu$, is approximately uniform with limits between $E_{\mathrm{min}}$ and $E_\nu$, with small corrections to the uniformity and the kinematic limits of order $m_e/E_\nu$ and $m_e$, respectively.   This cross section increases with energy as the DUNE flux is falling, and the event rate is expected to peak at $\sim$18 GeV. IMD could provide a constraint on the high energy tail of the $\nu_\mu$ flux; however, such a constraint would have little impact on the DUNE neutrino oscillation analyses. This process is discussed further in Section~\ref{sec:IMD}, and in less detail than $\nu e^-\to\nu e^-$ in this manuscript.

\section{MINERvA's neutrino-electron scattering flux measurement}
\label{sec:minerva-nue}

\noindent

The MINERvA experiment is the only accelerator experiment to date that has successfully used this technique~\cite{Park:2015eqa,Valencia:2019mkf} to significantly reduce its uncertainty on a predicted neutrino flux. MINERvA reconstructed these events in a segmented scintillator detector with neutrinos at energies similar to DUNE's. The first analysis with the low energy NuMI beam~\cite{Park:2015eqa} observed 127 total events including a predicted background of $30\pm 4$ events; a second, recent analysis with the medium energy NuMI beam~\cite{Valencia:2019mkf} found 1021 events with a predicted background of $212\pm13.5$ events. The background composition of the two analyses was different because of the event selection and beam energies.  In the medium (low) energy analysis, the background was approximately $28\%$ ($55\%$) $\nu_e$ charged-current interactions, primarily quasielastic like events $\nu_e n_{\mathrm{bound}}\to e^- p$, $54\%$ ($30\%$) neutral current interactions, primarily with a $\pi^0$ in the final state, and $18\%$ (15\%) $\nu_\mu$ charged current events, also primarily with a $\pi^0$ in the final state and a very low energy final state $\mu^-$.  In both analyses, backgrounds in the segmented scintillator were reduced by requiring an electron energy of $800$~MeV or greater, which is not a desirable selection for a DUNE near detector because of the physics interest in the low energy neutrino flux.  Because of the  angular resolution in the MINERvA segmented scintillator, with a granularity of $\sim$2~cm, MINERvA did not attempt to use angular information to reconstruct the incoming neutrino energy.  The systematic uncertainty on the observed rate in the MINERvA medium (low) energy measurement was $1.8\%$ ($5\%$), and was mostly due to uncertainties in the background reactions.  The uncertainty on background reactions, particularly the low $Q^2$ behavior of the $\nu_e$ quasielastic-like background events, is significantly lower in the medium energy analysis than in the low energy analysis, largely due to better knowledge of the low $Q^2$ behavior of neutrino reactions due to MINERvA data itself~\cite{Ruterbories:2018gub,Valencia:2019mkf}.  In the medium energy analysis, the electron reconstruction efficiency and electromagnetic energy scale of the detector were also noted contributors of systematic uncertainty, but were not dominant sources.  Both analyses had a $1$--$2\%$ uncertainty in the application of the event rate to the neutrino flux prediction from the fiducial mass of the detector.  As a result of these analysis, the fractional uncertainty on MINERvA's low energy flux between $2$ and $10$ GeV was reduced from $8.7\%$ to $6.0\%$, and the uncertainty at the focusing peak of the medium energy beam was reduced from $7.5\%$ to $3.9\%$.

\section{LBNF beam}
\label{sec:lbnf_beam}

The planned Deep Underground Neutrino Experiment (DUNE)~\cite{duneCDRphys, duneCDRdet} will operate in the Long Baseline Neutrino Facility (LBNF)~\cite{duneCDRlbnf} beamline at Fermilab. LBNF is designed to operate at an initial beam power of 1.2 MW, with a design capacity of 2.4 MW, more than three times the maximum intensity of the NuMI beamline (700 kW)~\cite{numi}. At 1.2 MW intensity, corresponding to $1.1 \times 10^{21}$ protons on target per year, and with a detector located 574m from the neutrino source, $\sim$120 $\nu$--$e^{-}$ events are expected per year per ton of argon. Hydrocarbon detectors have a higher ratio of electrons to nucleons, and therefore have higher event rates per unit mass. The expected rates per year per ton are $\sim$144 for CH and $\sim$152 for CH$_{2}$.

To create the LBNF neutrino beam, protons from the Main Injector strike a fixed target, producing pions and kaons, which are focused by a system of magnetic horns into a decay volume, where the mesons decay primarily into muons and muon-flavor neutrinos. The specific parameters of LBNF are optimized to maximize the sensitivity to CP violation in DUNE. The target is a long, thin rod, 2 m in length and 10 mm in diameter. Three horns are used in the optimized configuration of LBNF, compared to only two horns in NuMI. The main advantage of the third horn is a reduction in the high-energy tail of the flux. The target protrudes slightly into the first horn, while the second and third horns are located 3 m and 17 m downstream of the target, respectively. The decay pipe volume is cylindrical, 200 m along the axis, and with a radius of 2 m. It begins 20 m downstream of the target, and is angled downward at 6 degrees (101 mrad.) with respect to the horizontal, such that it points toward the on-axis far detector.

The horns are designed to focus positive mesons in forward horn current (FHC) mode, leading to a primarily $\nu_{\mu}$ beam. In reverse horn current (RHC) mode, the beam consists primarily of $\bar{\nu}_{\mu}$. The wrong-sign contamination is higher in RHC because the proton-carbon interactions produce more $\pi^{+}$ than $\pi^{-}$, but the contamination is primarily in the flux tail. Electron-flavor neutrinos make up 1\% of the total flux, and arise primarily from the decay chain $\pi \rightarrow \mu \rightarrow e$. At energies above 10 GeV, neutrinos from kaon decays dominate, and the $\nu_{e}$ contamination is larger from $K^{\pm} \rightarrow \pi^{0} e^{\pm} \nu_e$ and $K^{0}_{L} \rightarrow \pi^{\pm} e^{\mp} \nu_e$. The neutrino flux is peaked at 2.5 GeV, the oscillation maximum for a baseline of 1300km. 

\begin{figure}[tbp]
\centering
  \includegraphics[width=\columnwidth]{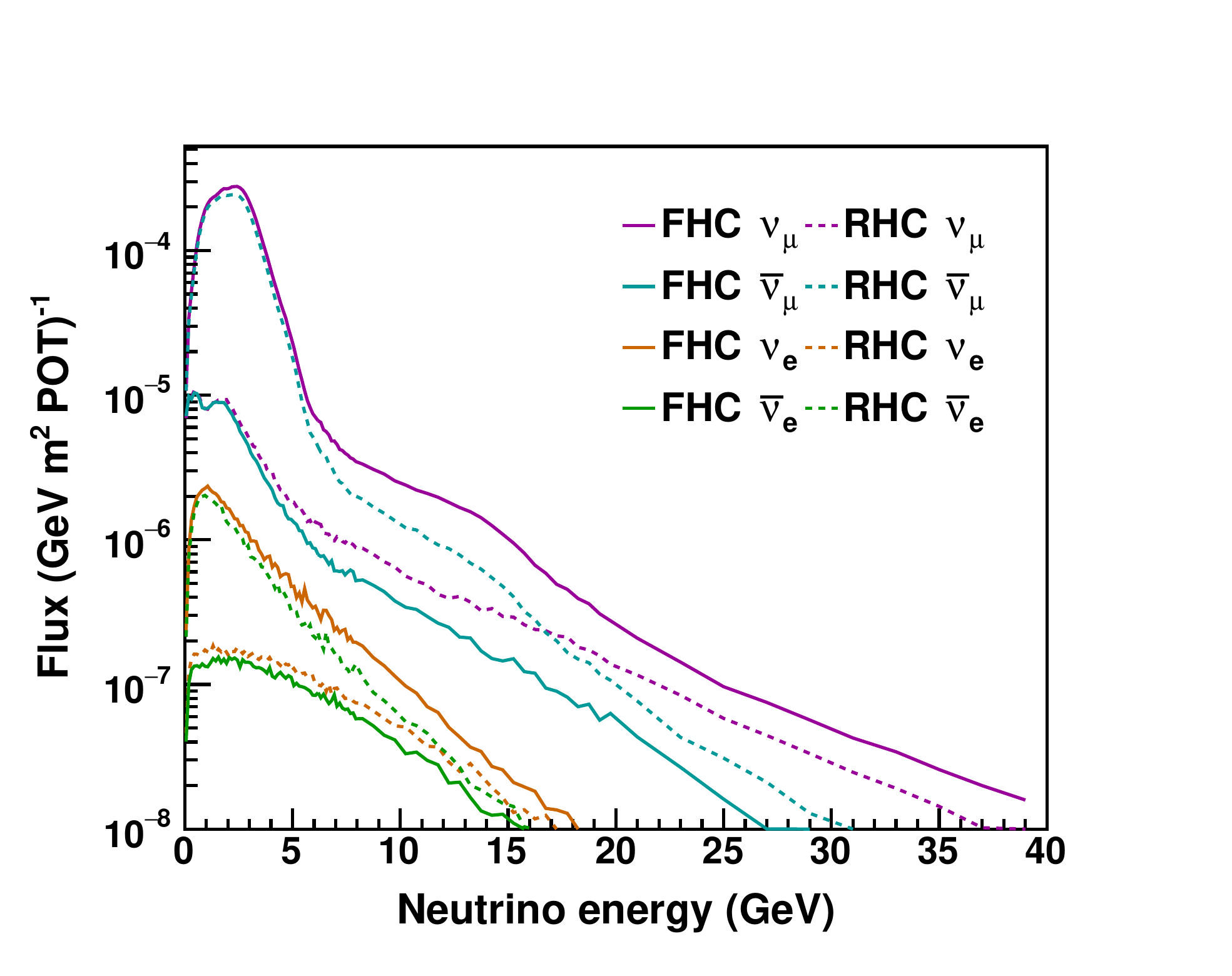} 
\caption{The DUNE flux prediction used in this analysis. The solid lines are the fluxes for each of the four neutrino species in FHC, while the dashed lines are the same for RHC.}
\label{fig:flux}
\end{figure}

The beamline is simulated with g4lbnf, a Geant4-based model. Proton interactions in the target, as well as the subsequent interactions of hadrons in the target and focusing system, are simulated using the ``QGSP\_BERT'' physics list, which combines the quark-gluon string with precompound (QGSP) model and a Bertini cascade at higher energies. This analysis is based on g4lbnf version v3r5p4, which is based on an 120 GeV proton beam. The energy spectra for all four neutrino species in both horn polarities used in this analysis are shown in Figure~\ref{fig:flux}.

Neutrinos can have non-zero angles with respect to the beam axis due to imperfect focusing, the finite width of the decay pipe, and the finite size of the detector. The mean neutrino angle at the LBNF near detector facility, 574m from the target along the beam axis, is approximately 1.5 mrad. The decay pipe geometry gives a maximum angle of 5.6 mrad. to the center of the near detector, corresponding to a decay at the edge of the decay pipe and nearest to the detector hall. 

\begin{figure}[tbp]
\centering
  \includegraphics[width=\columnwidth]{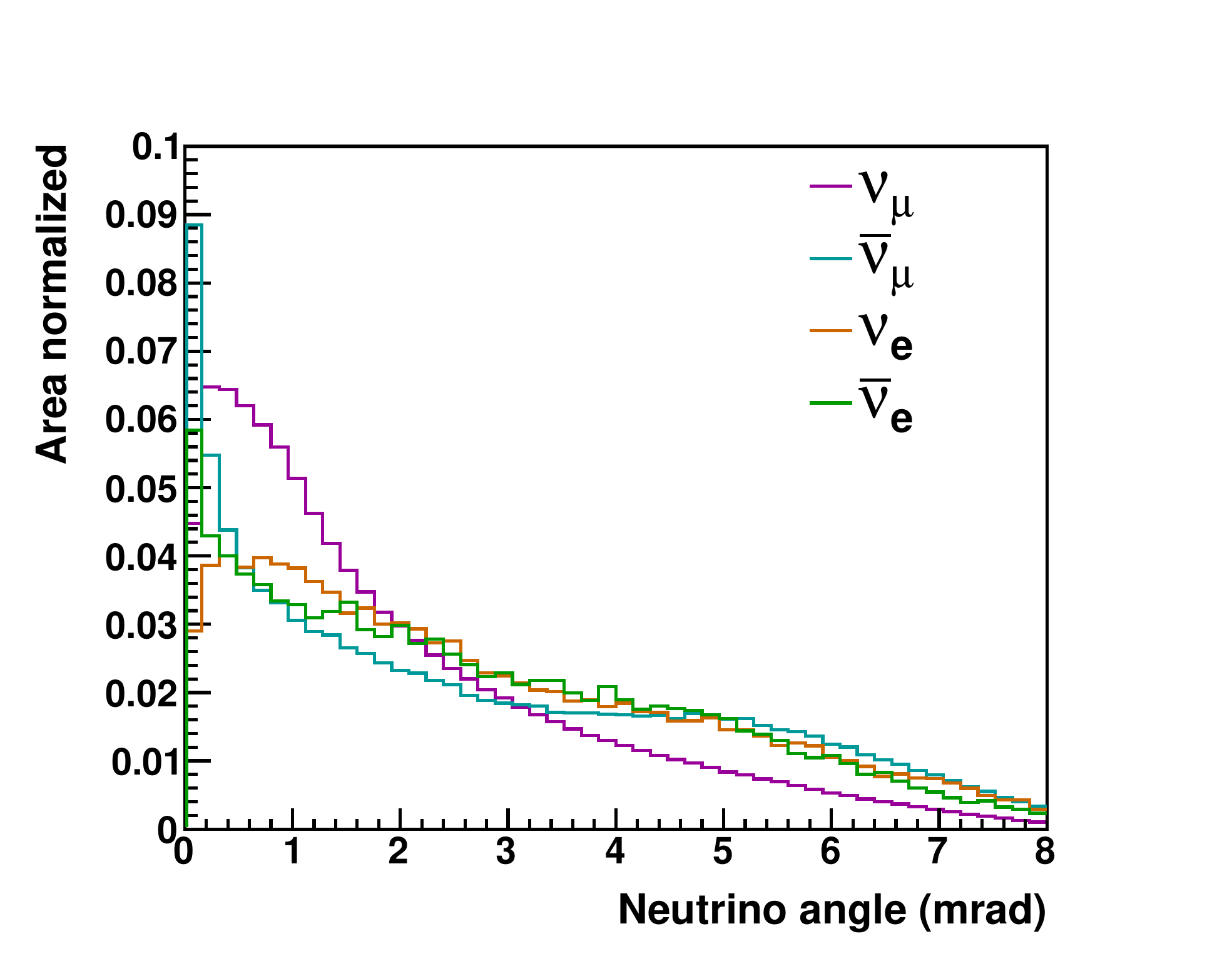} 
\caption{The angle of the neutrino beam at the LBNF near detector facility for each flavor in FHC. This accounts for the finite decay pipe but neglects the finite detector size. The respective curves are area normalized.}
\label{fig:nu_angle}
\end{figure}

Larger angles are possible for neutrinos originating from decays outside the decay pipe region (88\% of neutrinos intersecting the near detector originate in the decay pipe, and an additional 8\% in the target hall). Muon decays can produce neutrinos with much shorter baselines. The area-normalized angular distribution for FHC is shown in Figure~\ref{fig:nu_angle} for the four neutrino species. The antineutrino distributions are more sharply peaked at zero angle because of very forward pions passing down the center of the horns, which are not defocused. These pions are generally higher in energy and give very forward decays. The larger high-angle tails for antineutrino and $\nu_{e}$ are due to muon decays.

The reconstruction of the neutrino angle is critical in an analysis of $\nu e^{-} \rightarrow \nu e^{-}$. The angle due to the finite detector width can be corrected by taking the neutrino angle to be the line connecting the mean decay position to the reconstructed interaction vertex. In LBNF, the mean decay position is approximately the center of the decay pipe due to approximate cancellation of the exponentially decreasing pion flux and quadratically increasing detector solid angle. The angular distribution due to the decay pipe width cannot be corrected, and effectively smears the distribution of $\theta_e$, the electron angle with respect to the neutrino.

The largest uncertainties on the flux prediction as a function of neutrino energy are due to hadron production on the target, and tolerances of the focusing system. The method for evaluating these uncertainties is similar to that of MINERvA~\cite{minervaFlux}. There are strong positive correlations between the flux predictions at the near and far detectors, between forward and reversed horn current, and between muon and electron (anti)neutrinos. Because of these correlations, a near detector \blue{flux constraint also constrains the far detector flux. In principle, an FHC (RHC) constraint would also constraint the RHC (FHC) flux, although in this analysis we investigate constraints in both modes.}

Uncertainties on the distribution of the incoming neutrino angle could bias an extraction of the energy spectrum from $\nu e^{-} \rightarrow \nu e^{-}$. The effect of hadron production uncertainties on the angle are found to be negligibly small. Varying the horn focusing parameters gives sub-percent changes to the angular distribution. One exception is the uncertainty on the width of the decay pipe, which determines the endpoint of the distribution in Fig.~\ref{fig:nu_angle} by specifying the maximum possible angle from the decay pipe to the near detector. This uncertainty produces large changes to the flux at very high angle, but affects less than 1\% of the total flux.

\section{Near detector technologies studied}
\label{sec:dune-nd}

For this study, we evaluate the expected performance of the DUNE near detector for $\nu e^{-} \rightarrow \nu e^{-}$. The primary design considered is a liquid argon time projection chamber (LAr TPC), roughly based on the ArgonCube concept~\cite{pixelLAr}. \blue{We also consider a high-resolution tracking detector (HRT), which has significantly improved energy and angular resolution compared to the LAr TPC, but a lower mass because it would require a gaseous rather than liquid argon target. This is meant to represent a generic tracking detector with a large number of high resolution spatial measurements per unit of $dE/dx$, but the performance parameters are roughly what could be achieved with a straw tube tracker like the one described in the reference design of the DUNE conceptual design report~\cite{duneCDRdet}, or with a gaseous argon TPC with good angular resolution. The purpose of including it in the study is to determine whether a lower mass detector with superior energy and angular resolution can provide a stronger constraint.} In addition, we consider a solid plastic scintillator (labelled ``CH'') detector, which is essentially what could be achieved by putting the MINERvA detector~\cite{minervaNIM} in the DUNE beam. In this analysis, we assume a 5 year exposure on each detector, with each horn current (the FHC and RHC analyses are performed separately). The LAr fiducial mass is assumed to be 30 tons, which can easily be achieved with a total active volume of $4 \times 3 \times 5$ m. Alternatively, this can represent a fiducial mass of 60 tons in a scenario where only 50\% of the total exposure is taken on-axis. \blue{The HRT and plastic scintillating detector are assumed to have a 5 ton fiducial mass. For comparison, the HRT and CH analyses are also repeated with a 30 ton mass equal to the LAr.}

There are three detector parameters that impact performance in the $\nu$--$e^{-}$ channel: the electromagnetic energy resolution, the angular resolution for forward electrons, and the threshold for rejecting events with other final-state charged particles, such as low-energy protons. Angular resolution is the most important metric for $\nu$--$e^{-}$ because the signal kinematic limit and the neutrino energy reconstruction both depend on $E_{e}\theta_{e}^{2}$. In this section, the procedure used to determine the expected angular resolution as a function of electron energy is described in detail for the liquid argon ND concept. The procedure is qualitatively similar for the HRT and CH detectors.

The liquid argon concept for the DUNE ND is based on ArgonCube. The detector consists of an array of optically segmented TPC modules in a common cryostat. The module size has a cross section of $1 \times 1$~m$^{2}$, with a central cathode dividing the TPC into two drift regions with a maximum drift distance of 50cm. Charge is read out by an array of pads, instead of the projective wires in the DUNE far detector design. We consider a pad size of $3 \times 3$mm, similar to what was used in initial demonstrations of the pad readout technology~\cite{pixelLAr}, for a total of $\sim$10$^{5}$ channels per m$^{2}$. With a maximum drift length of 50 cm, transverse diffusion is estimated to be $0.8$~mm, based on 13 cm$^{2}$/s at 1 kV/cm~\cite{gushchin, lngDet, Sorensen}. 

The angular resolution is determined by the position resolution of the detector, and by multiple scattering of electrons in liquid argon. Forward-going electrons from $\nu$--$e^{-}$ elastic scattering are nearly parallel to the readout plane, and will intersect individual rectangular pads at nearly right angles. The 2D electron angle in the plane perpendicular to the drift direction depends on the pad coordinate; for 3mm pitch the position resolution is 3mm~$/\sqrt{12} = 0.87$mm. A potential aliasing effect exists for tracks nearly parallel to a row of pads, which can be mitigated by staggering the pads in successive rows. Improved position and thus angular resolution can be achieved with a triangular pad design. Because diffusion is small due to the short drift, collected charge for a forward-going particle will be shared among two adjacent triangles, and the relative charge collected on each triangle is proportional to the position of the electron. This feature is used by MINERvA to achieve 3mm position resolution with 1.7cm scintillator strip pitch. The 2D angle in the drift plane is determined by timing. The neutrino interaction time is determined by the detection of scintillation photons. The position resolution in the timing direction is expected to be significantly better than the pad size. 

The radiation length in liquid argon is 14cm. For a 3mm pad pitch, this corresponds to $N$ = 47 position measurements per radiation length, $X_{0}$. The resolution due to measurement, $\sigma_{\mathrm{meas}}$, and multiple coulomb scattering, $\sigma_{\mathrm{MS}}$, can be calculated as

\begin{align}
  \label{eq:sigmas}
  \begin{aligned}
    \sigma_{\mathrm{meas}} &= \sqrt{\frac{12 N}{(N+1)(N+2)}}\frac{\sigma_{x}}{L}, \\
    \sigma_{\mathrm{MS}} &= \frac{0.015~\mathrm{GeV}}{p}\sqrt{\frac{L}{X_{0}}},
  \end{aligned}
\end{align}

\noindent
where $L$ is the track length. The measurement resolution decreases with track length, while the multiple scattering term increases. The optimal track length to fit is approximately one radiation length. Other hard scattering processes also contribute. To quantify this effect, electrons are simulated in liquid argon with Geant4 10.2. The electron position is determined at 3mm intervals and smeared by a Gaussian function with a sigma of 1 mm. An uncertainty is placed on each position measurement based on the average multiple scattering according to Equation~\ref{eq:sigmas}. The resulting points are fit to a straight line. The reconstructed angle is then compared to the true angle to determine the resolution.

\begin{figure}[tbp]
\centering
  \subfloat[1 GeV] {\includegraphics[trim={0 0 0 1cm}, clip, width=0.95\columnwidth]{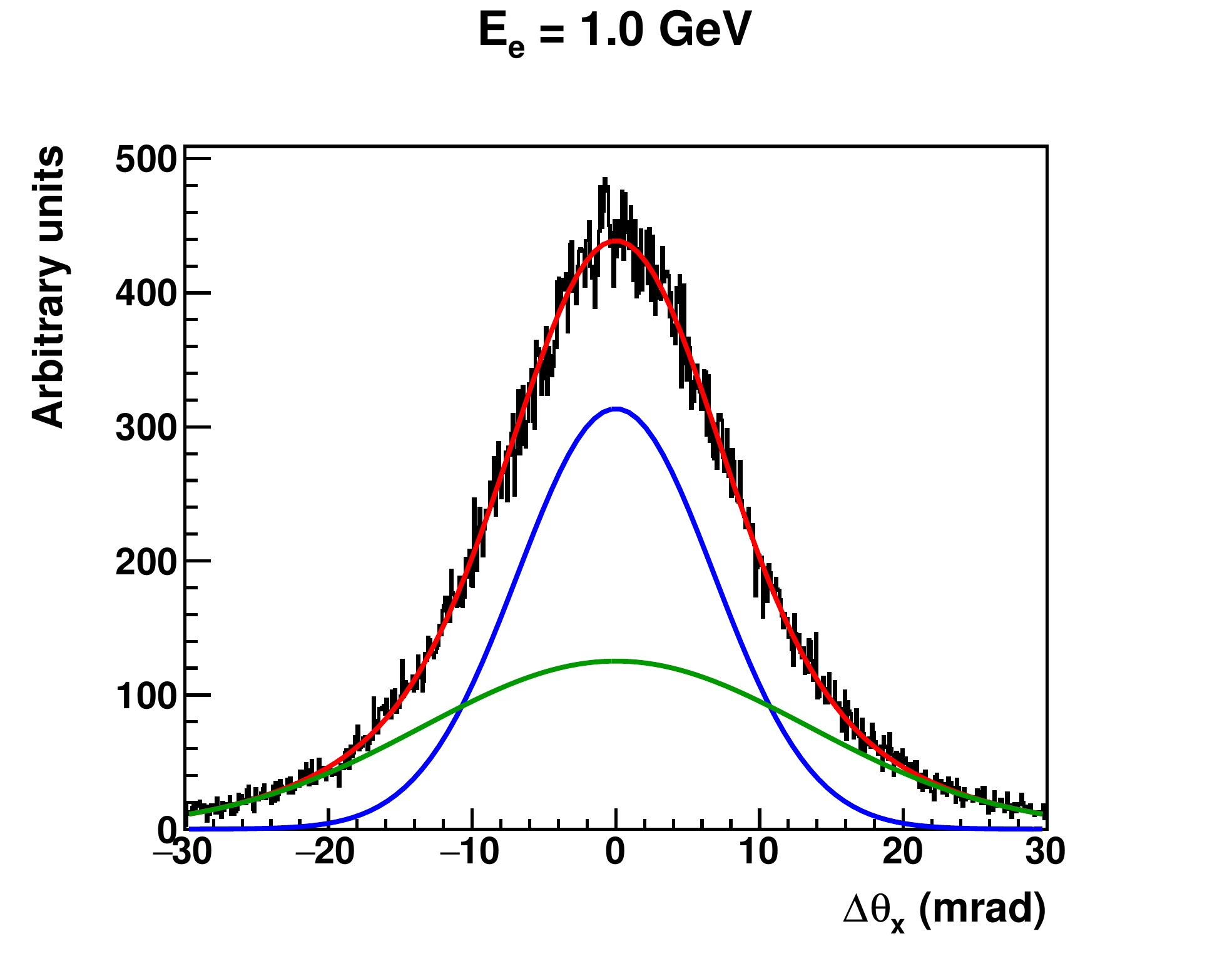}}\\
  \subfloat[5 GeV] {\includegraphics[trim={0 0 0 1cm}, clip, width=0.95\columnwidth]{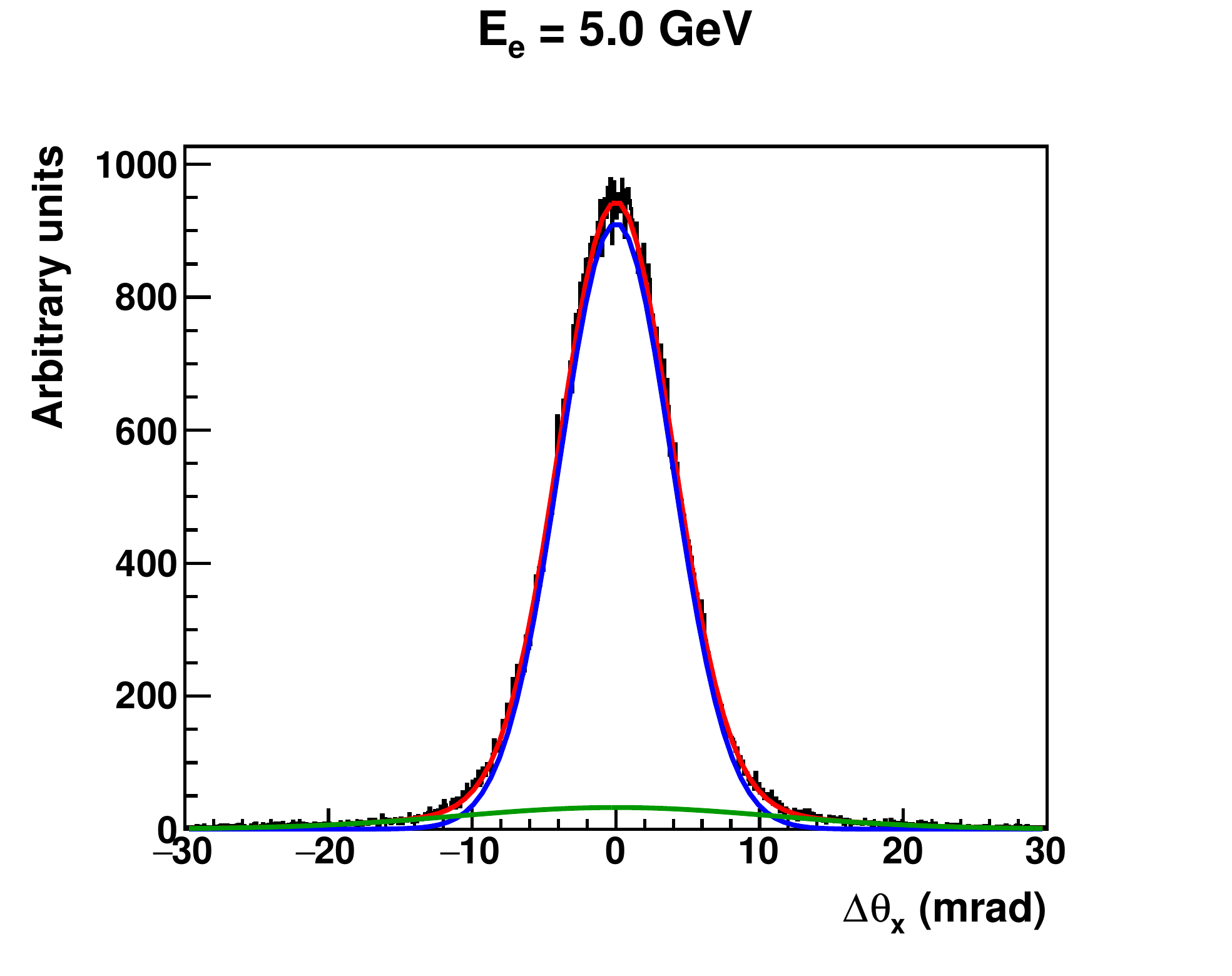}}
\caption{The difference between reconstructed and true two-dimensional angle $\theta_{e,x}$, for electrons at 1 and 5 GeV in the LAr TPC detector. The distributions are fit to a double Gaussian, where the inner (blue) fit accounts primarily for multiple coulomb scattering and the outer (green) fit accounts for hard scattering.}
\label{fig:angular_res}
\end{figure}

\begin{figure}[tbp]
\centering
  \includegraphics[width=\columnwidth]{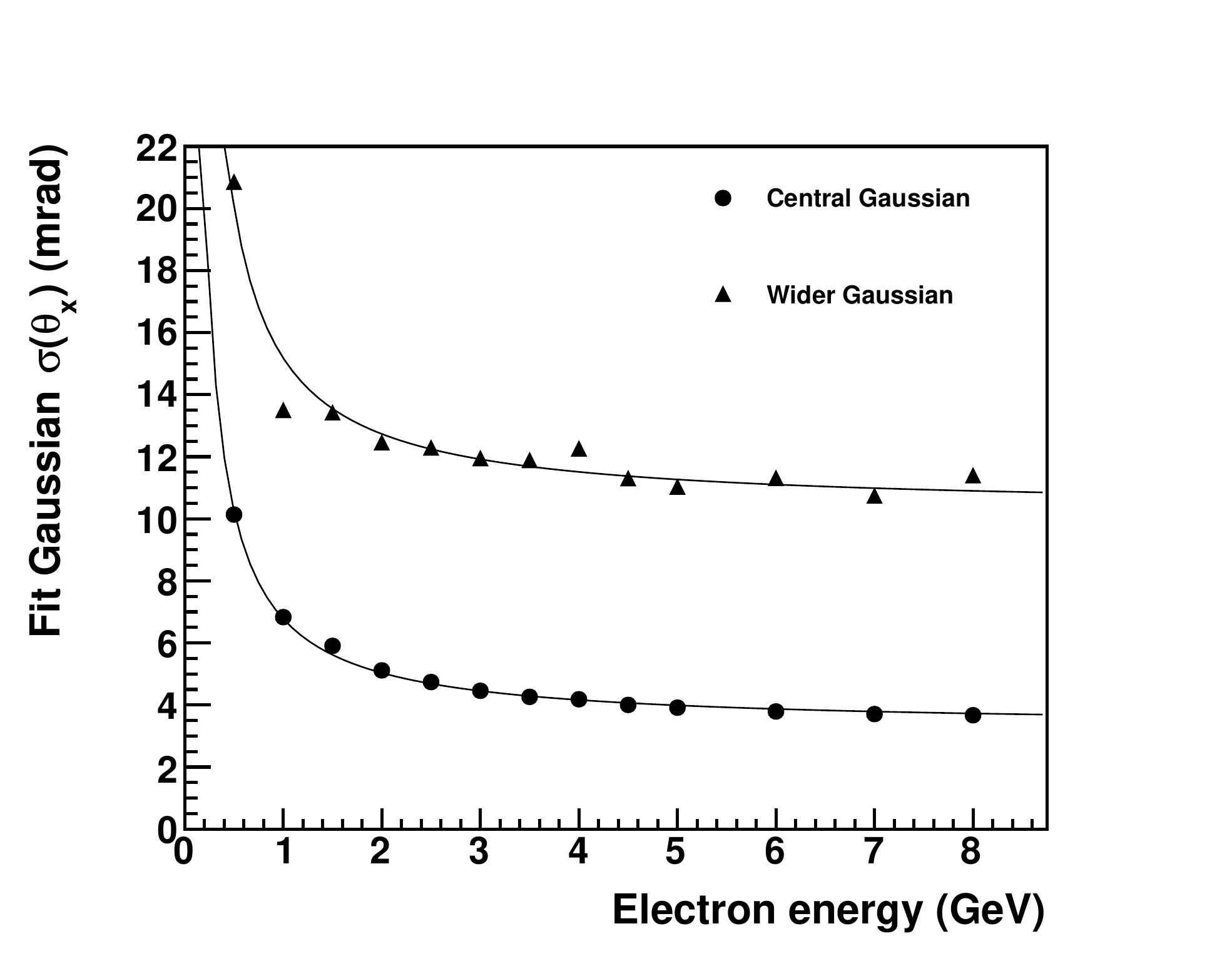}
\caption{Electron angular resolution is parameterized by a double Gaussian, with sigmas that depend on the electron energy. Angle residuals are fit at each energy, as in Figure~\ref{fig:angular_res}. Circles show to the central Gaussian sigma, and triangles show the wide Gaussian.}
\label{fig:res_vs_E}
\end{figure}

This procedure is repeated for electrons of varying momentum. Figure~\ref{fig:angular_res} shows the residual of the two-dimensional angle, $\Delta\theta_{e,x}$, for electrons at 1 GeV and 5 GeV. The energy dependence is parameterized as a function of electron energy according to a double Gaussian. The widths of each Gaussian, and the relative normalization, are determined from fits to these distributions. The dependence on electron energy is fit assuming constant and $1/E_{e}$ terms. The central multiple scattering term plateaus at 4 mrad., which is essentially the measurement limit. The dependence on electron energy is shown in Figure~\ref{fig:res_vs_E}.

\blue{The expected angular resolution functions for the HRT and plastic scintillator detectors are determined similarly. For the HRT, a position measurement is made in each two-dimensional projection every 8 cm, with an assumed transverse resolution of 200 $\mu m$. The angular resolution function is determined by the same procedure as described above, but in a volume with density 0.1 g/cm${^3}$, which is approximately the average density for a straw-tube or high-pressure gaseous detector.  An electron density equivalent to that of CH$_2$ is assumed. The resulting angular resolution in the HRT is significantly better than in LAr, reaching $\sim$1.5 mrad. at high energy.}

The scintillator detector is based on MINERvA~\cite{minervaNIM}. MINERvA tracker planes are spaced by 2 cm, with at least 4 cm between two planes of the same two-dimensional orientation. The assumed position resolution per plane is 3 mm. The angular resolution achieved with these assumptions is somewhat worse than in LAr, $\sim$8 mrad. at high energy, and consistent with the angular resolution of MINERvA.

\begin{figure}[tbp]
\centering
  \includegraphics[width=0.9\columnwidth]{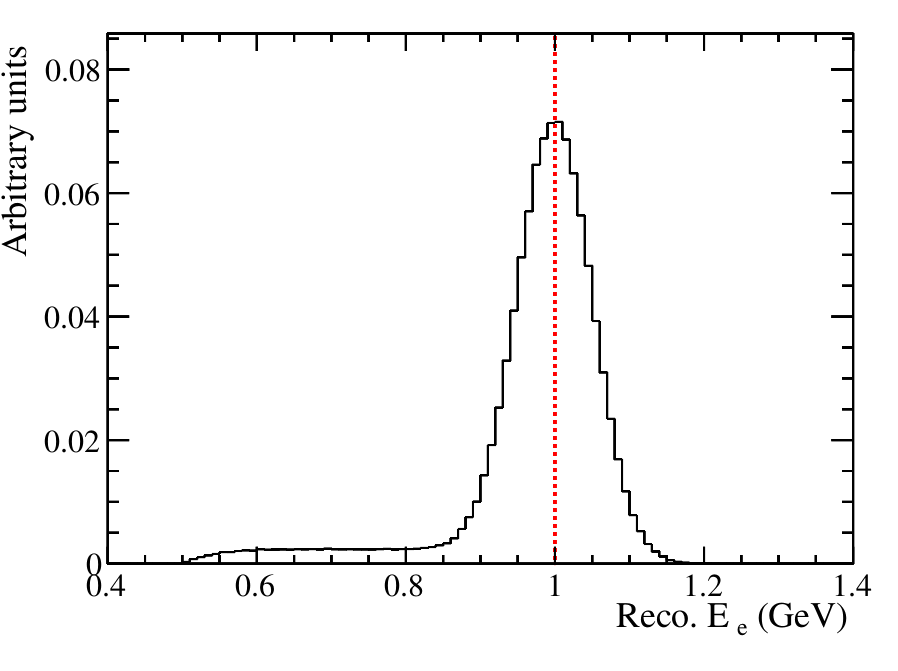}
\caption{Electron energy resolution is parameterized by a Gaussian centered on the true electron energy with a width of 5\%, and a low-side tail, which affects 10\% of events and is close to uniform for reconstructed energies in the range $\eetrue/2$--$\eerec$.}
\label{fig:eres}
\end{figure}
\blue{For all detector types considered above, the reconstructed electron energy is treated in the same way. It is parameterized by a Gaussian centered on the true electron energy with a width of 5\%, and a low-side tail, which affects 10\% of events and has a fairly arbitrary $P(\eerec) \propto 1-(4 \eerec/\eetrue - 3)^6$ form, which serves to smear events in this tail between $\eetrue/2$--$\eerec$ without a step function, and includes a generic misreconstruction effect in the analysis. Figure~\ref{fig:eres} shows an illustrative example, of the $\eerec$ distribution for an electron with $\eetrue = 1$ GeV. Systematic studies which vary the size of the low-side tail, and shift the peak of the distribution (to mimic an energy scale bias) are considered in Section~\ref{sec:systematic_uncertainties}.}

\blue{Finally, for the reference ``perfect'' detector, the electron energy and angular reconstruction are assumed to be perfect, and perfect background rejection is assumed. The intention with the perfect detector is to show the inherent limitations to the technique due to the relatively low $\nu$--$e^-$ event rate and the divergence of the neutrino beam at the near detector location. For the perfect detector, CH$_2$ was used as the target as it has the highest electron density.
}

\section{Study framework}
\label{sec:study-framework}

The aim of this study is to test how well the flux normalization and shape uncertainties can be constrained from the reconstructed electron energy and angle with respect to the nominal beam axis ($E_{e}$, $\theta_{e}$) for different potential DUNE near detector designs. We simulate $\nu$--$e^{-}$ scattering events with the GENIE event generator~\cite{Andreopoulos:2009rq,Andreopoulos:2015wxa}, which uses the tree level cross section given in Section~\ref{sec:nue_scat_intro} and include radiative corrections to the cross section. The full DUNE three-horn optimized flux is used, including the beam divergence as described in Section~\ref{sec:lbnf_beam}. Detector resolutions, as described in Section~\ref{sec:dune-nd}, are used to smear the GENIE prediction as a function of $E_{e}$, $\theta_{e}$. Additionally, backgrounds from $\nu_{e}$--$^{40}$Ar interactions which produce electrons or single photons from $\pi^0$ decays are included in the study, as described in Section~\ref{sec:backgrounds}. The fitting framework is described in Section~\ref{sec:fitter}, the main results are shown in Section~\ref{sec:results}, and bias tests to check the robustness of the result to different input assumptions are described in Sections~\ref{sec:bias} and~\ref{sec:systematic_uncertainties}.

\subsection{Selection and backgrounds}
\label{sec:backgrounds}
The selection for this analysis is very simple. The signal processes will produce a single very forward-going electron, and no other particles at the vertex. We consider two types of backgrounds: $\nu_e$--$^{40}$Ar interactions which produce a forward-going electron; and events in which a single photon from a neutrino--nucleus produced $\pi^{0}$ is reconstructed. These are referred to as the $\nu_{e}$ and $\gamma$ backgrounds. We apply a cut on the extra energy deposited at the vertex of $E_{\mathrm{extra}} \leq 20$ MeV ($\leq 30$ MeV for the CH detector) for $\nu_e$--$^{40}$Ar interactions, and a cut of $E_{\gamma} \leq 50$ MeV on the second photon energy for the $\pi^{0}$ production background. Additionally, the $\pi^0$ background is suppressed by a factor of 0.1 to account for the $\gamma$/$e^{\pm}$ separation capabilities of the detectors considered. The extra energy of 20 MeV corresponds to a proton with a range of 5 mm in liquid argon, which will deposit energy on more than one pad. \blue{Under these assumptions, the $\nu_e$ backgrounds are always significantly larger than the $\pi^0$ backgrounds.  As noted in Section~\ref{sec:dune-nd}, for the ``perfect'' detector options, perfect background rejection is assumed.}

Although it was shown in Ref.~\cite{Park:2015eqa} that a cut on $E_{e}\theta_{e}^2$ provides good separation between signal and background events, we do not make such a cut in this analysis. Instead, we only consider events with a reconstructed electron angle $\theta_{e}^{\mathrm{reco}} \leq 60$ mrad., and perform the fit in $E_{e}^{\mathrm{reco}}$--$\theta_{e}^{\mathrm{reco}}$ space, in which signal and background are reasonably well separated. Because of this separation, and the fit method used (described below), an $E_{e}\theta_{e}^2$ cut is not necessary.

\subsection{Fitting framework}
\label{sec:fitter}
In this analysis, the simulated data, including backgrounds and all resolution effects described in Section~\ref{sec:dune-nd}, are binned into \eerec--\therec~bins, and scaled to the expected event rate given the detector mass and exposure time relevant for each ND options described in Section~\ref{sec:dune-nd}. Bins are 4 mrad. wide in $\theta_{e}$, with 15 bins in the range 0--60 mrad. In $E_{e}$, there are 45 bins in the range 0--60 GeV, where the bin edges are defined such that the central value $\pm$5\% lies inside the bin, with a minimum bin width of 0.2 GeV, motivated by the expected $E_{e}$ resolution of $\sim$5\% in our simple model. Although the binning is somewhat arbitrary, changes to the binning (2x finer binning) had no significant effect on the fits described below.

\begin{table*}[tbp]
\tabcolsep7.5pt
\caption{Number of $E_{\nu}$ template, and template binning for all detectors and beam configurations considered. The binning was decided by requiring $\geq$500 events/template in the GENIE prediction. The predicted event rates given are the total number of neutrino-electron events in five years of running in the nominal 1.2 MW beam for each detector.}
\label{tab:binning}
\begin{center}
  \small
  {\renewcommand{\arraystretch}{1.2}
    \begin{tabular}{c|c|c|c|p{8cm}}
\hline
\hline
Beam & Detector & Rate & N. bins & $E_\nu$ binning (GeV) \\
\hline
\multirow{7}{*}{FHC} & 5t CH & 4479 & 6 & 0, 1.875, 2.5, 3.125, 3.875, 5.875, 100 \\
& 5t HRT/perfect & 4753 & 7 & 0, 1.875, 2.375, 2.875, 3.5, 4.5, 8.5, 100 \\
& 30t LAr & 22458 & 28 & 0.0, 1.25, 1.5, 1.75, 2.0, 2.125, 2.25, 2.375, 2.5, 2.625, 2.75, 2.875, 3.0, 3.125, 3.25, 3.5, 3.75, 4.0, 4.25, 4.625, 5.125, 5.875, 6.875, 8.5, 10.0, 12.0, 14.5, 18.5, 100 \\
& 30t CH & 26873 & 32 & 0, 1.125, 1.375, 1.625, 1.75, 1.875, 2.0, 2.125, 2.25, 2.375, 2.5, 2.625, 2.75, 2.875, 3.0, 3.125, 3.25, 3.375, 3.5, 3.75, 4.0, 4.25, 4.5, 4.875, 5.375, 6.125, 7.0, 8.5, 10.0, 11.5, 13.5, 16.0, 100 \\
& 30t HRT/perfect & 28519 & 34 & 0, 1.125, 1.375, 1.625, 1.75, 1.875, 2.0, 2.125, 2.25, 2.375, 2.5, 2.625, 2.75, 2.875, 3.0, 3.125, 3.25, 3.375, 3.5, 3.625, 3.875, 4.125, 4.375, 4.75, 5.125, 5.625, 6.375, 7.25, 8.5, 10.0, 11.5, 13.0, 15.0, 18.5, 100 \\
\hline
\multirow{7}{*}{RHC} & 5t CH & 3168 & 4 & 0, 2.125, 2.875, 3.875, 100 \\
& 5t HRT/perfect & 3362 & 4 & 0, 2.125, 2.875, 3.875, 100 \\
& 30t LAr & 15885 & 18 & 0, 1.25, 1.625, 1.875, 2.125, 2.375, 2.625, 2.875, 3.125, 3.375, 3.625, 4.0, 4.375, 5.0, 6.0, 7.75, 10.5, 14.0, 100 \\
& 30t CH & 19008 & 23 & 0, 1.25, 1.5, 1.75, 2.0, 2.125, 2.25, 2.375, 2.5, 2.625, 2.75, 2.875, 3.125, 3.375, 3.625, 3.875, 4.25, 4.75, 5.5, 6.625, 8.5, 11.0, 14.0, 100 \\
& 30t HRT/perfect & 20172 & 25 & 0, 1.25, 1.5, 1.75, 2.0, 2.125, 2.25, 2.375, 2.5, 2.625, 2.75, 2.875, 3.0, 3.25, 3.5, 3.75, 4.0, 4.375, 4.875, 5.625, 6.75, 8.5, 10.5, 13.0, 17.0, 100 \\
\hline
\hline
\end{tabular}}
\end{center}
\end{table*}
\begin{figure*}[tbp]
  \centering       
\ifnotoverleaf
  \subfloat[$E_{1}$ (0--1.875 GeV)]      {\includegraphics[width=0.25\textwidth]{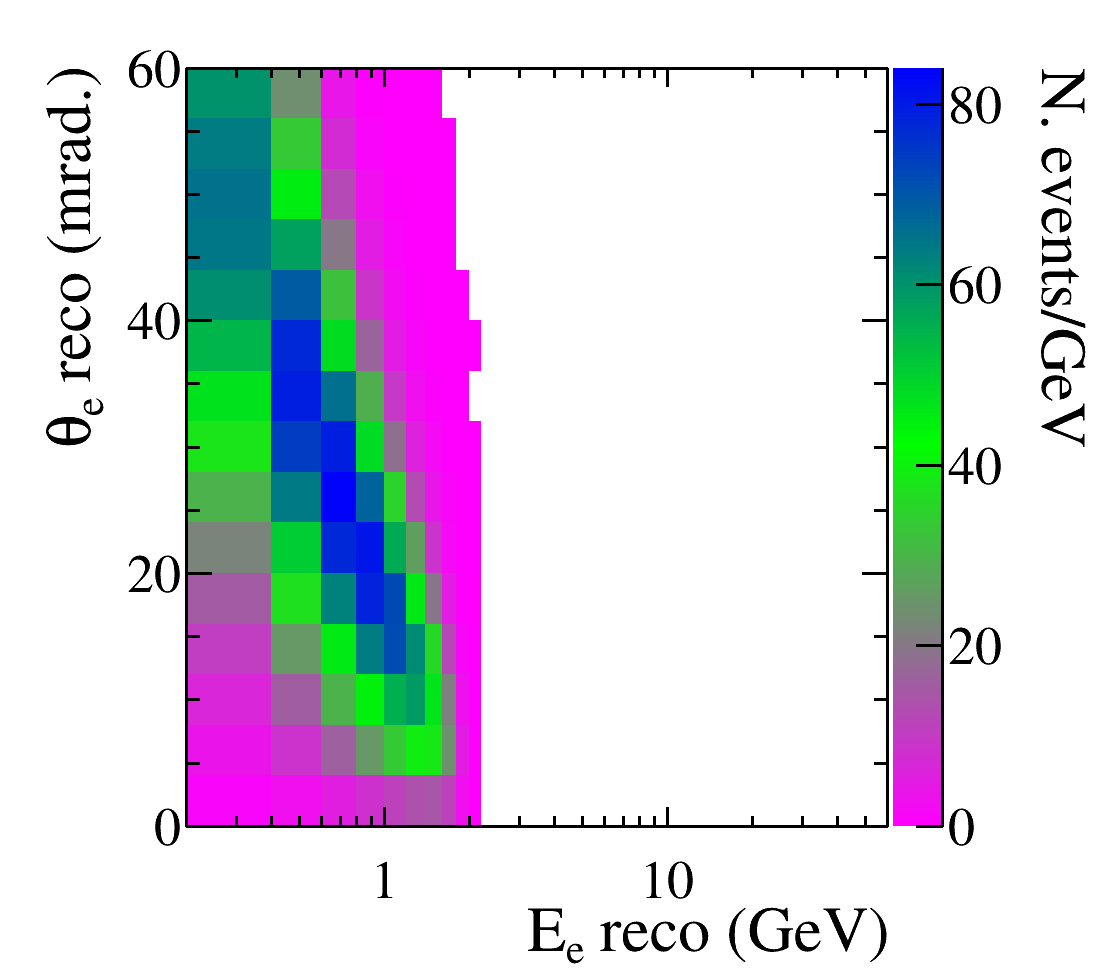}}
  \subfloat[$E_{2}$ (1.875--2.5 GeV)]    {\includegraphics[width=0.25\textwidth]{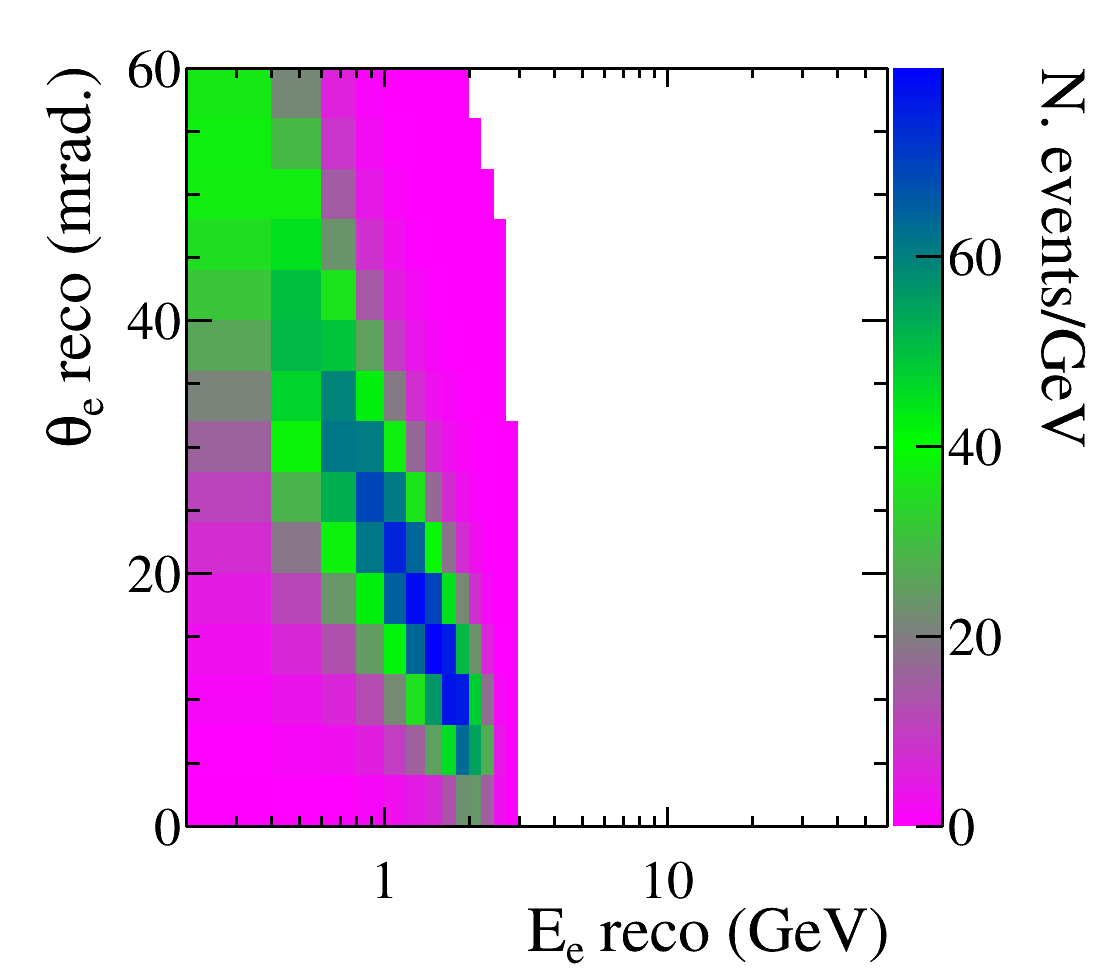}}
  \subfloat[$E_{3}$ (2.5--3.125 GeV)]    {\includegraphics[width=0.25\textwidth]{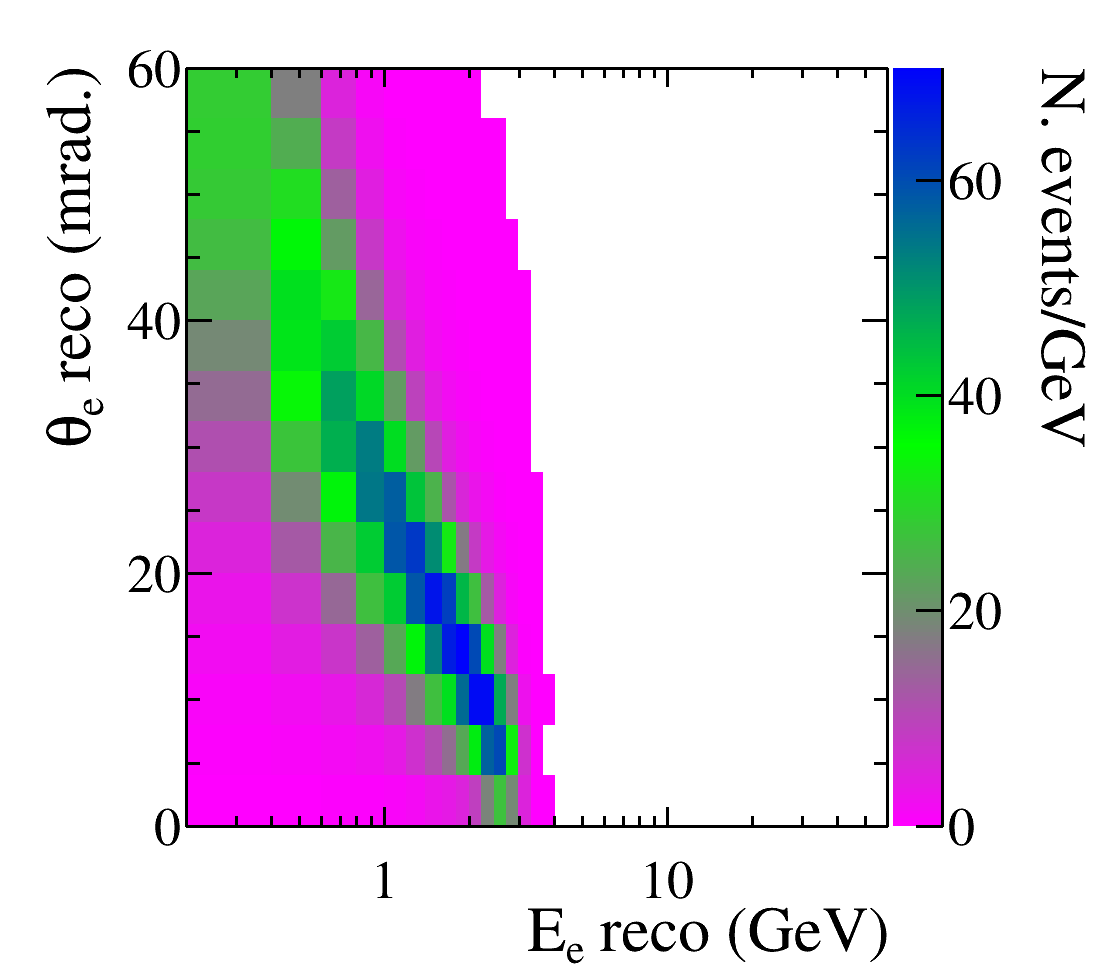}}
  \subfloat[$E_{4}$ (3.125--3.875 GeV)]  {\includegraphics[width=0.25\textwidth]{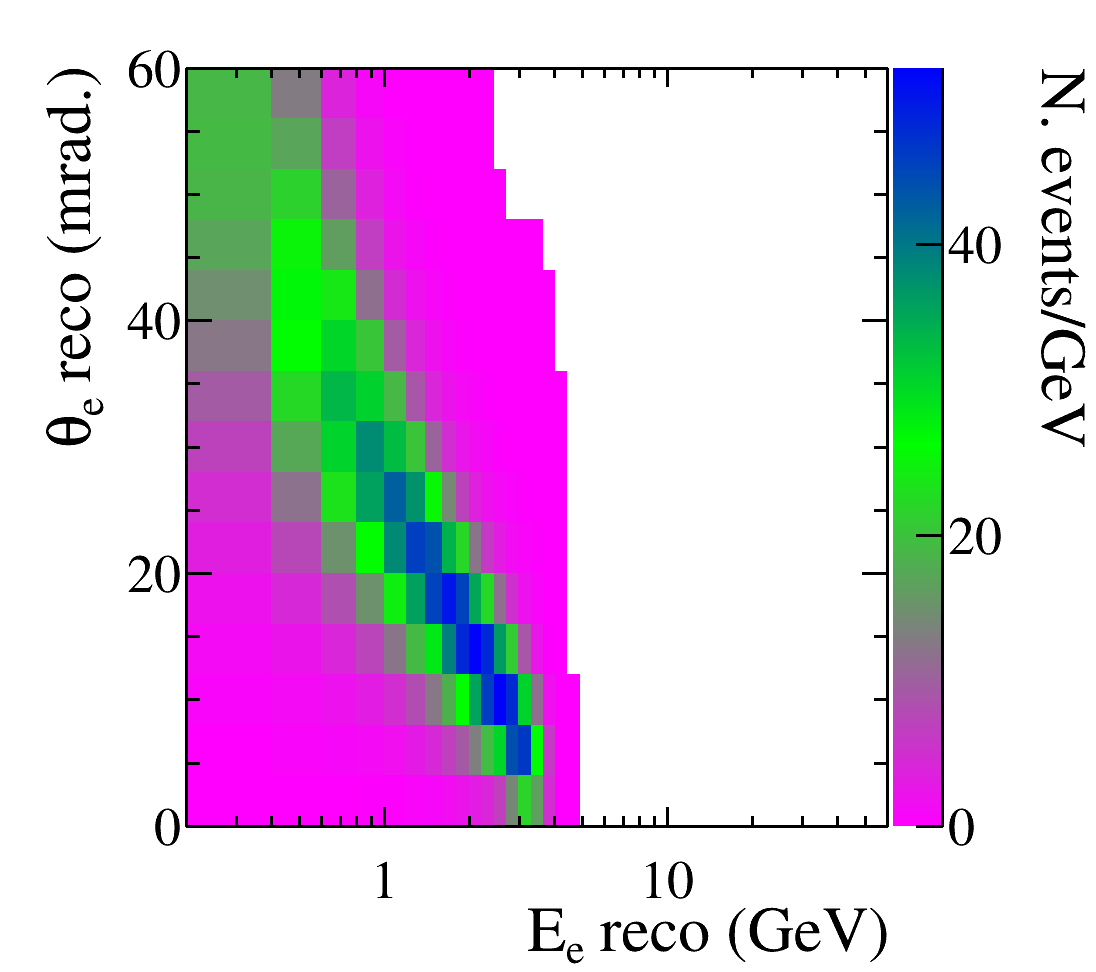}}\\
  \subfloat[$E_{5}$ (3.875--5.875 GeV)]  {\includegraphics[width=0.25\textwidth]{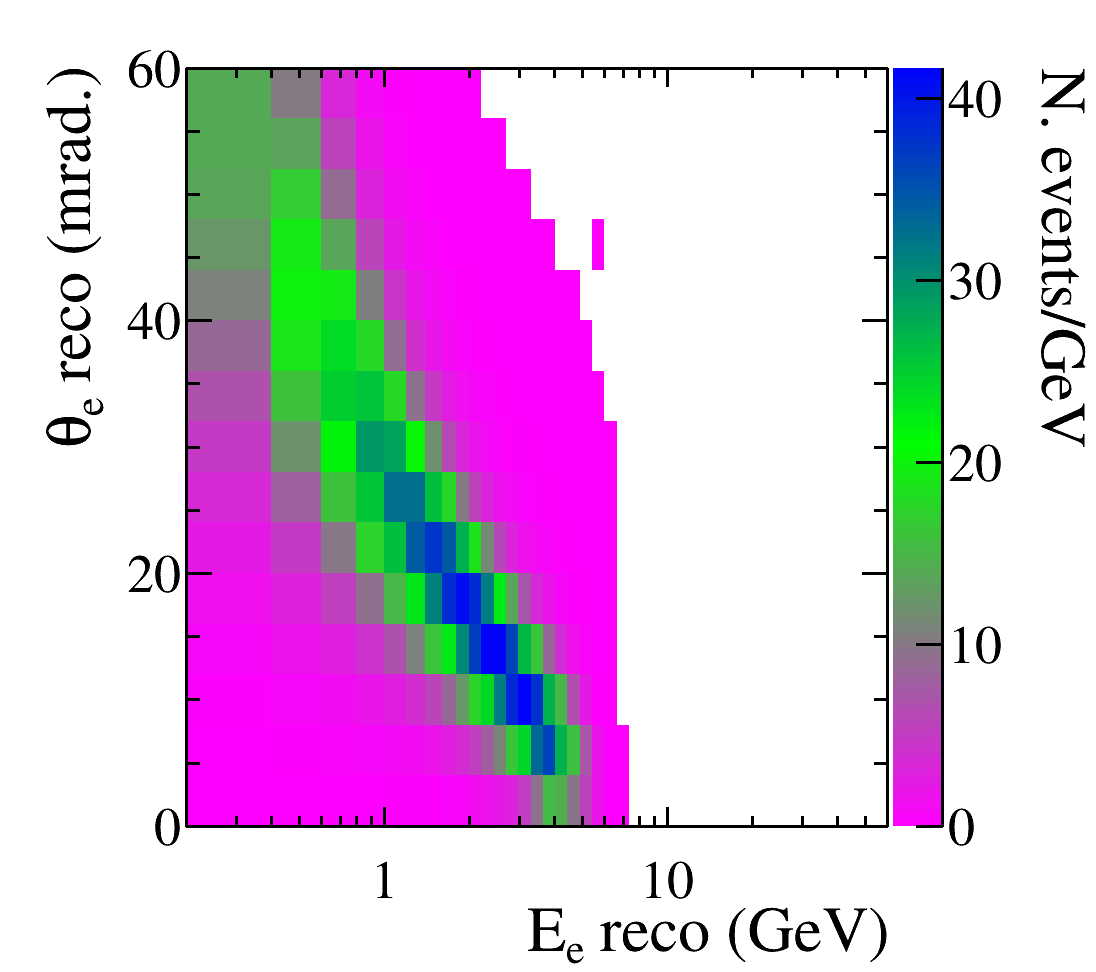}}
  \subfloat[$E_{6}$ (5.875--100 GeV)]    {\includegraphics[width=0.25\textwidth]{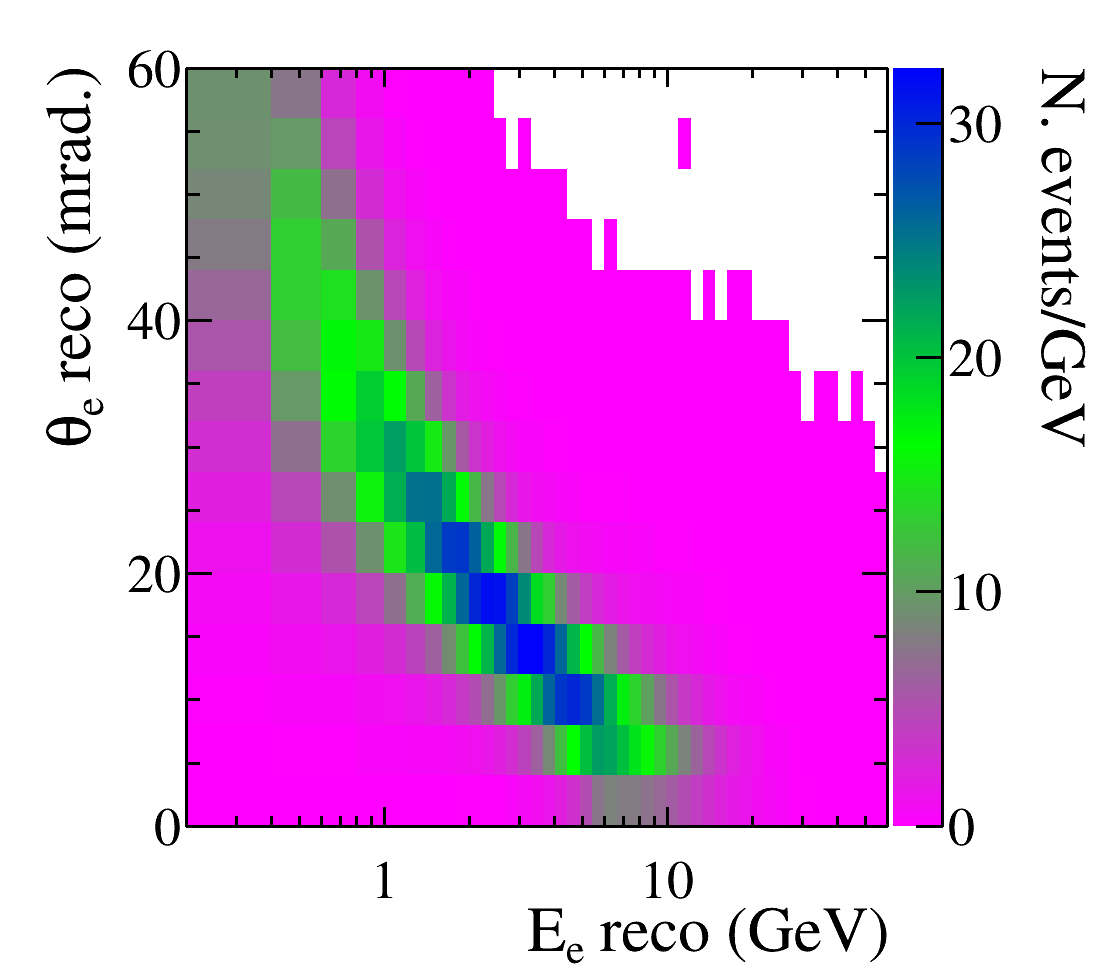}}
  \subfloat[$\nu_{e}$]                   {\includegraphics[width=0.25\textwidth]{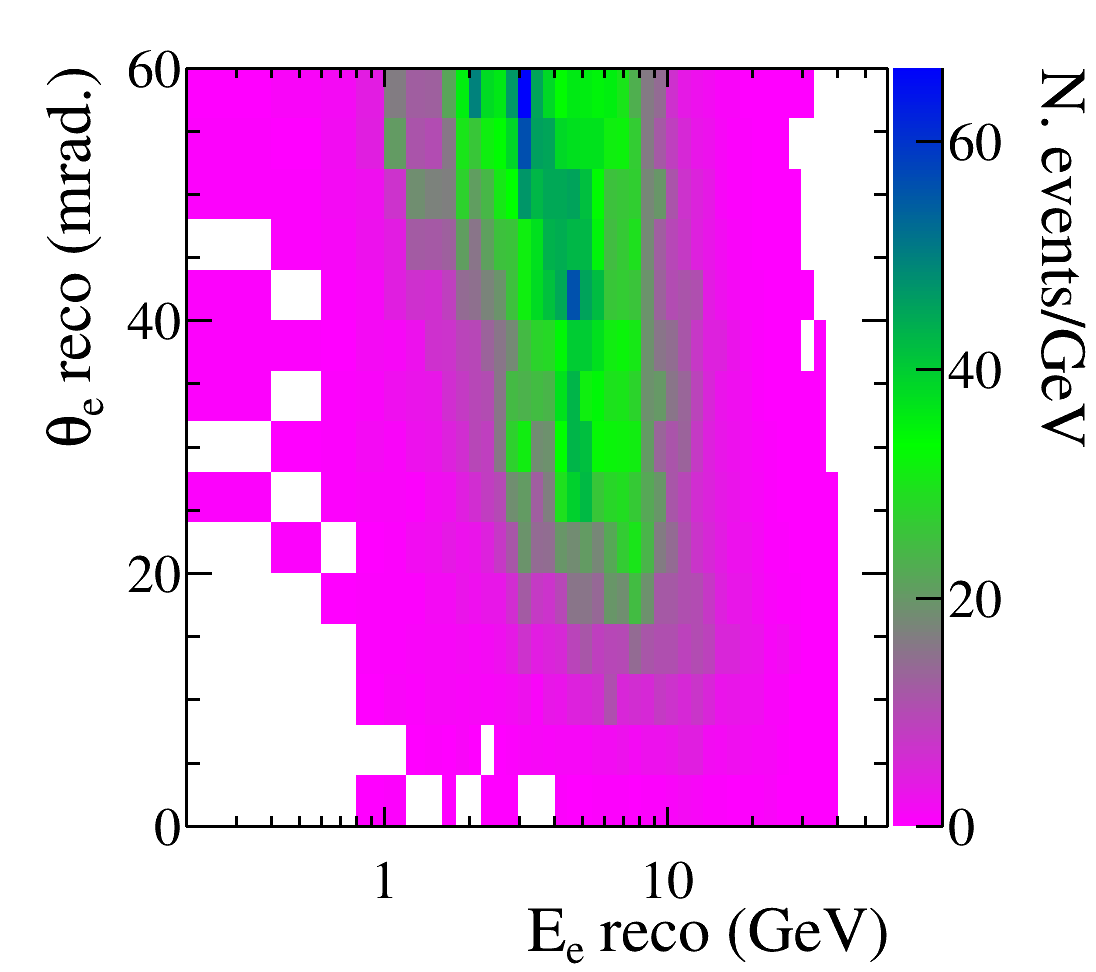}}
  \subfloat[$\gamma$]                   {\includegraphics[width=0.25\textwidth]{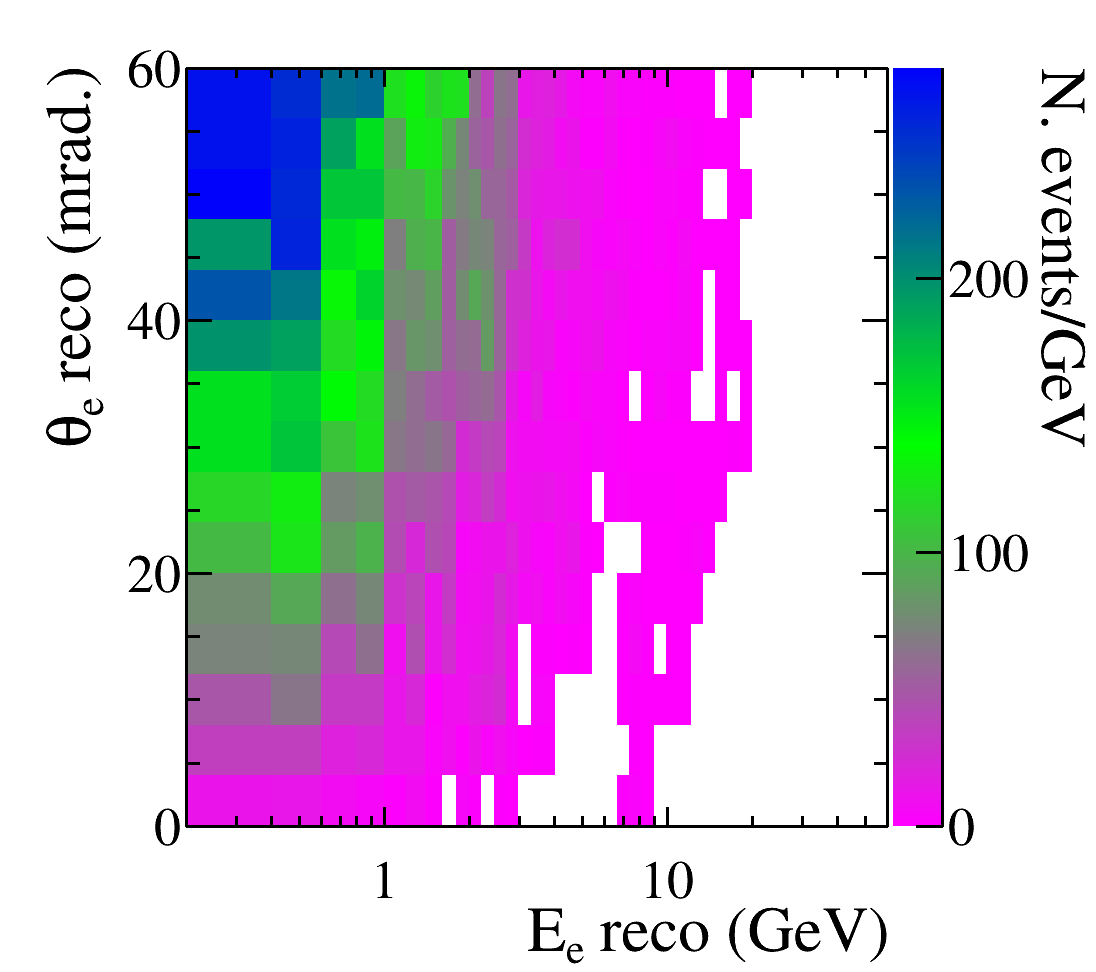}}
\fi
  \caption{Example fit templates for neutrino-electron elastic scattering in six bins of true neutrino energy, and for two background categories, for the 5t CH detector in FHC mode. Each template shows the expected event spectrum as a function of electron energy and angle, for neutrinos in a given energy range.}
  \label{fig:example_templates}
\end{figure*}

\begin{figure}[tbp]
  \centering
\ifnotoverleaf
  \subfloat[Thrown simulated data]  {\includegraphics[width=0.33\textwidth]{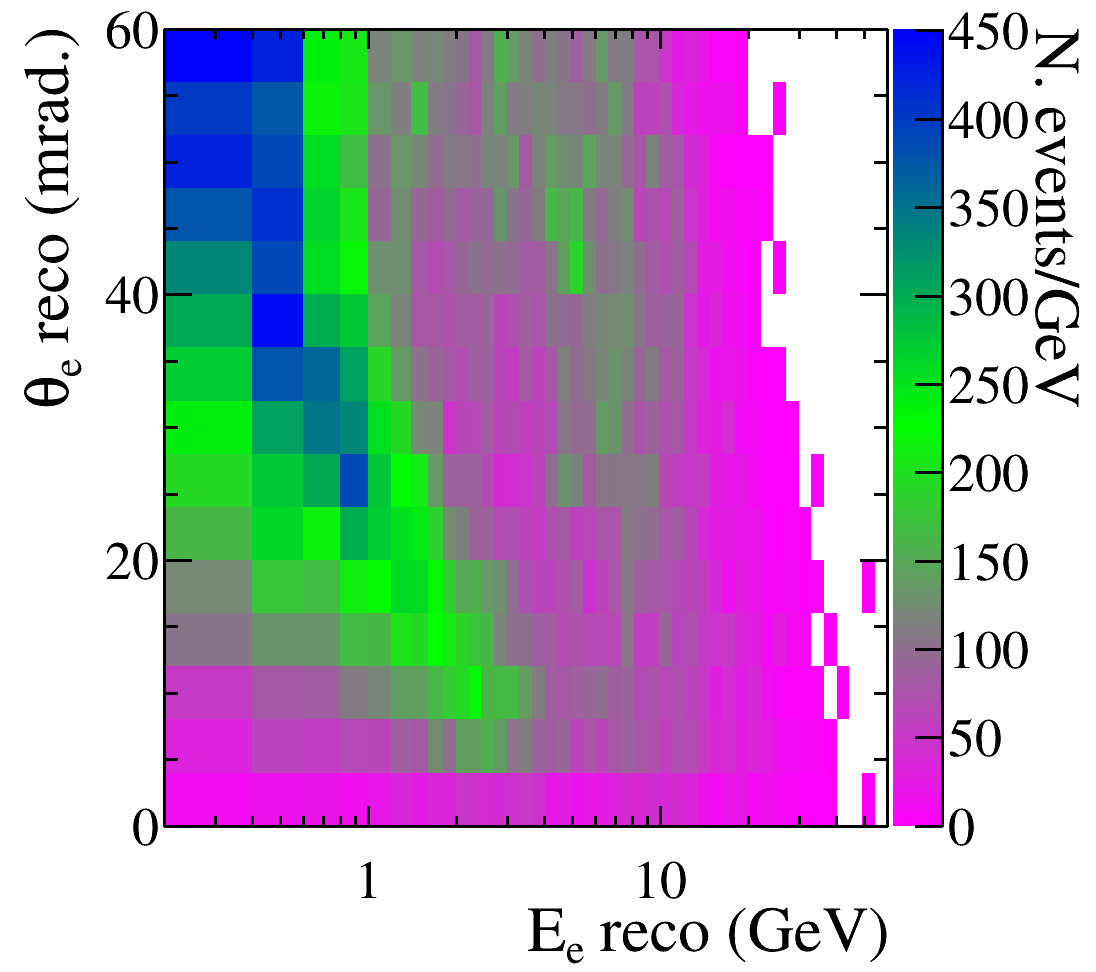}}\\
  \subfloat[Best fit]               {\includegraphics[width=0.33\textwidth]{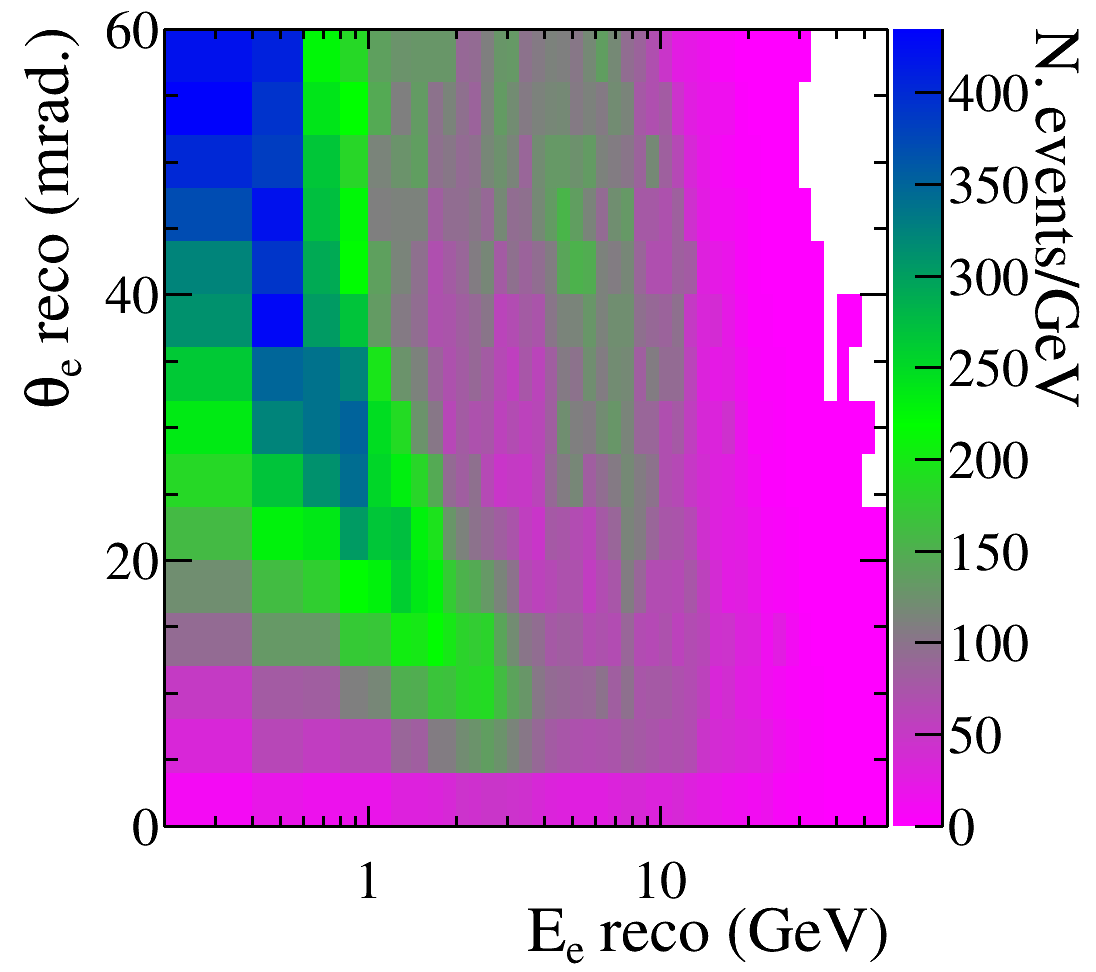}}\\
  \subfloat[Thrown - fit]           {\includegraphics[width=0.33\textwidth]{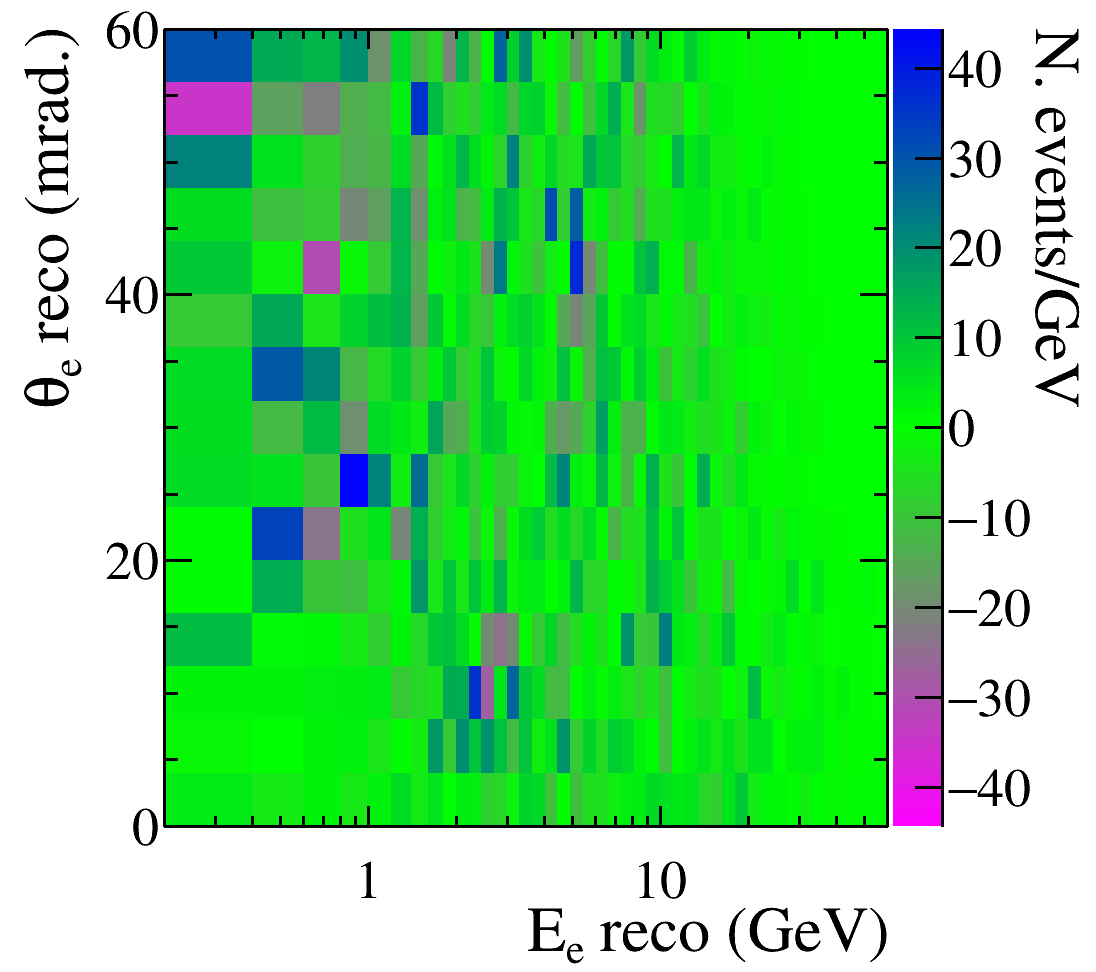}}
  \fi
  \caption{Example fit to simulated nominal FHC data in the LAr TPC detector with a single ``statistical throw'', as it is described in the text, showing the prefit, postfit, and residual event rates.}
  \label{fig:example_fit}
\end{figure}
\begin{figure}[tbp]
  \centering
\ifnotoverleaf
\subfloat[FHC]  {\includegraphics[width=0.4\textwidth]{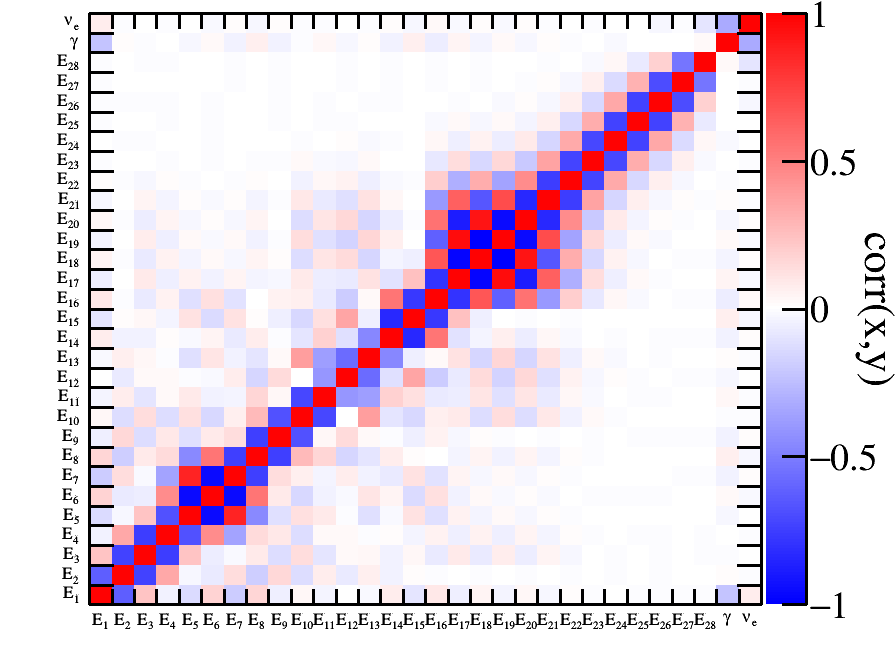}}\\
  \subfloat[RHC]  {\includegraphics[width=0.4\textwidth]{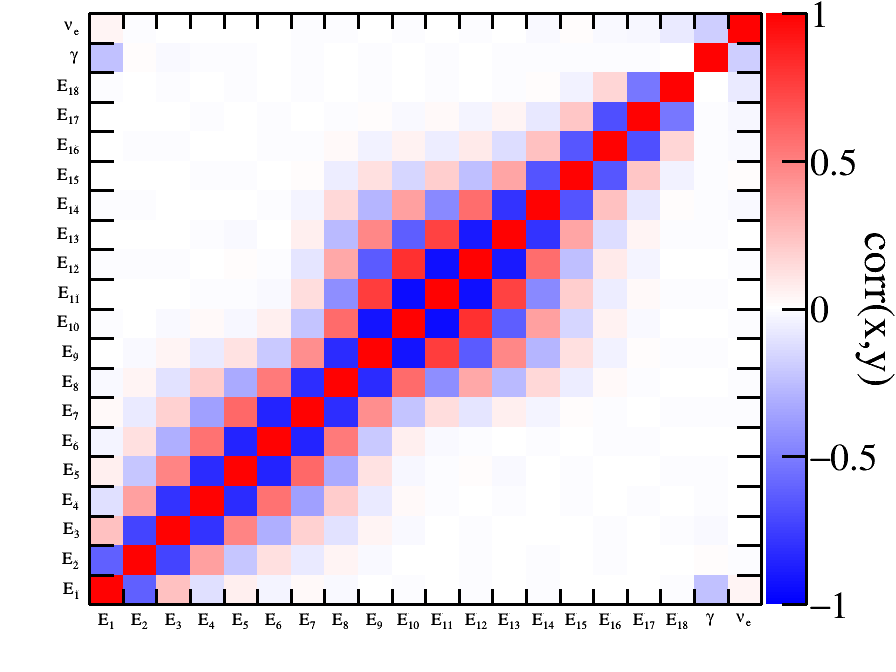}}
\fi
  \caption{Postfit correlation matrices between fit parameters shown for simulated LAr TPC fits with statistical variations about the nominal model in both FHC and RHC.}
  \label{fig:example_covar}
\end{figure}

We use a simple template fitting approach, to fit the simulated data with Monte Carlo, as we would for real data. Each template is binned in the same reconstructed $\theta_{e}$-$E_{e}$ bins as the simulated data, and is integrated over a true $E_{\nu}$ range. By varying the normalizations of these templates to the fit to the simulated data, a constraint on true $E_{\nu}$ can be extracted. The $E_{\nu}$ binning is chosen by merging DUNE flux bins such that each merged bin contains a minimum of 500 events, to ensure that the template normalization parameters can be approximated with a Gaussian. The binning for each of the ND scenarios for FHC and RHC is summarized in Table~\ref{tab:binning}. Note that each template gives a 2D reconstructed \eerec--\therec distribution, which has been integrated over the nominal flux distribution between the $E_{\nu}^{\mathrm{min}}$ and $E_{\nu}^{\mathrm{max}}$ boundaries of the template, and over all neutrino flavors (as the data cannot distinguish them), again using the nominal relative fractions of each flavor given by the nominal neutrino flux. These two necessary assumptions are a potential source of bias in the analysis, which will be discussed in detail later. By using an adaptive binning based on the expected statistics, we balance the impact of this bias with the statistical error. Two additional templates are included for the $\nu_e$ and $\gamma$ backgrounds, the normalizations for which are also unconstrained parameters in the fit. Example templates are shown in Figure~\ref{fig:example_templates} for the 5t CH detector in FHC.

In the fit, we use the ``L-BFGS-B'' algorithm~\cite{Zhu:1997:ALF:279232.279236}, from the SciPy optimize package~\cite{scipy} to minimize the Poisson-Likelihood:
\begin{equation}
  \chi^{2} = 2\sum^{N}_{i=1}\left[ \mu_{i}(\vec{\mathbf{x}}) - n_{i} + n_{i}\ln\frac{n_{i}}{\mu_{i}(\vec{\mathbf{x}})}\right],
\end{equation}
\noindent where $\mu_{i}(\vec{\mathbf{x}})$ is the MC prediction, which is a function of the template normalizations, $\vec{\mathbf{x}}$, and $n_{i}$ is the number of data events in the $i$th bin. We exclude bins with $E_{e} < 0.5$ GeV from the fit to account for a detector threshold. Modifying the value chosen for the detector threshold had a minimal effect on the fit because there are relatively few events in the very forward, low $E_{e}$ region (which corresponds to very low $E_{\nu}$), and the lowest $E_{\nu}$ template extends well past the threshold in all cases.

An example fit, using the nominal LAr design, where a ``statistical throw'' has been performed on the simulated data, is shown in Figure~\ref{fig:example_fit}. In each bin, the number of events is drawn randomly from a Poisson distribution; this acts as a very basic sanity check for the fitter. The fitted \eerec--\therec distribution approximates the simulated data well. In Figure~\ref{fig:example_covar}, we show the output correlation matrix from this fit. The ``checkerboard'' pattern is striking, neighboring bins are strongly anticorrelated, which indicates that the neighboring templates have a very similar effect on the fitted distribution. This is not a problem; indeed, using such fine $E_{\nu}$ binning maximizes the flux constraining power of the fit, and minimizes the potential for bias. However, it does mean that the postfit distributions of individual fits are difficult to interpret by-eye. Figure~\ref{fig:example_covar} also shows that the correlations are small between the signal template parameters, $E_{i}$, and the two background templates, labeled $\gamma$ and $\nu_e$. This indicates that the fit is able to distinguish signal and background very well, which is unsurprising given the different regions of \eerec--\therec space they occupy, as shown in Figure~\ref{fig:example_templates}. The independence of the signal and background templates will be checked more quantitatively with bias studies in Sections~\ref{sec:bias} and~\ref{sec:systematic_uncertainties}.

Although the postfit distribution is difficult to interpret by-eye, it can be used to constrain the flux by assigning weights to possible fluxes. The probability of a possible flux being consistent with the measured $\nu$--$e^{-}$ data can be calculated using~\cite{Park:2015eqa}
\begin{align}
      P(\vec{N}|\vec{M}) = &\frac{1}{(2\pi)^{\frac{\kappa}{2}}}\frac{1}{|\Sigma_{\mathrm{N}}|^{\frac{1}{2}}}   \nonumber \\
       & \times \exp \left[\frac{1}{2}(\vec{N}
      -\vec{M})^{T}\Sigma^{-1}_{\mathrm{N}}(\vec{N}-\vec{M}) \right],
  \label{eq:probability}
\end{align}
\noindent where $\Sigma_{\mathrm{N}}$ is the data covariance (see the correlation matrix in Figure~\ref{fig:example_covar}), $|\Sigma_{\mathrm{N}}|$ is its determinant, $\vec{N}$ is the postfit $E_{\nu}$ template normalizations, $\vec{M}$ is the model rebinned to match the template binning, and $\kappa$ is the number of $E_{\nu}$ templates.

The probability calculated in Equation~\ref{eq:probability} can be used to constrain the flux covariance matrix provided by the beam group, $\xi_{ij}$, to show the impact of the $\nu$--$e^-$ constraint. The postfit covariance matrix $\Xi_{ij}$ can be calculated
\begin{equation}
  \Xi_{ij} = \frac{1}{N_{k}} \Sigma_{k}\left[ P(\vec{N}|\vec{M})_{k} \left(M_{ik} - \overline{M}_{i}\right)\left(M_{jk} - \overline{M}_{j}\right)\right],
  \label{eq:postfit_covar}
\end{equation}
\noindent using $k$ throws of the original matrix, where the weighted average in the $i$th bin is $\overline{M}_{i} = 1/N_{k} \left[\Sigma_{k}  P(\vec{N}|\vec{M})_{k} M_{ik}\right]$. A comparison of the pre- and post-fit covariance matrices can be used to investigate how well the $\nu$--$e^{-}$ sample can constrain the flux. The vector $\overline{M}_{i}$ give the central values of the post-fit, and deviations of these from the true value are a useful measure of bias in this procedure. \blue{Note that because the LBNF beam is a mix of different flavors (see Figure~\ref{fig:flux}), all with different $\nu$--$e^{-}$ cross sections, it is not possible to constrain the flux completely independently of an input flux model, as some assumptions have to be made about the relative contributions from each flavor, if not their spectra. The technique described in this work could be used to constrain any flux model, but we choose to show the additional constraint which can be applied to the Geant4-based DUNE flux simulation, as it is the most sophisticated set of assumptions about the relative contributions from each flavor that we have available.}

\subsection{Results}
\label{sec:results}

\begin{figure*}[tbp]
  \centering
\ifnotoverleaf
  \subfloat[FHC pre-fit]  {\includegraphics[width=0.45\textwidth]{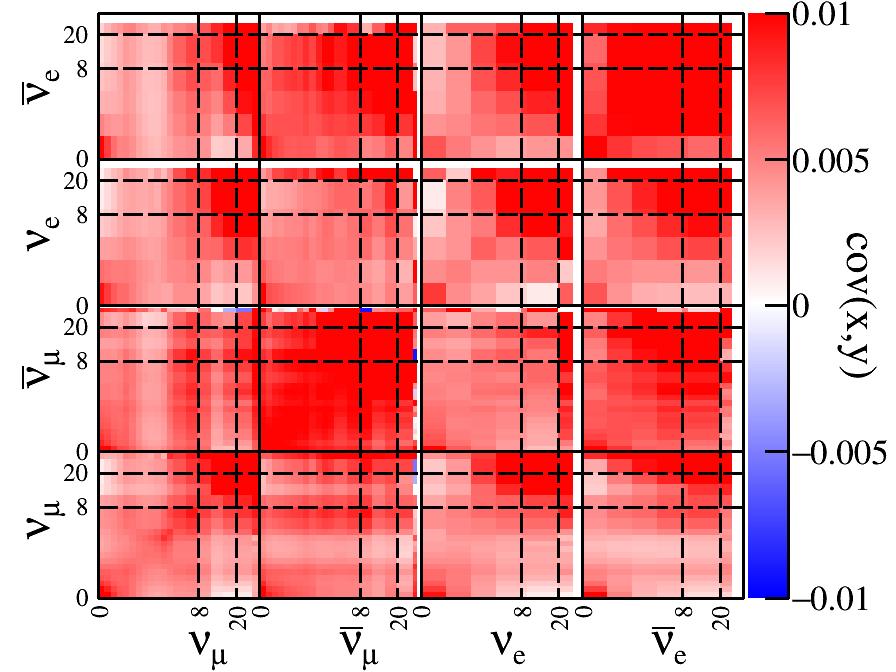}}
  \subfloat[FHC post-fit] {\includegraphics[width=0.45\textwidth]{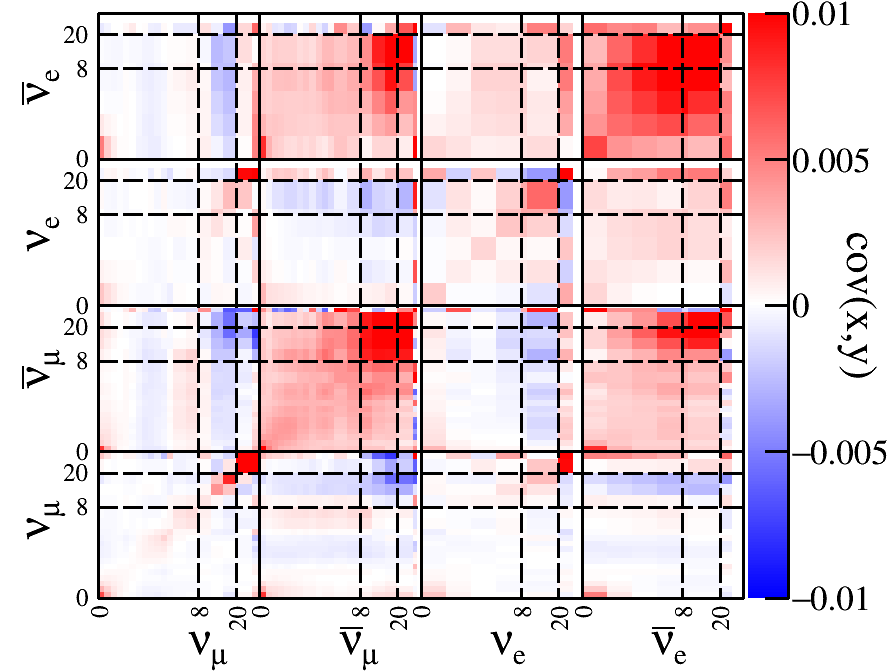}}\\
  \subfloat[RHC pre-fit]  {\includegraphics[width=0.45\textwidth]{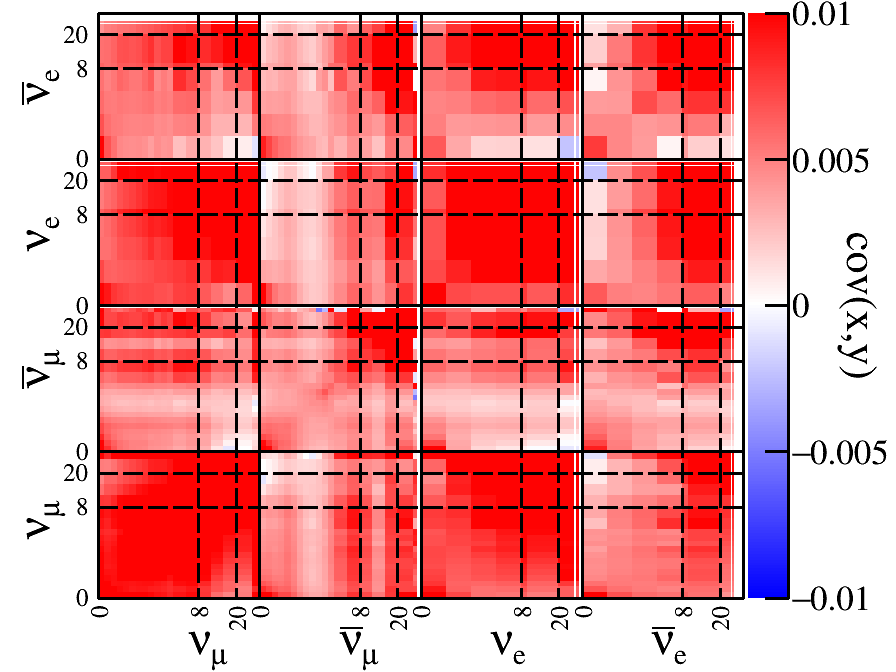}}
  \subfloat[RHC post-fit] {\includegraphics[width=0.45\textwidth]{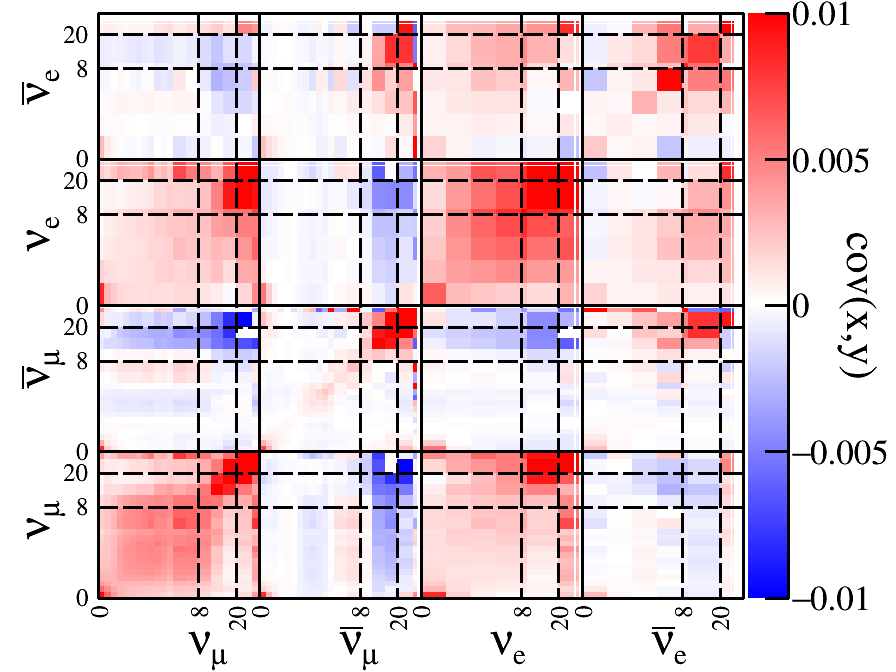}}
\fi
\caption{Pre- and post-fit FHC and RHC flux covariance matrices for the nominal LAr detector configuration.}
  \label{fig:LAR_nominal_covariances}
\end{figure*}

Using the output data covariance from each fit (see the LAr example correlation matrices in Figure~\ref{fig:example_covar}), and Equations~\ref{eq:probability} and~\ref{eq:postfit_covar}, the constraint on the DUNE flux prediction for each beam and detector configuration can be calculated. In each case, we constrain the full flux covariance, as shown for the nominal LAr configuration for both FHC and RHC in Figure~\ref{fig:LAR_nominal_covariances}, along with the pre-fit covariance matrices provided by the beam group for comparison. It is clear that the uncertainties are much smaller for the $\nu_{\mu}$ and $\nu_{e}$ ($\bar{\nu}_{\mu}$ and $\bar{\nu}_{e}$) in FHC (RHC) after the fit. The correlations between flavors has also been decreased significantly, although anticorrelations are introduced because the $\nu$--$e^{-}$ sample cannot distinguish flavors, so decreasing the contribution from one flavor can be increased by increasing the contribution from another. Note that the relationship between flavors is already limited by the input beam covariance matrix. The relationship between flavors is more complicated for the RHC case, with stronger correlations obvious in the postfit covariance shown in Figure~\ref{fig:LAR_nominal_covariances}. Although this is expected by the larger wrong-sign contamination in RHC relative to FHC, the fitting technique described here seems to work for both.

\begin{figure*}[tbp]
  \centering
\ifnotoverleaf
  \subfloat[FHC]  {\includegraphics[width=0.95\textwidth]{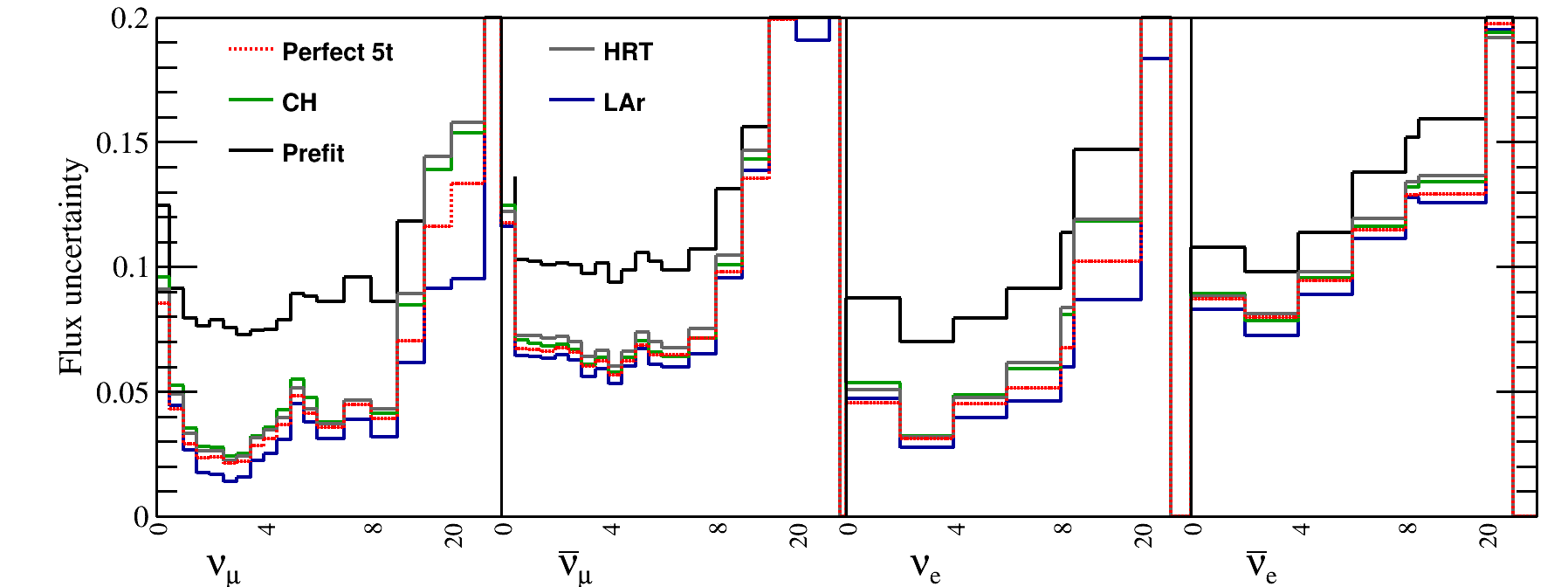}}\\
  \subfloat[RHC]  {\includegraphics[width=0.95\textwidth]{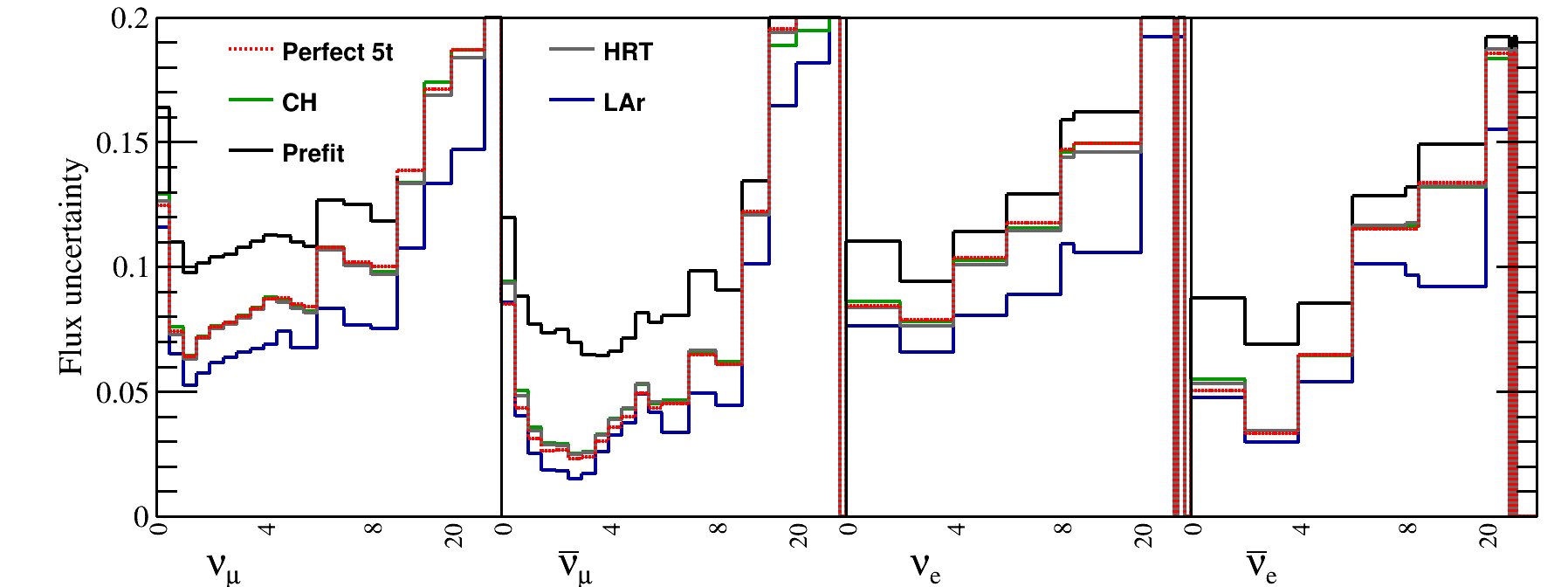}}
\fi
\caption{Bin-by-bin flux uncertainties as a function of neutrino energy and flavor in both FHC and RHC, shown for the pre-fit and post-fit for all detector configurations considered in this work. These are correspond to the square-root of the diagonal elements of the flux covariance matrices, examples of which are shown in Figure~\ref{fig:LAR_nominal_covariances}.}
  \label{fig:nominal_det_constraint}
\end{figure*}
Although interesting, the covariance matrices shown in Figure~\ref{fig:LAR_nominal_covariances} are difficult to interpret by eye, so for the remainder of this work, only the diagonal elements of the covariance will be considered, although we note that the full covariance is calculated each time when producing these plots. The diagonal elements of the covariance matrices are shown for all detector configurations, for both FHC and RHC in Figure~\ref{fig:nominal_det_constraint}, with the diagonal elements of the pre-fit flux covariances for comparison. It can be seen that for both FHC and RHC modes, the uncertainty on the dominant flavor ($\nu_{\mu}$ and $\bar{\nu}_{\mu}$ respectively) in the flux peak is reduced from $\sim$8\% to $\sim$1--3\%, depending on the detector configuration used.

\begin{figure}[tbp]
  \centering
  \ifnotoverleaf
  \subfloat[FHC]  {\includegraphics[width=0.45\textwidth]{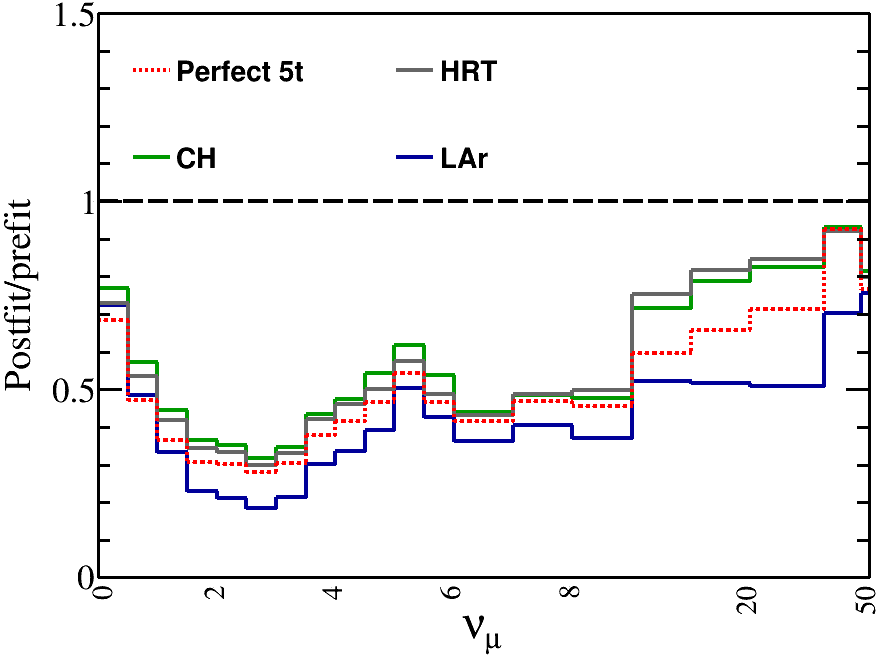}}\\
  \subfloat[RHC]  {\includegraphics[width=0.45\textwidth]{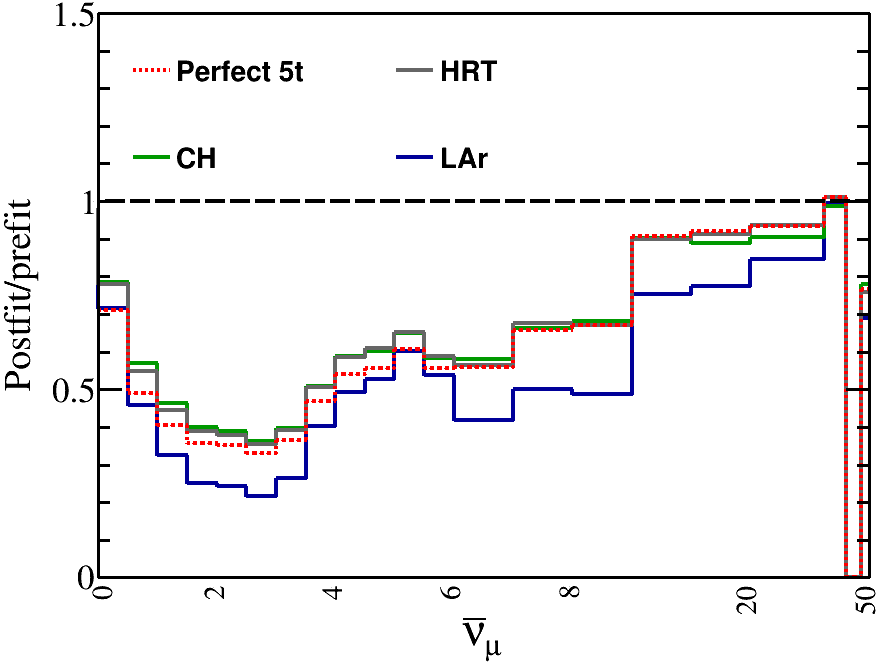}}
  \fi
  \caption{Bin-by-bin flux uncertainties as a function of neutrino energy for $\nu_{\mu}$ (FHC) and $\bar{\nu}_{\mu}$ (RHC), shown as a ratio with the pre-fit flux uncertainty for all detector configurations considered in this work.}
  \label{fig:nominal_det_constraint_ratio}
\end{figure}
In order to make it easier to compare the different detectors considered, a ratio is taken with respect to the diagonals of the nominal beam covariance matrix, in Figure~\ref{fig:nominal_det_constraint_ratio}, and only the dominant $\nu_{\mu}$ ($\bar{\nu}_{\mu}$) flavor contributions are shown for FHC (RHC) as they are most interesting. \blue{It is clear from Figure~\ref{fig:nominal_det_constraint_ratio} that the flux constraining power of the 5t detectors is significantly weaker than for the 30t LAr detector, indicating that the improved reconstruction performance of the CH or HRT does not add significant strength to the flux constraining power of the analysis, except perhaps in the lowest energy bins well below the flux peak. This is also true for the perfect 5t detector, so is not simply a consequence of our assumptions about the HRT/CH performance, it seems that statistics are paramount for this analysis. As expected, the HRT does better than the CH detector, although this may be partially due to the higher electron density in CH$_2$ than CH.} 

\begin{figure}[tbp]
  \centering
  \ifnotoverleaf
  \subfloat[FHC]  {\includegraphics[width=0.45\textwidth]{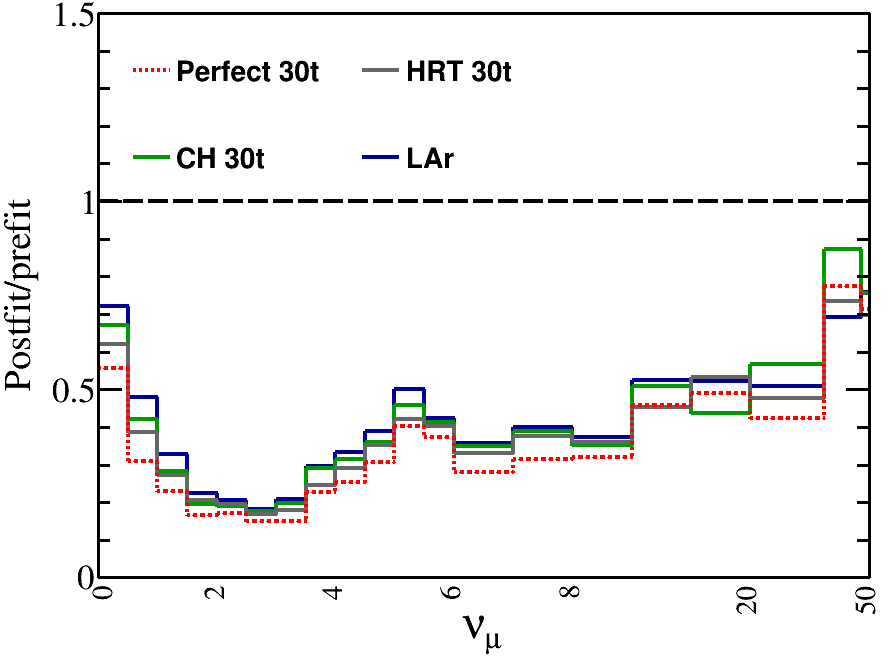}}\\
  \subfloat[RHC]  {\includegraphics[width=0.45\textwidth]{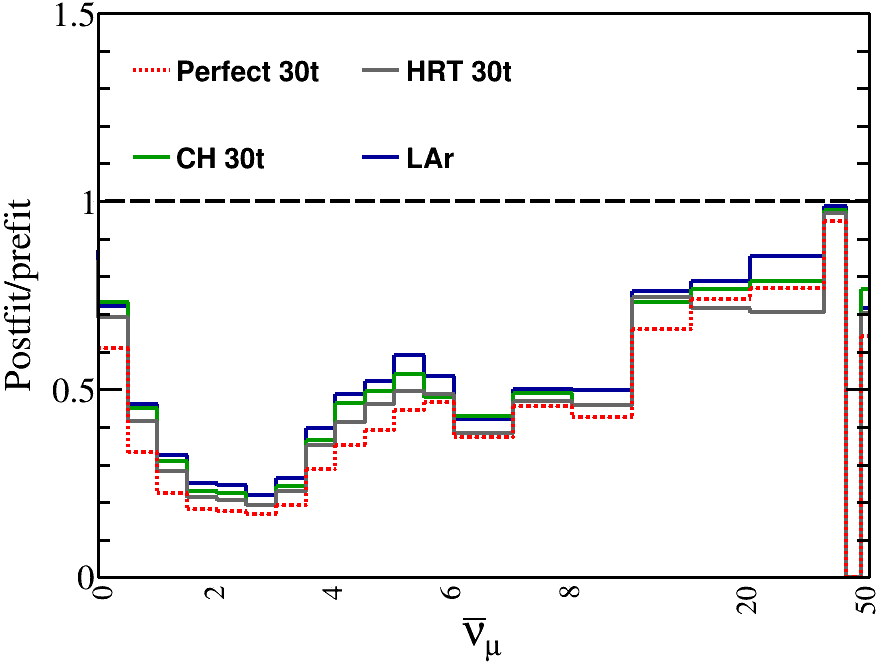}}
  \fi
  \caption{Bin-by-bin flux uncertainties as a function of neutrino energy for $\nu_{\mu}$ (FHC) and $\bar{\nu}_{\mu}$ (RHC), shown as a ratio with the pre-fit flux uncertainty for all equal mass (30t) detector configurations considered in this work.}
  \label{fig:nominal_det30_constraint_ratio}
\end{figure}
\blue{Figure~\ref{fig:nominal_det30_constraint_ratio} is the same as Figure~\ref{fig:nominal_det_constraint_ratio}, but with 30t versions of all detector technologies. As expected, with equal masses, the LAr detector performs the least well, which is due to the lower electron density, and worse resolution than the other detector options, but the improvements to the flux constraint seen for the other detector options are fairly small compared with LAr, which supports the conclusion that the most important factor is the statistics gained with larger masses. That said, at a certain point higher statistics would not help, as the intrinsic divergence of the beam would become the limiting factor for the analysis.  We have not determined at which LAr or perfect detector mass the limit of statistical improvement occurs in the present study.}

\begin{figure}[tbp]
  \centering
  \ifnotoverleaf
  \subfloat[FHC]  {\includegraphics[width=0.45\textwidth]{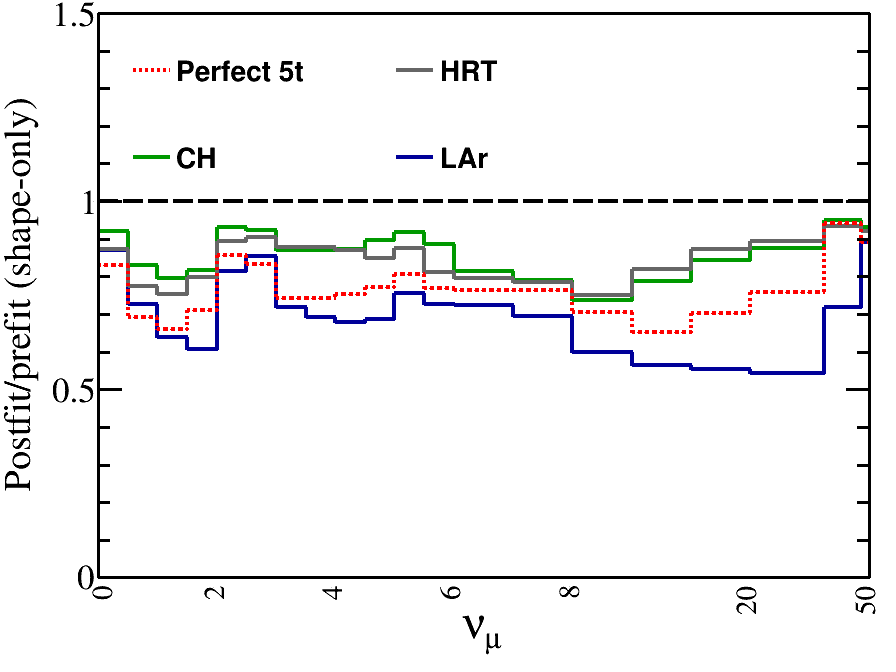}}\\
  \subfloat[RHC]  {\includegraphics[width=0.45\textwidth]{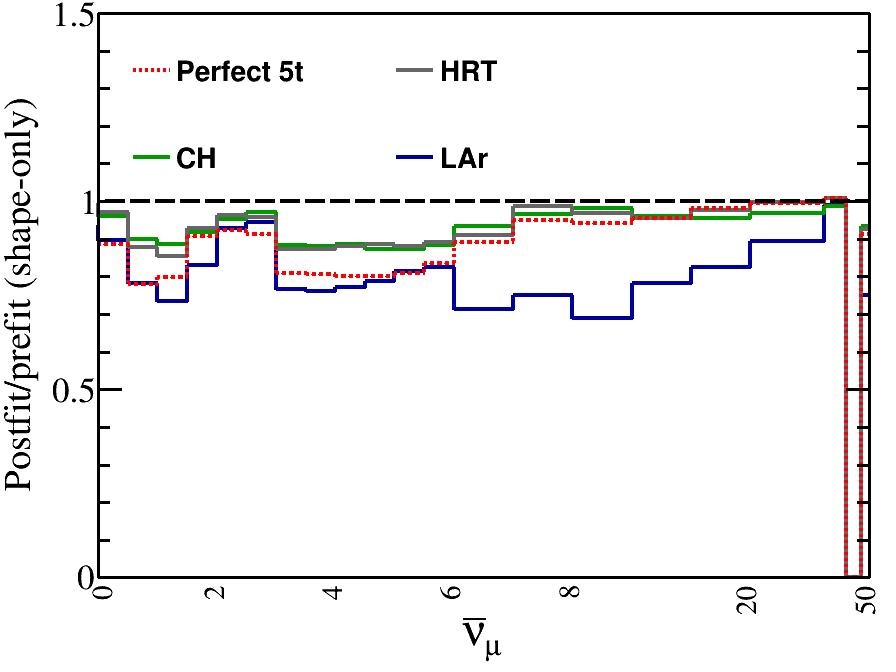}}
  \fi
  \caption{Bin-by-bin shape-only flux uncertainties as a function of neutrino energy for $\nu_{\mu}$ (FHC) and $\bar{\nu}_{\mu}$ (RHC), shown as a ratio with the pre-fit flux uncertainty for all detector configurations considered in this work.}
  \label{fig:nominal_det_constraint_ratio_shape}
\end{figure}
By making a shape-only version of the pre- and post-fit flux covariance matrices, by normalizing each flux throw such that the integral is the same, and forming the postfit flux covariance with Equation~\ref{eq:postfit_covar}, and then taking a ratio as in Figure~\ref{fig:nominal_det_constraint_ratio}, the ability for the $\nu$--$e^-$ sample to improve the flux shape can be investigated. Such a plot is shown in Figure~\ref{fig:nominal_det_constraint_ratio_shape}, from which it is clear that there is only a marginal improvement in the understanding of the flux shape relative to the input beam covariance matrix. The shape-only uncertainties are still $\sim$70\% ($\sim$80\%) of the nominal shape uncertainties for FHC (RHC). It seems that the power to constrain the flux normalization is largely responsible for the improvements seen in Figures~\ref{fig:nominal_det_constraint} and~\ref{fig:nominal_det_constraint_ratio}.

\begin{figure}[tbp]
  \centering
  \ifnotoverleaf
  \subfloat[FHC]  {\includegraphics[width=0.45\textwidth]{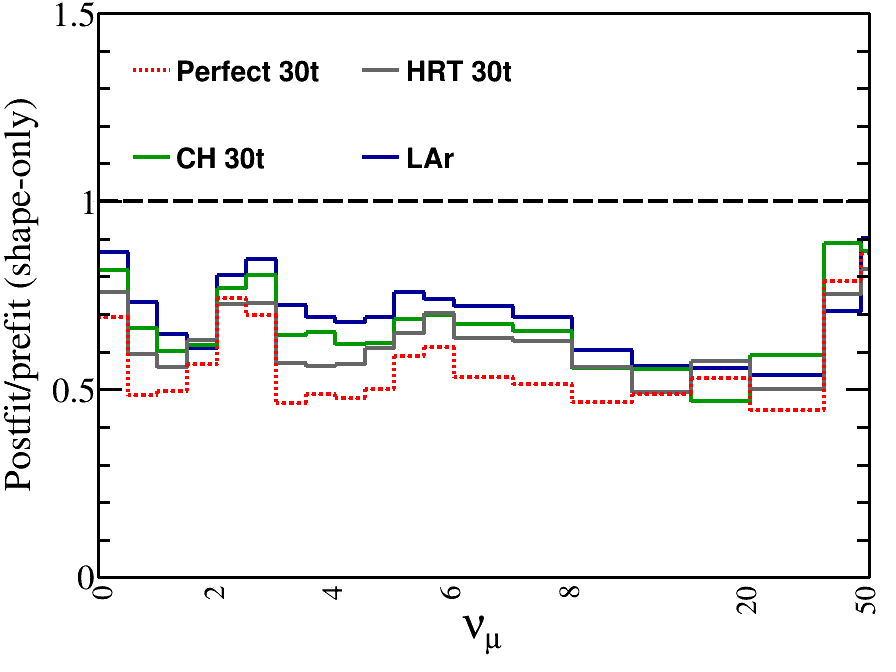}}\\
  \subfloat[RHC]  {\includegraphics[width=0.45\textwidth]{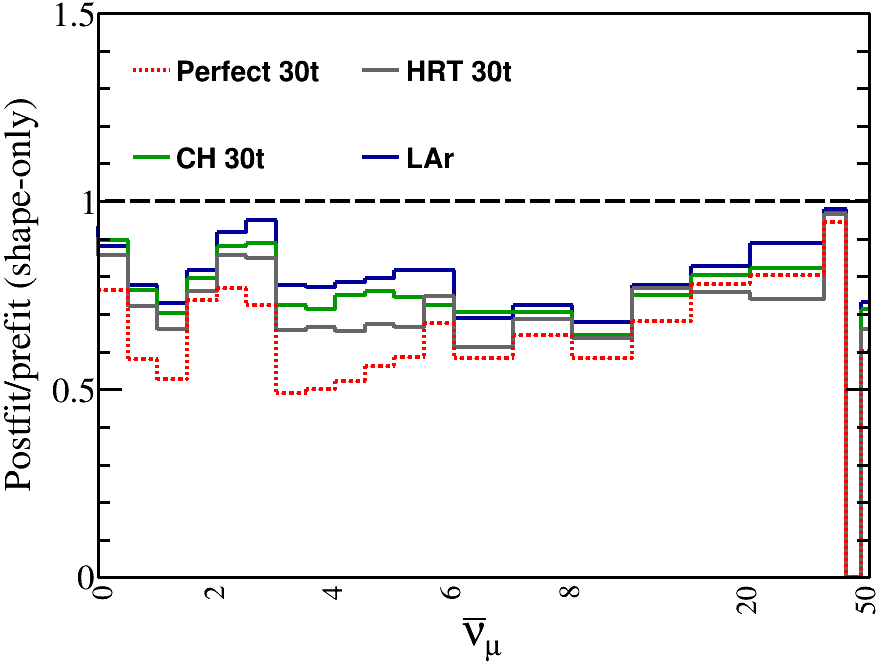}}
  \fi
  \caption{Bin-by-bin shape-only flux uncertainties as a function of neutrino energy for $\nu_{\mu}$ (FHC) and $\bar{\nu}_{\mu}$ (RHC), shown as a ratio with the pre-fit flux uncertainty for all equal mass (30t) detector configurations considered in this work.}
  \label{fig:nominal_det30_constraint_ratio_shape}
\end{figure}
\blue{Improved detector resolutions would be expected to have a larger impact on the shape constraint than on the total flux normalization, and indeed, we can see from Figure~\ref{fig:nominal_det_constraint_ratio_shape} that the 5t perfect detector is nearly comparable to the performance of the 30t LAr detector. In the equal mass case, shown in Figure~\ref{fig:nominal_det30_constraint_ratio_shape}, we see that the 30t better resolution detectors do substantially better than the LAr, but that even for these large detectors, the shape uncertainties are still 50\% of their nominal.}

\begin{figure*}[tbp]
  \centering
  \ifnotoverleaf
  \subfloat[FHC]  {\includegraphics[width=0.95\textwidth]{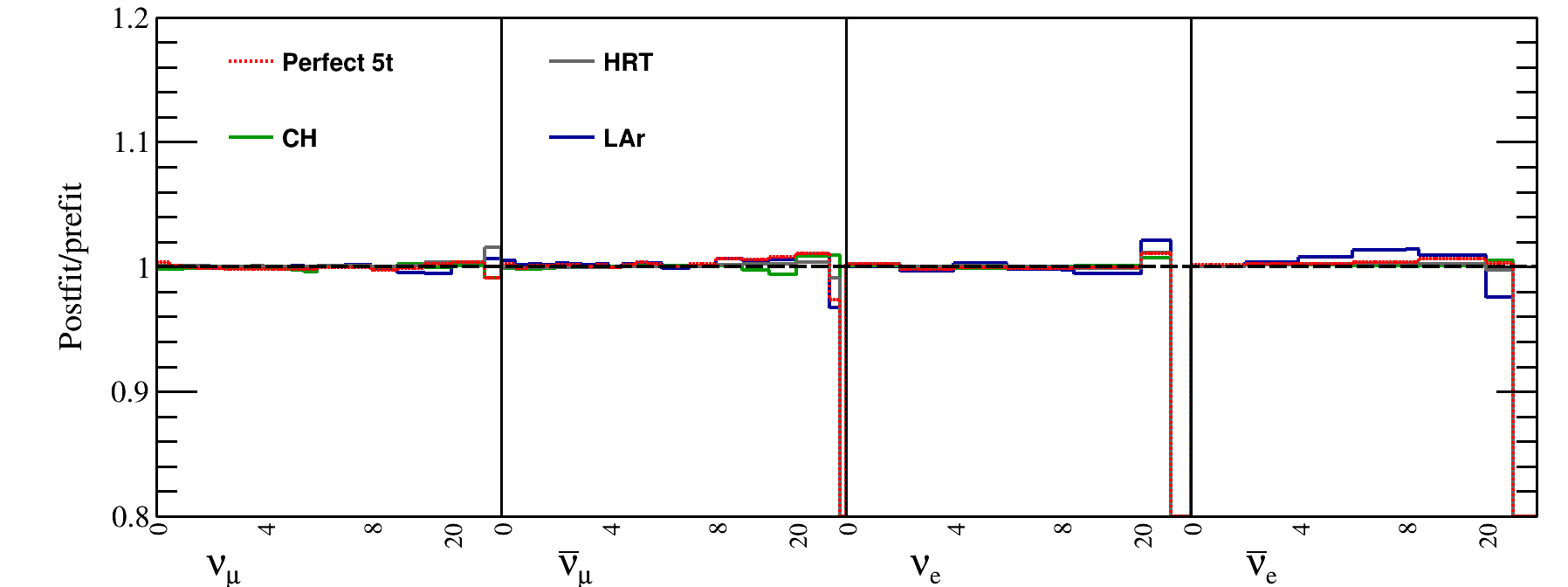}}\\
  \subfloat[RHC]  {\includegraphics[width=0.95\textwidth]{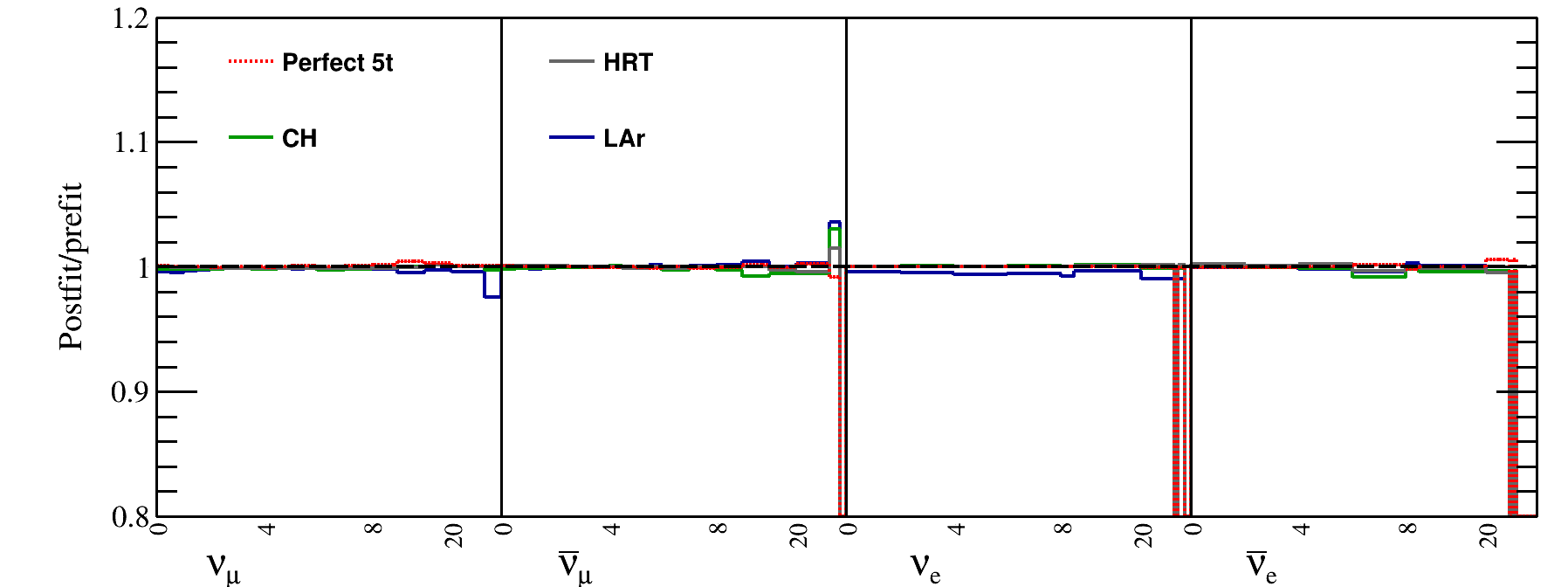}}
  \fi
  \caption{Weighted average flux values as a function of neutrino energy and flavor in both FHC and RHC, shown as a ratio with the pre-fit flux for all detector configurations considered in this work.}
  \label{fig:nominal_det_central}
\end{figure*}
In addition to the postfit covariance, another interesting quantity is the weighted average flux values, $\overline{M}_{i}$, obtained when calculating the covariance matrix in Equation~\ref{eq:postfit_covar}. These are shown for all detector configurations considered in both FHC and RHC modes in Figure~\ref{fig:nominal_det_central}. Large deviations from the pre-fit flux would indicated a bias in some $E_{\nu}$ region; no such deviations are observed. More sophisticated studies of possible bias in the procedure are performed in Section~\ref{sec:bias}.

\subsection{Bias tests}
\label{sec:bias}
Some bias in the fit results is expected for two reasons. Firstly, the $E_{\nu}$ binning used in the fit is coarser than the binning of the flux covariance matrix, at least for some regions of $E_{\nu}$, so variations within those bins cannot be correctly handled by the fit and will introduce small biases. However, the $E_{\nu}$ binning is limited by the statistics of the $\nu$--$e^-$ scattering sample, so biases due to the coarse binning should be small relative to the statistical uncertainty in the fit. Secondly, the fit cannot distinguish between flavors, and all flavors contribute to the $\nu$--$e^-$ sample, but with different cross sections (see Section~\ref{sec:nue_scat_intro}). The fit therefore implicitly relies on the expected relationship between flavors, and throws of the flux covariance matrix which change this relationship cannot be dealt with perfectly by the fit. \blue{Indeed, one of the weaknesses inherit in using a $\nu$--$e^{-}$ sample as a flux constraint for an accelerator neutrino beamline is that assumptions have to be made about the relative contribution of each flavor}.

Although the weighted average flux values obtained when calculating the covariance matrix in Equation~\ref{eq:postfit_covar}, and shown for the nominal detector configurations in Figure~\ref{fig:nominal_det_central} should give an indication of any bias, and the results suggest that any bias is relatively small, the size of the bias can be calculated more quantitatively by modifying the ``data'' in the fits, and fitting with the nominal Monte Carlo simulation. If there is an implicit bias towards the input MC, then the fit will not reproduce these modified data distributions. The level of agreement, including uncertainties, between the best fit flux distribution and the input ``true'' distribution can be assessed with the test-statistic
\begin{equation}
  \chi^{2} = \sum_{i=0}^{N}\sum_{j=0}^{N} \left(\nu_i^{\mathrm{TRUE}} - \nu_i^{\mathrm{FIT}}\right) M_{ij}^{-1}  \left(\nu_j^{\mathrm{TRUE}} - \nu_j^{\mathrm{FIT}}\right)
  \label{eq:bias_chi2}
\end{equation}
\noindent where, the indices $i$ and $j$ are over the true $E_{\nu}$ bins used in the fit, and the matrix $M_{ij}$ is the post-fit covariance matrix between fit parameters (as in Figure~\ref{fig:example_covar}), with the background parameters removed.

\begin{figure}[tbp]
  \centering
  \ifnotoverleaf
  \subfloat[FHC]  {\includegraphics[width=0.45\textwidth]{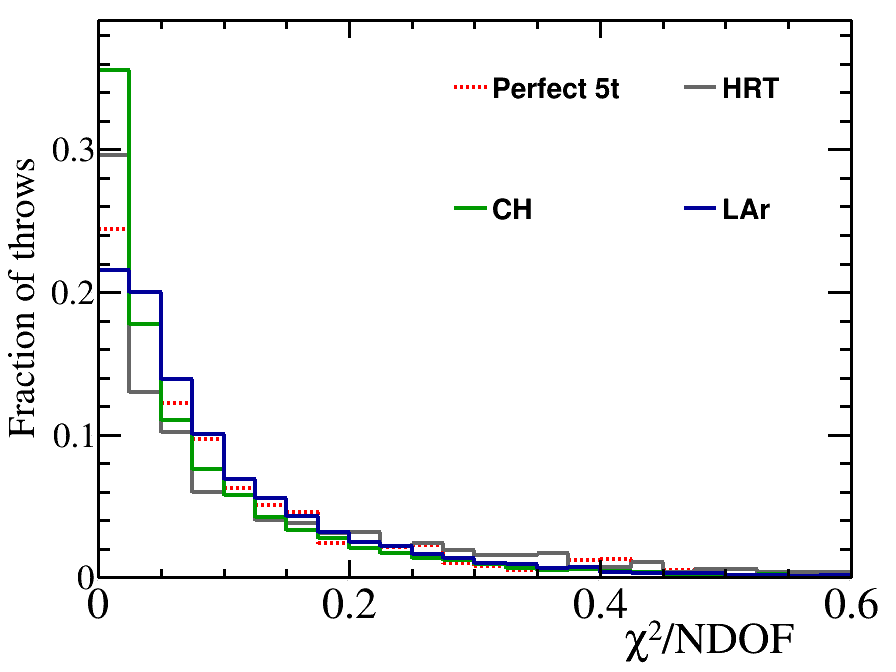}}\\
  \subfloat[RHC]  {\includegraphics[width=0.45\textwidth]{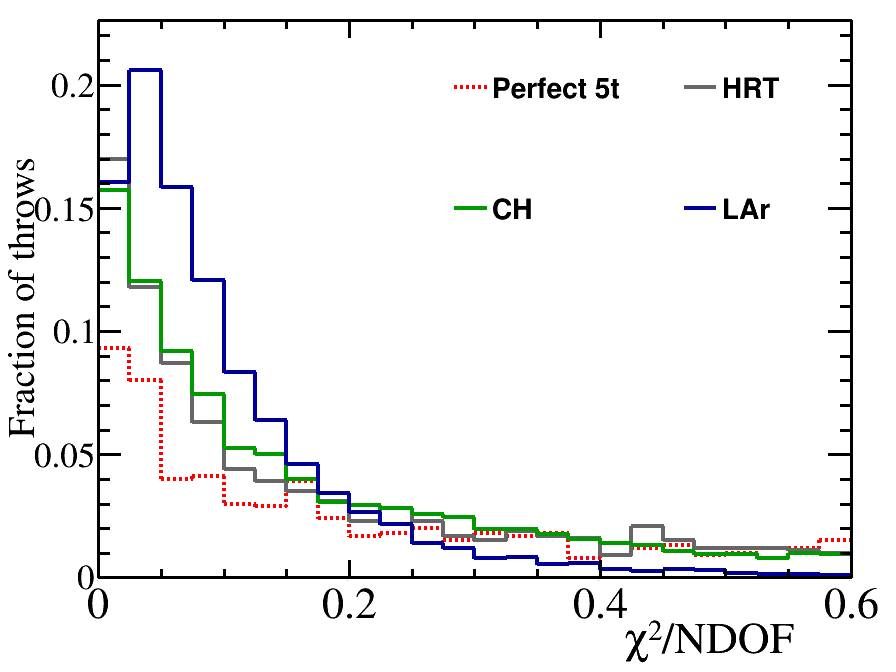}}
  \fi
  \caption{$\chi^{2}$/NDOF for the true and post-fit values and post-fit uncertainties for 10,000 independent throws of the input flux covariance matrix, shown for all detector configurations considered, in both FHC and RHC.}
  \label{fig:nominal_det_bias}
\end{figure}
To provide a meaningful point of reference, the bias in the fitting technique described in this work was assessed using throws of the input flux covariance matrix. Throws of that matrix, produced through Cholesky decomposition, describe the expected variation in the neutrino flux given all prior known information about the beam. Figure~\ref{fig:nominal_det_bias} shows the distribution of $\chi^{2}$-values, calculated with Equation~\ref{eq:bias_chi2}, obtained for 10,000 independent throws of the input covariance matrix. It is clear from Figure~\ref{fig:nominal_det_bias} that the biases seen for both FHC and RHC are indeed small for all detector configurations tested. \blue{Biases are larger in RHC than FHC, which is probably due the relative beam purities. The biases are larger (most noticeable in RHC) for the 5t detectors than the 30t LAr detector because of the small number of templates used in the fits (see Table~\ref{tab:binning}), which was confirmed by checking that the bias disappears for the 30t versions of those detectors.} Improvements might be seen by modifying the fit to not expect that fit parameters should be Gaussian, which would relax the 500 expected events per bin requirement which set the template binning. However, reducing the required number of events per bin to 300 did not significantly change the result, so any improvement to alleviate that small bias is outside the scope of this work. \blue{This bias can be understood as being due to the width of the templates, and the inability of the fit to deal with changes to the assumed flux distribution within each template.}

\begin{figure*}[tbp]
  \centering
  \ifnotoverleaf
  \subfloat[FHC pre-fit]  {\includegraphics[width=0.45\textwidth]{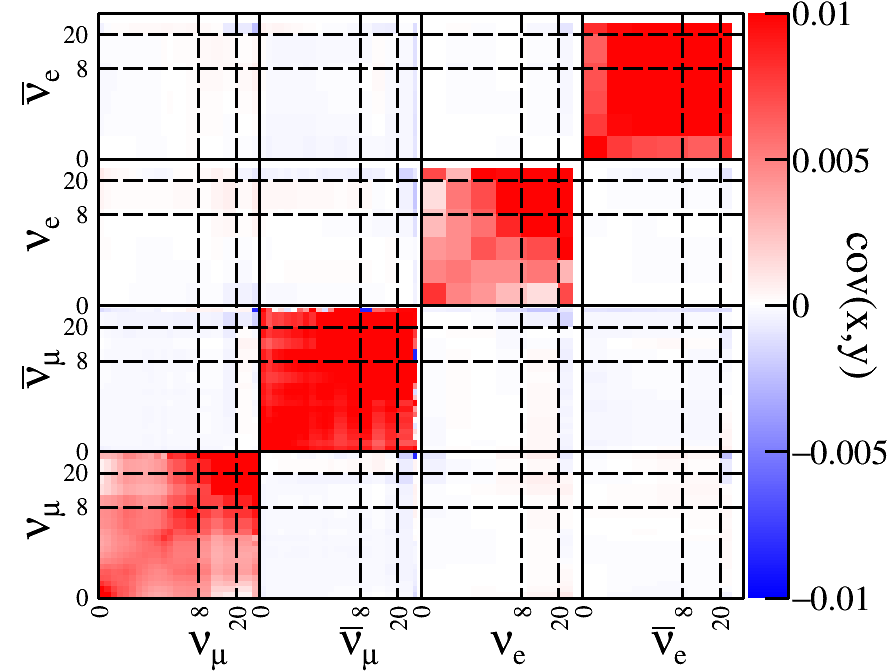}}
  \subfloat[FHC post-fit] {\includegraphics[width=0.45\textwidth]{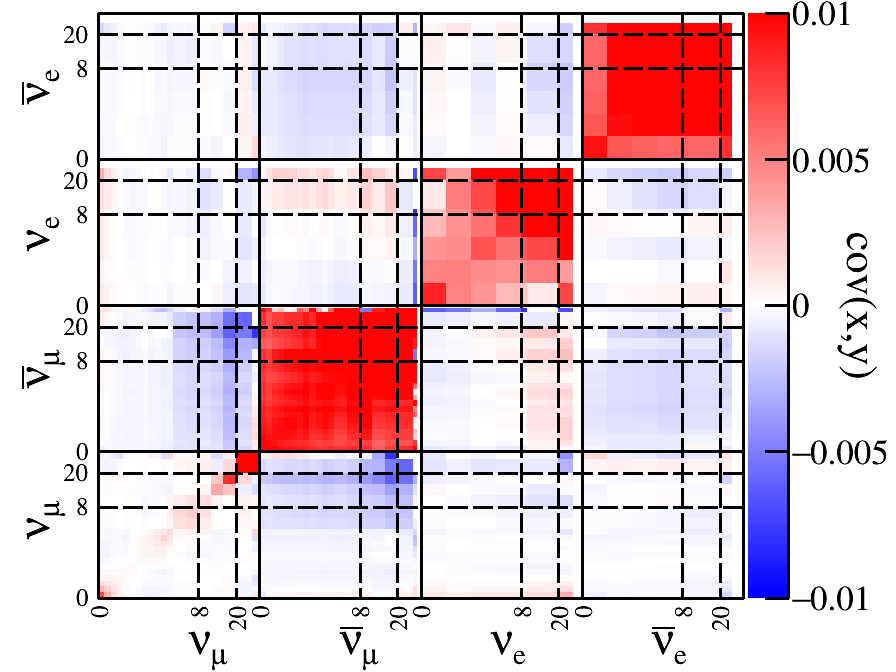}}\\
  \subfloat[RHC pre-fit]  {\includegraphics[width=0.45\textwidth]{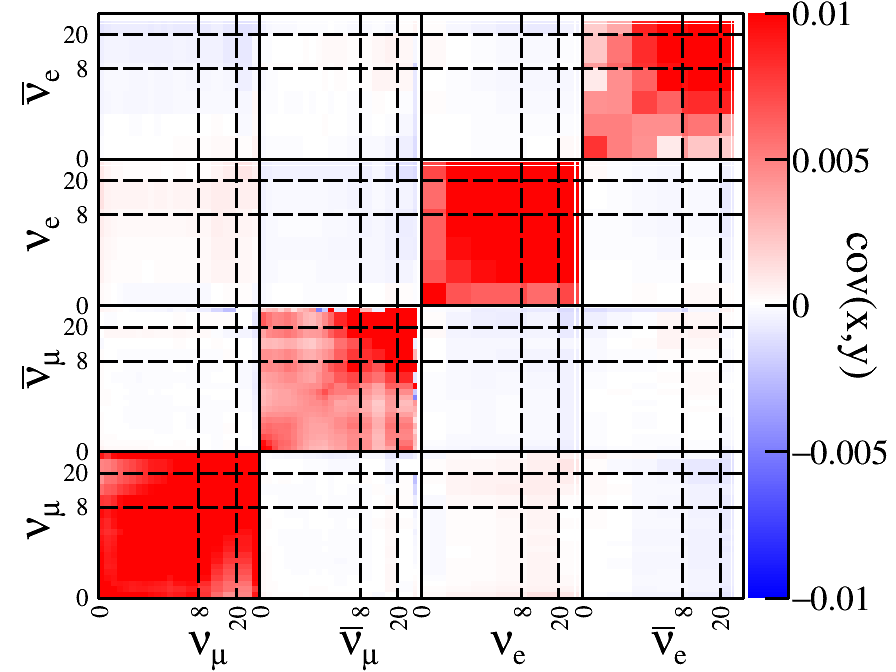}}
  \subfloat[RHC post-fit] {\includegraphics[width=0.45\textwidth]{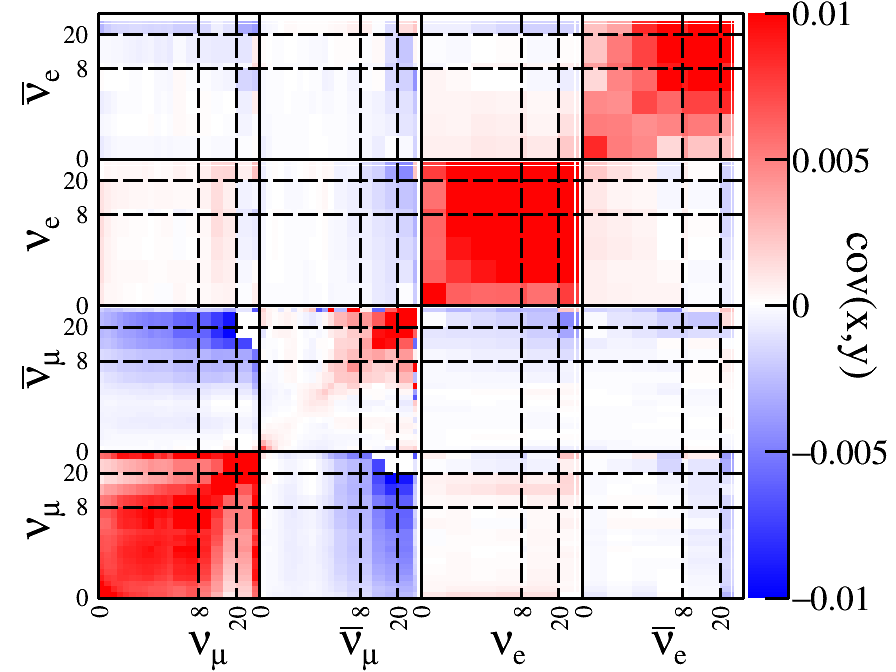}}
  \fi
  \caption{Pre- and post-fit FHC and RHC flux covariance matrices for the nominal LAr detector configuration with correlations between flavors removed.}
  \label{fig:LAR_noflavcorr_covariances}
\end{figure*}

\begin{figure}[tbp]
  \centering
  \ifnotoverleaf
  \subfloat[FHC]  {\includegraphics[width=0.45\textwidth]{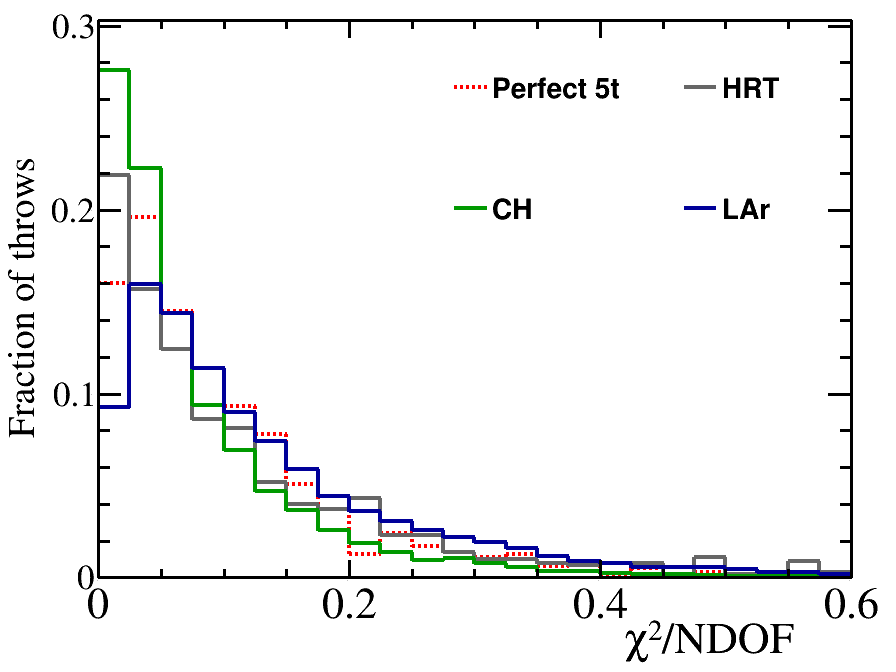}}\\
  \subfloat[RHC]  {\includegraphics[width=0.45\textwidth]{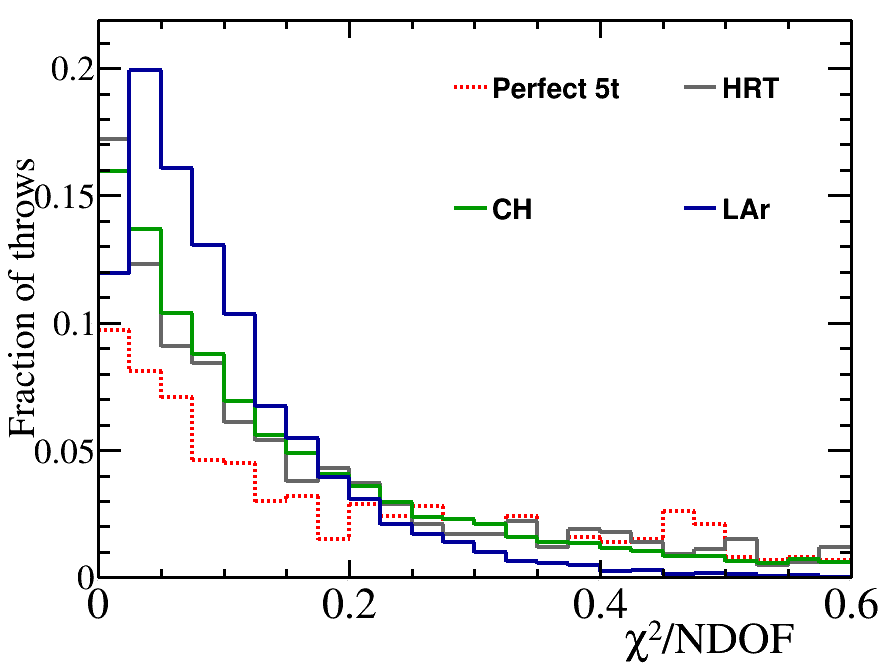}}
  \fi
  \caption{$\chi^{2}$/NDOF for the true and post-fit values and post-fit uncertainties for 10,000 independent throws of the input flux covariance matrix without correlations between flavors, shown for all detector configurations considered, in both FHC and RHC.}
  \label{fig:noflavcorr_det_bias}
\end{figure}
In order to test how sensitive the fit is to changes in relative contribution from each flavor, the post-fit covariance matrix and bias tests were reproduced using a modified version of the input beam covariance matrix, where the covariances between flavors was removed. The pre- and post-fit covariance matrices, are shown in Figure~\ref{fig:LAR_noflavcorr_covariances}, which can be compared with Figure~\ref{fig:LAR_nominal_covariances}, which included the flavor correlations. \blue{The distributions of $\chi^{2}$-values for each detector type, calculated with Equation~\ref{eq:bias_chi2}, obtained for 10,000 independent throws of the modified covariances matrices from Figure~\ref{fig:LAR_nominal_covariances} are shown in Figure~\ref{fig:noflavcorr_det_bias}. The biases have uniformly increased slightly with respect to those shown in Figure~\ref{fig:nominal_det_bias}, indicating that the fit relies on the expected flavor composition of the beam as expected, although the bias in the fitted distributions are not significant if those assumptions no longer hold true. Although this test shows a small bias, there is still an implicit reliance on the shape of each flavor contribution to the flux. It should be noted that if the correlation between flavors {\it and} the bin-to-bin correlation between each flavor were removed, the fit would not perform well at all because it would have no way to break the ambiguity between the different flavor contributions to the rate.}

\begin{figure}[tbp]
  \centering
  \ifnotoverleaf
  \subfloat[FHC]  {\includegraphics[width=0.45\textwidth]{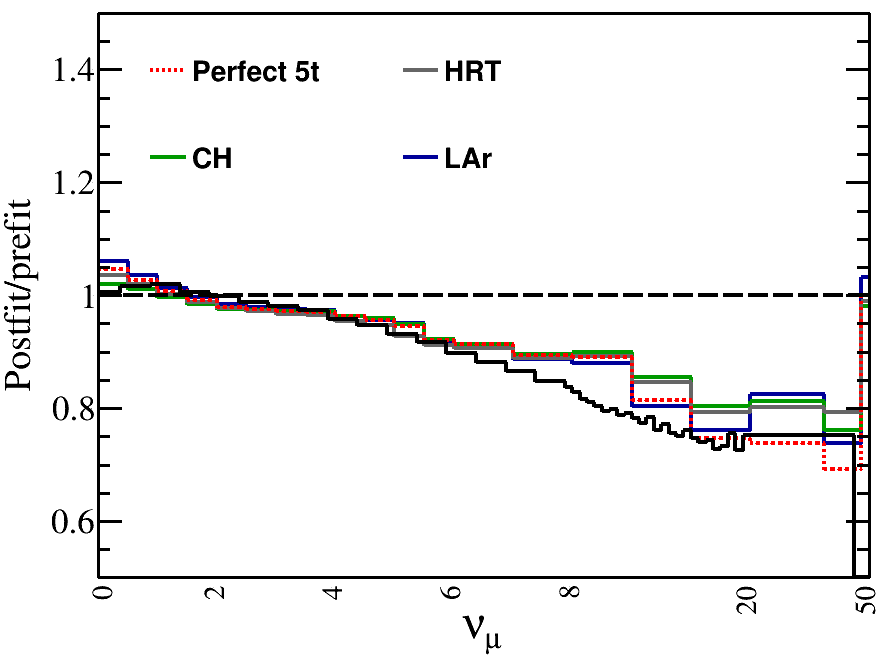}}\\
  \subfloat[RHC]  {\includegraphics[width=0.45\textwidth]{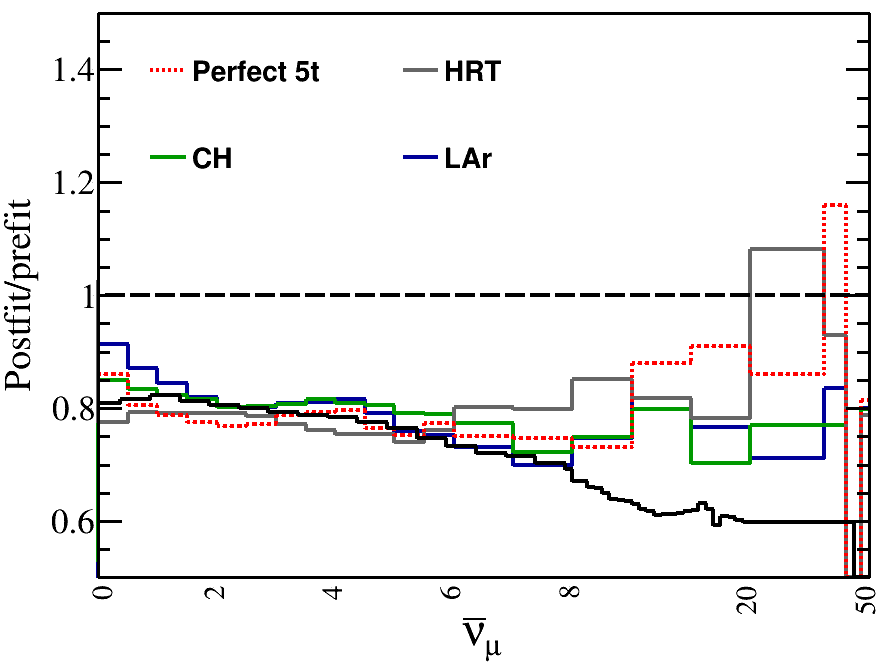}}  
  \fi
  \caption{Weighted average flux values as a function of neutrino energy for the $\nu_{\mu}$ ($\bar{\nu}_{\mu}$) component of the FHC (RHC) flux, shown as a ratio with the pre-fit fluxes where the ``data'' has been modified by the reweighting functions shown as the black lines.}
  \label{fig:crazy_det_central}
\end{figure}
The bias studies shown in Figure~\ref{fig:nominal_det_bias} show that the fit converges on the true input flux in an unbiased way for the variations expected given our prior understanding of the beam. However, as well as reducing the flux uncertainty, we would hope that the $\nu$--$e^-$ sample could provide an independent check that the flux is correctly modeled, and correctly fit, and identify, distortions to the flux distribution in data which are not covered by the uncertainties in the input flux covariance matrix. To that end, a ``crazy'' flux distribution was produced independently for each mode by increasing the target density by 30\%, well outside its tolerance. Figure~\ref{fig:crazy_det_central} shows how well the fit performs when trying to reproduce the ``crazy'' fake data sets in both FHC and RHC modes. The weighted average flux values obtained by all of the detector configurations tested roughly reproduces the crazy flux reweighting as a function of $E_{\nu}$ (indicated by the black line in Figure~\ref{fig:crazy_det_central}). \blue{Although the agreement is not perfect, the post-fit distributions follow the distorted flux well for all detector options in both FHC and RHC mode, indicating that $\nu$--$e^{-}$ samples would be able to diagnose a large flux bias in both modes, which is reassuring. The ability to fit out the flux distortion worsens at high energies, which is likely to be due to the sparse $E_{\nu}$ templates used in the fit (see Table~\ref{tab:binning}), but is good across the bulk of the DUNE flux distribution.} This constitutes another important result and highlights the necessity of making $\nu$--$e^-$ scattering measurements at DUNE. Even though the shape of the flux uncertainty cannot be significantly reduced from the uncertainty obtained from the beam simulation, it can diagnose unmodeled issues with flux prediction {\it in situ}.

\FloatBarrier
\subsection{Systematic uncertainties}
\label{sec:systematic_uncertainties}
In the eventual application of this analysis to DUNE, various systematic uncertainties will need to be included or marginalized over in the fit. These will include detector uncertainties, signal and background cross section uncertainties. In order to quantify the effect that such uncertainties have on the analysis, and the size of the uncertainties on the post-fit flux distribution, a series of fake data studies have been performed. In these studies, the Monte Carlo remains unchanged as described above, but instead of using the same Monte Carlo simulation as the ``fake data'' in the fit, the Monte Carlo is modified to form an alternative fake data set, and the fit is repeated to quantify the effect that this unknown effect has on the output. We investigate a number of systematic sources, described in this section, using the nominal LAr detector configuration.

\begin{figure}[tbp]
  \centering
  \ifnotoverleaf
  \includegraphics[width=0.45\textwidth]{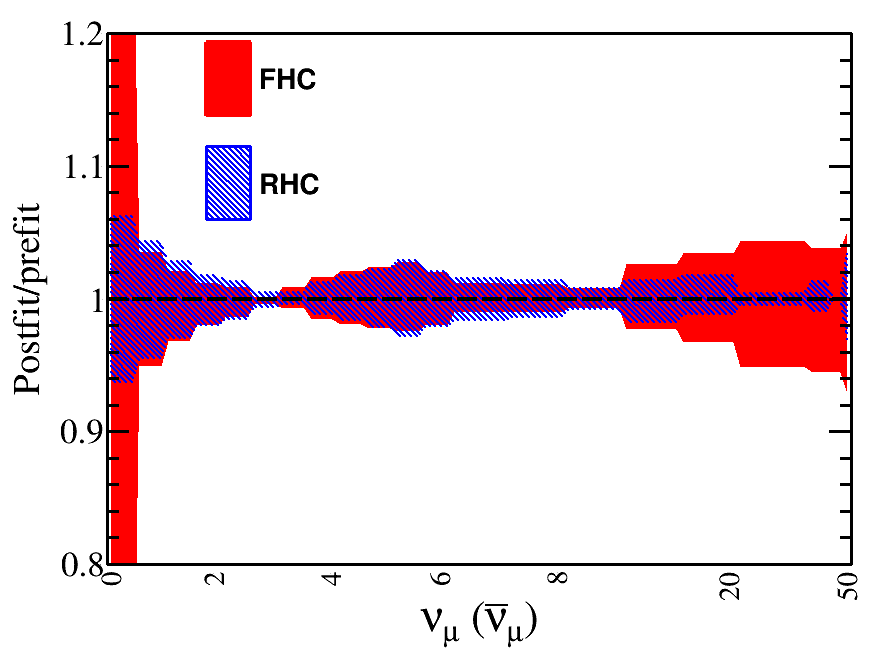}
  \fi
  \caption{Weighted average flux values as a function of neutrino energy for the $\nu_{\mu}$ ($\bar{\nu}_{\mu}$) component of the FHC (RHC) flux, shown as a ratio with the pre-fit flux for fake data studies corresponding to a $\pm$5\% shift in the fraction of the reconstructed electron energy in the low energy tail.}
  \label{fig:elow_central}
\end{figure}

\begin{figure}[tbp]
  \centering
  \ifnotoverleaf
  \includegraphics[width=0.45\textwidth]{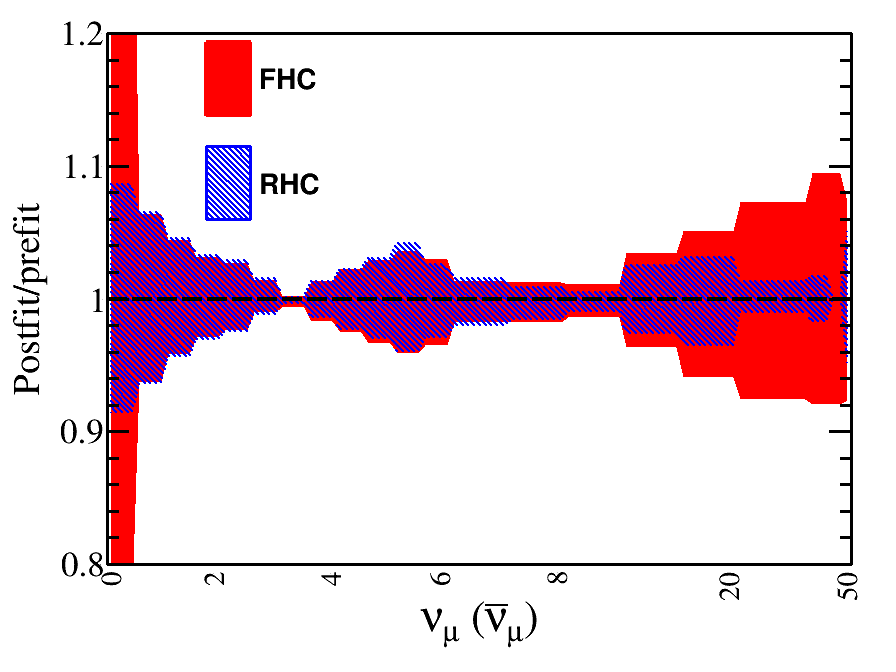}
  \fi
  \caption{Weighted average flux values as a function of neutrino energy for the $\nu_{\mu}$ ($\bar{\nu}_{\mu}$) component of the FHC (RHC) flux, shown as a ratio with the pre-fit flux for fake data studies with a 2\% uncertainty in the electron energy scale.}
  \label{fig:escale_central}
\end{figure}
Here we investigate two sources of systematic error in the energy reconstruction. Firstly, we investigate whether mismodeling the fraction of the reconstructed electron energy in the low-energy tail significantly biases the results. In this analysis, the nominal fraction of events in the low energy tail is 10\%, and we have assigned a very conservative $\pm$5\% systematic uncertainty. In Figure~\ref{fig:elow_central}, the error band due to that systematic uncertainty is show on the weighted average flux values (the values of $\overline{M}_{i}$ produced by Equation~\ref{eq:postfit_covar}), relative to the nominal case (a line at $y = 1$), for the nominal LAr detector in both FHC and RHC. The second energy reconstruction systematic is a 2\% uncertainty in the reconstructed electron energy. Again, in Figure~\ref{fig:escale_central}, the error band due to that systematic uncertainty is show on the weighted average flux values, for the nominal LAr detector  in both FHC and RHC.

Both, conservative, energy reconstruction uncertainties considered introduce percent-level changes in the average flux value across most of the energy range, and subpercent-level changes around the flux peak, which are small relative to the size of the postfit flux uncertainty (see Figure~\ref{fig:nominal_det_constraint} for comparison) in both FHC and RHC modes. We note also that the variations shown here did not significantly change the average $\chi^{2}$ value obtained in the 10,000 flux throws relative to the nominal case shown in Figure~\ref{fig:nominal_det_bias}, indicating that although there are shape variations to the weighed average flux distribution in Figures~\ref{fig:escale_central} and~\ref{fig:elow_central}, these are within the expected post-fit shape uncertainty in the flux. For larger shifts to the energy reconstruction systematics, larger biases are seen, as would be expected.

\begin{figure}[tbp]
  \centering
  \ifnotoverleaf
  \includegraphics[width=0.45\textwidth]{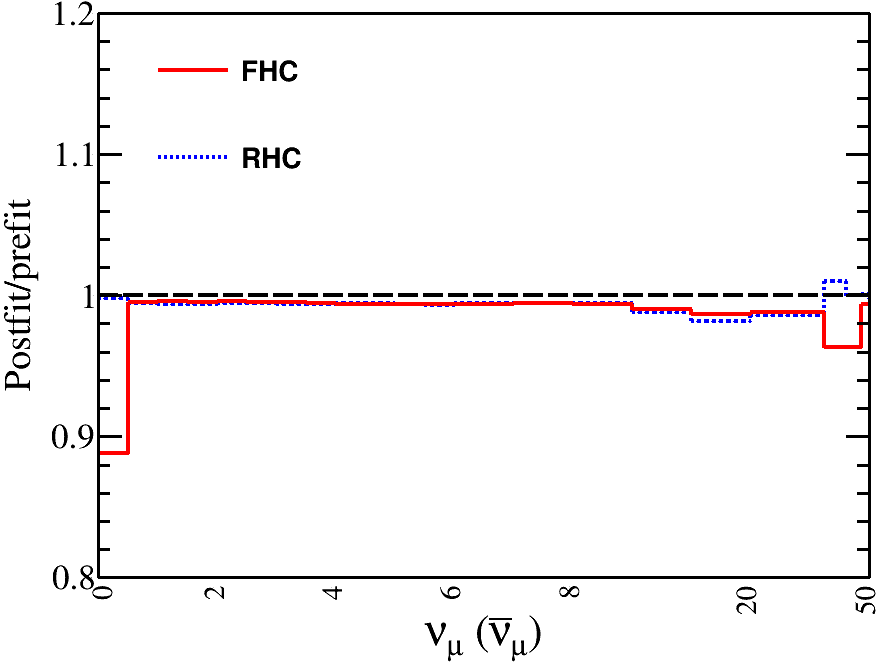}
  \fi
  \caption{Weighted average flux values as a function of neutrino energy for the $\nu_{\mu}$ ($\bar{\nu}_{\mu}$) component of the FHC (RHC) flux, shown as a ratio with the pre-fit flux for fake data studies with a 2 mrad. shift to the beam pointing along the x-axis.}
  \label{fig:xbias_central}
\end{figure}

Although the beam pointing uncertainty is included in the input flux covariance matrix, it is interesting to ask what the effect on the analysis would be if the beam direction were mismodeled by some constant amount. Fake data studies were performed where a bias of 2 mrad. was introduced in the x-axis of the beam direction, by shifting the reconstructed electron angle w.r.t the nominal beam direction in the MC used in the fit. Figure~\ref{fig:xbias_central} shows the weighted average flux values for the nominal LAr detector in both FHC and RHC modes. Here, no error band is produced as a bias of $\pm$2 mrad. in any direction would have the same effect as the response is symmetric around the beam axis (at least for the rather simple detector reconstruction considered here). The effect from the beam pointing bias is small for both beam modes. As for the energy reconstruction systematics considered, the beam pointing uncertainty did not increase the average $\chi^{2}$ for the 10,000 flux throws considered, relative to the nominal case shown in Figure~\ref{fig:nominal_det_bias}, as is expected given that the prefit flux uncertainty includes beam pointing uncertainties, and this analysis is not expected to be able to strongly constrain them. Larger deviations of 5 mrad. were also tested, although are not shown, for which a strong bias is seen in the best fit $\chi^{2}$ distribution.

\begin{figure}[tbp]
  \centering
  \ifnotoverleaf
  \subfloat[FHC]  {\includegraphics[width=0.45\textwidth]{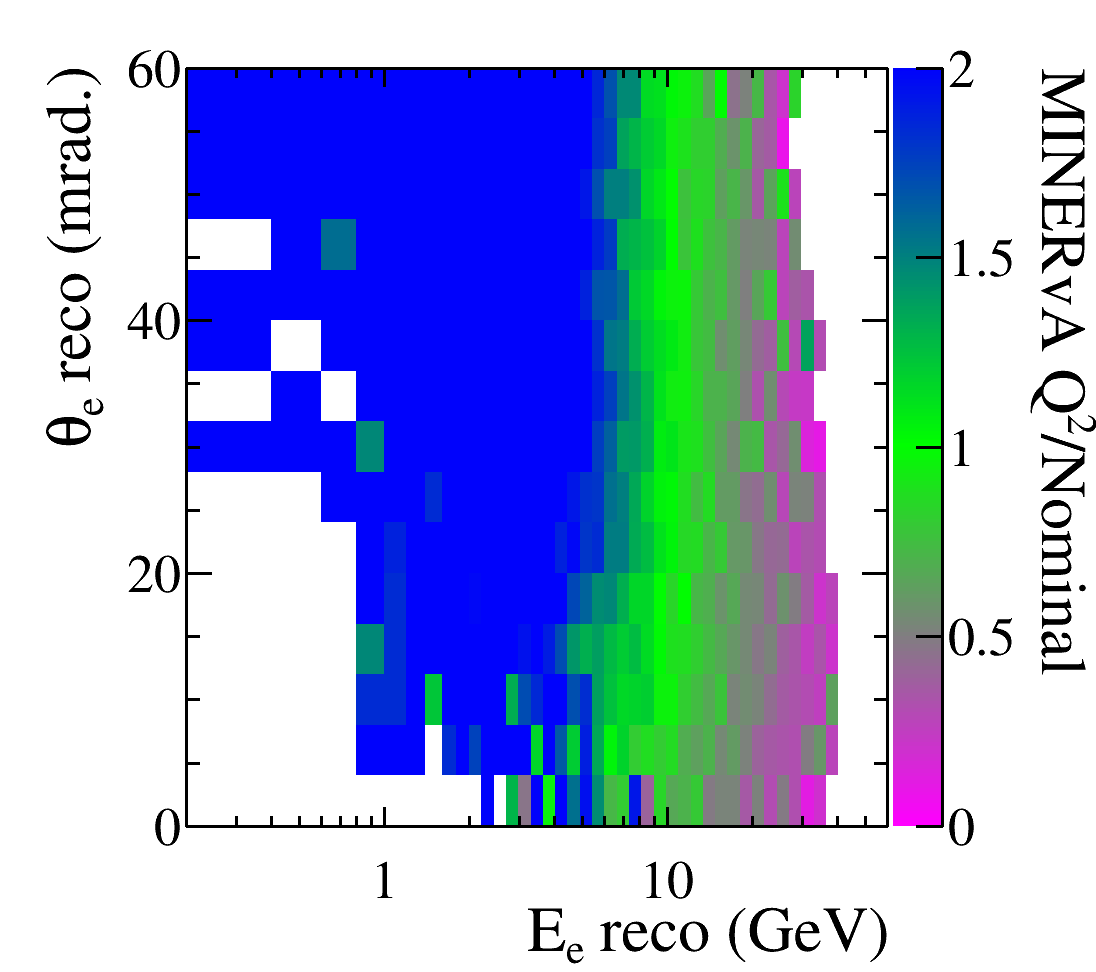}}\\
  \subfloat[RHC]  {\includegraphics[width=0.45\textwidth]{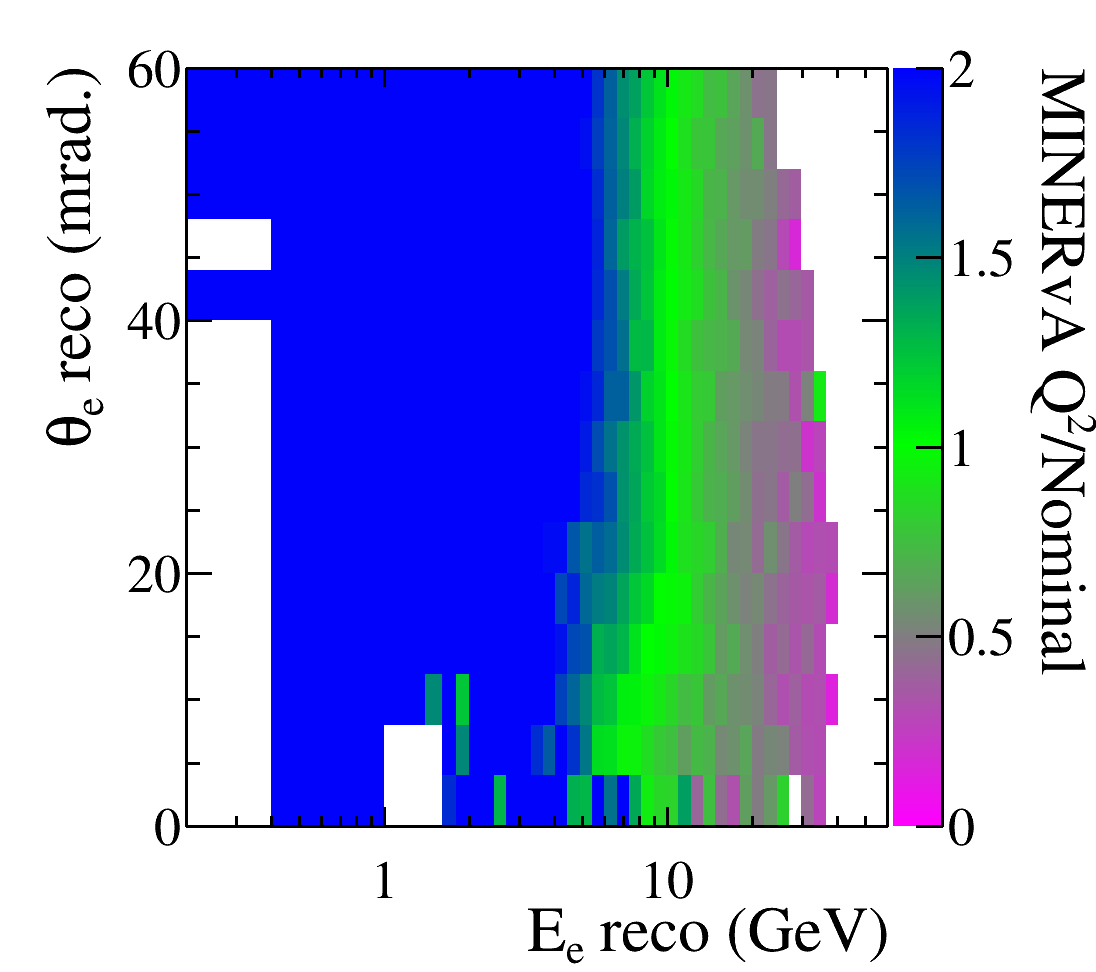}}
  \fi
  \caption{The effect of the observed MINERvA data--MC $Q^{2}$ distortion for charged current quasielastic-like events shown for the $\nu_{e}$ ($\bar{\nu}_{e}$) template in FHC (RHC), shown as a ratio to the nominal template.}
  \label{fig:minq2_templates}
\end{figure}

\begin{figure}[tbp]
  \centering
  \ifnotoverleaf
  \includegraphics[width=0.45\textwidth]{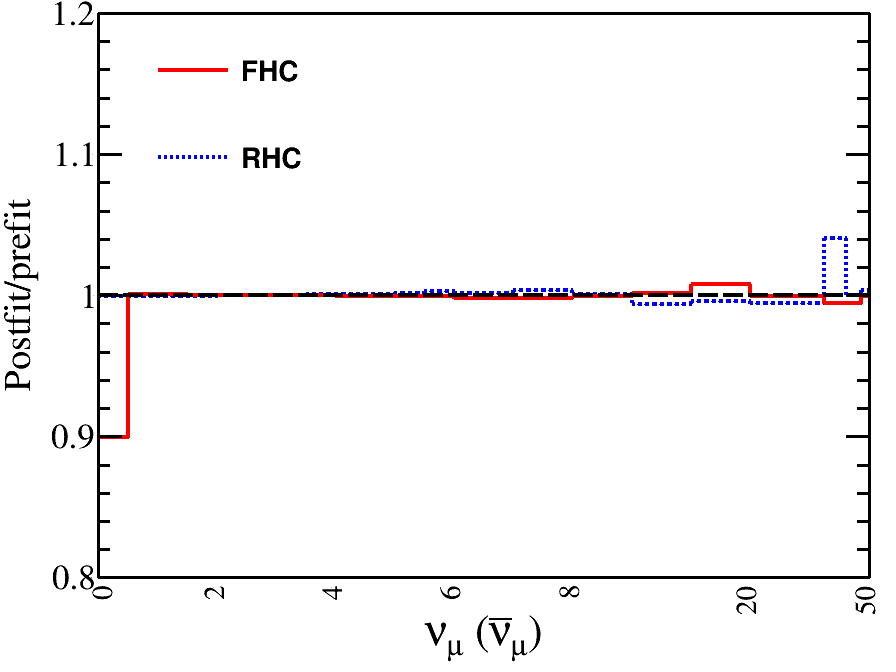}
  \fi
  \caption{Weighted average flux values as a function of neutrino energy for the $\nu_{\mu}$ ($\bar{\nu}_{\mu}$) component of the FHC (RHC) flux, shown as a ratio with the pre-fit flux for fake data studies with and without the MINERvA $Q^{2}$ distortion applied to the $\nu_{e}$ ($\bar{\nu}_{e}$) template.}
  \label{fig:minq2_lar_central}
\end{figure}

Changes to the background predictions could also affect the result, although correlations between background and signal templates in the postfit covariances were weak (see the example in Figure~\ref{fig:example_covar}), it may be expected that such changes will not have a large effect on the result. MINERvA have observed a deviation between Monte Carlo expectation and data as a function of reconstructed $Q^{2}$ (see Ref.~\cite{Ruterbories:2018gub} for a definition) for $\nu_{\mu}$--CH and $\bar{\nu}_{\mu}$--CH charged current quasielastic-like events. In particular, there is a strong suppression at low $Q^{2}$ values, which might correspond to the region of overlap between the signal and background templates. 
MINERvA has concluded, however, that the majority of this disagreement is due not to truly quasielastic events which would create background to a neutrino-electron scattering analysis, but rather to higher recoil processes such as pion production where the pion is observed in nuclear final state interactions~\cite{Ruterbories:2018gub,Valencia:2019mkf}.
In order to investigate this effect on the analysis, the input $\nu_{e}$--$^{40}$Ar and $\bar{\nu}_{e}$--$^{40}$Ar background events are modified from the nominal GENIE prediction by reweighting according to the observed MINERvA ratio, as a function of true $Q^{2}$. The effect that the $Q^{2}$ modification has on the $\nu_e$ background template is shown, as a ratio with the nominal template, in Figure~\ref{fig:minq2_templates}, for both FHC and RHC modes. Figure~\ref{fig:minq2_lar_central} shows the effect on the weighted average flux values, with the MINERvA $Q^{2}$ distortion applied, for the nominal LAr detector in both FHC and RHC modes. In both cases, the bias on the weighted average flux distribution is very small except at high energies or in very low statistics bins.

\subsection{Impact of radiative corrections}

\begin{figure}[tbp]
  \centering
  \ifnotoverleaf
  \includegraphics[width=0.45\textwidth]{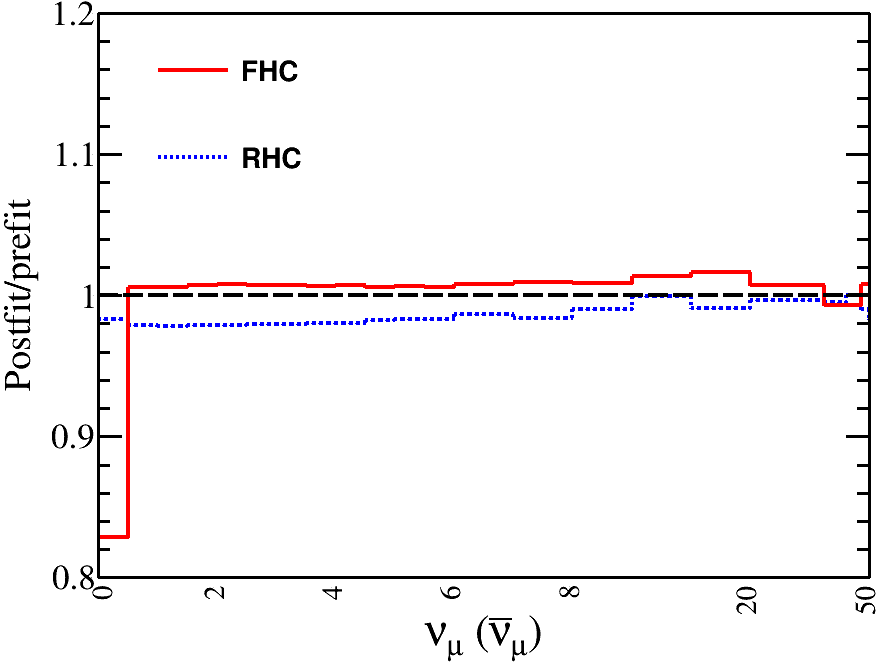}
  \fi
  \caption{Weighted average flux values as a function of neutrino energy for the $\nu_{\mu}$ ($\bar{\nu}_{\mu}$) component of the FHC (RHC) flux, shown as a ratio with the pre-fit flux for the fake data studies with and without radiative corrections applied.}
  \label{fig:NoRadCorr_central}
\end{figure}

Although the signal $\nu$--$e^-$ scattering process is well understood in principle, the radiative corrections applied in this analysis (described in Appendix~\ref{sec:radiative}) are not included in GENIE, and may be refined by a more careful calculation later. To quantify the importance of the radiative corrections in the analysis, the study was repeated without radiative corrections applied, in the same manner as the previous studies into the effect of systematic uncertainties, although it is explicitly not a systematic uncertainty. Figure~\ref{fig:NoRadCorr_central} shows the weighted average flux values produced with 10,000 throws of the input flux covariance, without radiative corrections applied (the default GENIE prediction), for the nominal LAr detector in FHC and RHC modes. The effect of removing the radiative corrections is to increase (decrease) the weighted average flux value in FHC (RHC) by 1--2\%. Interestingly, the average $\chi^{2}$ value from the 10,000 flux throws considered is slightly increased relative to the nominal case shown in Figure~\ref{fig:nominal_det_bias}, the increase is larger than the systematic uncertainties considered in Section~\ref{sec:systematic_uncertainties}, although not significantly. This can be understood because the absolute rate is well constrained by the analysis presented here, so a normalization-only change in the flux (e.g., a fully correlated shift between all energy bins) is more strongly constrained than some variations in the flux shape.

\section{Inverse Muon Decay}
\label{sec:IMD}

\begin{figure}[tbp]
  \centering
\ifnotoverleaf
  \subfloat[Rate by neutrino energy]  {\includegraphics[width=0.45\textwidth]{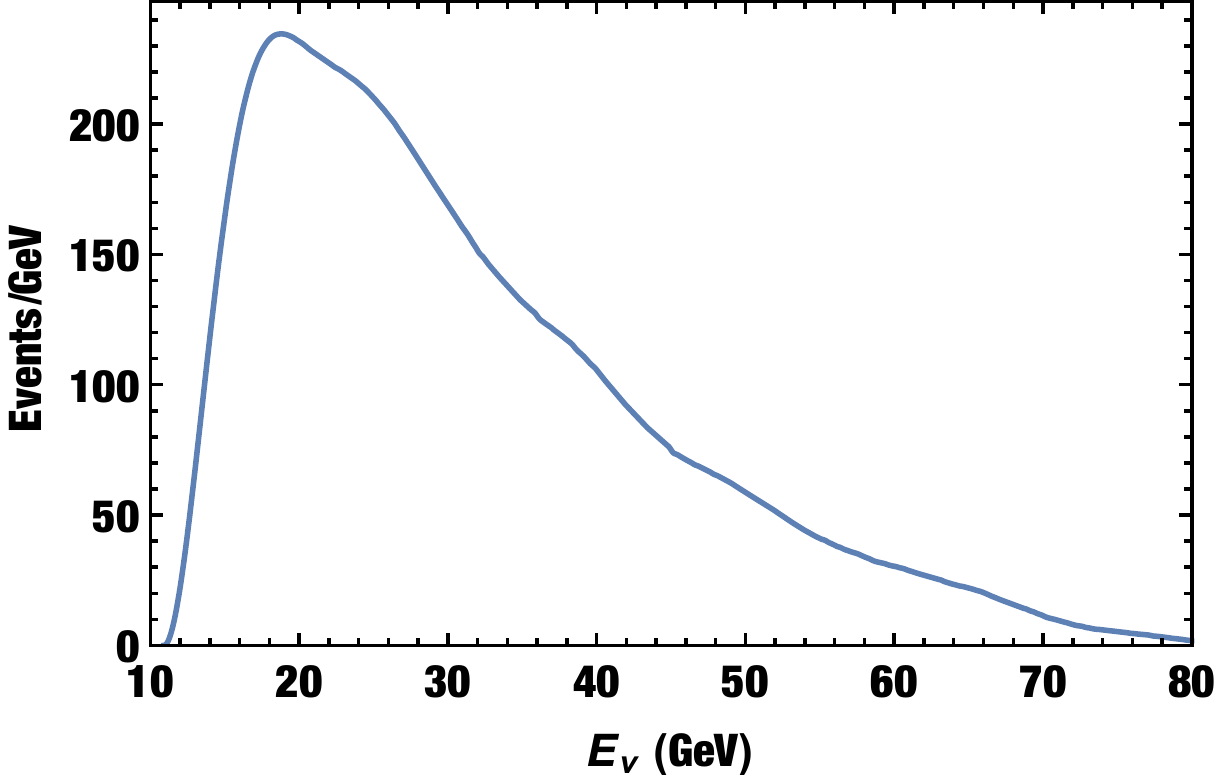}}\\
  \subfloat[Rate by muon energy]               {\includegraphics[width=0.45\textwidth]{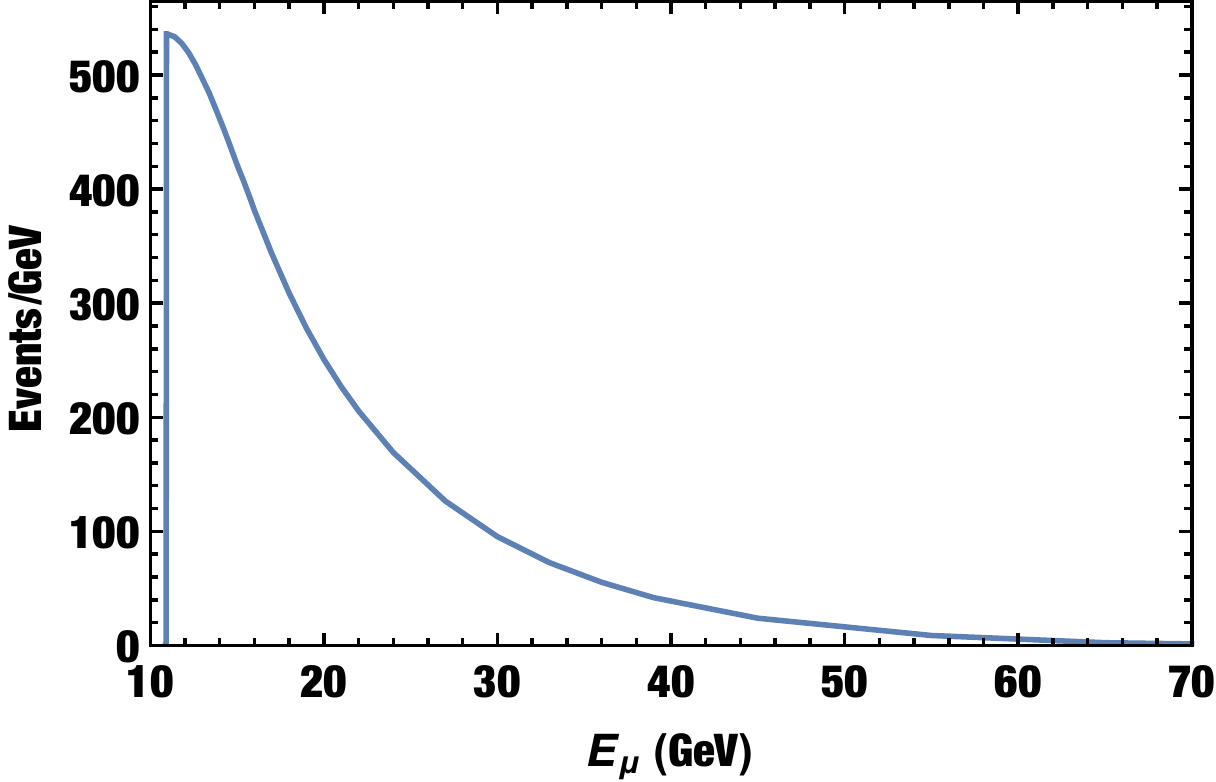}}
  \fi
  \caption{The rate of inverse muon decay events shown as a function of neutrino energy (top) and produced muon energy (bottom) in a five year exposure of the neutrino beam for a $30$ ton reference detector}
  \label{fig:IMD-rate}
\end{figure}

\begin{figure}[tbp]
  \centering
\ifnotoverleaf
  \includegraphics[width=0.45\textwidth]{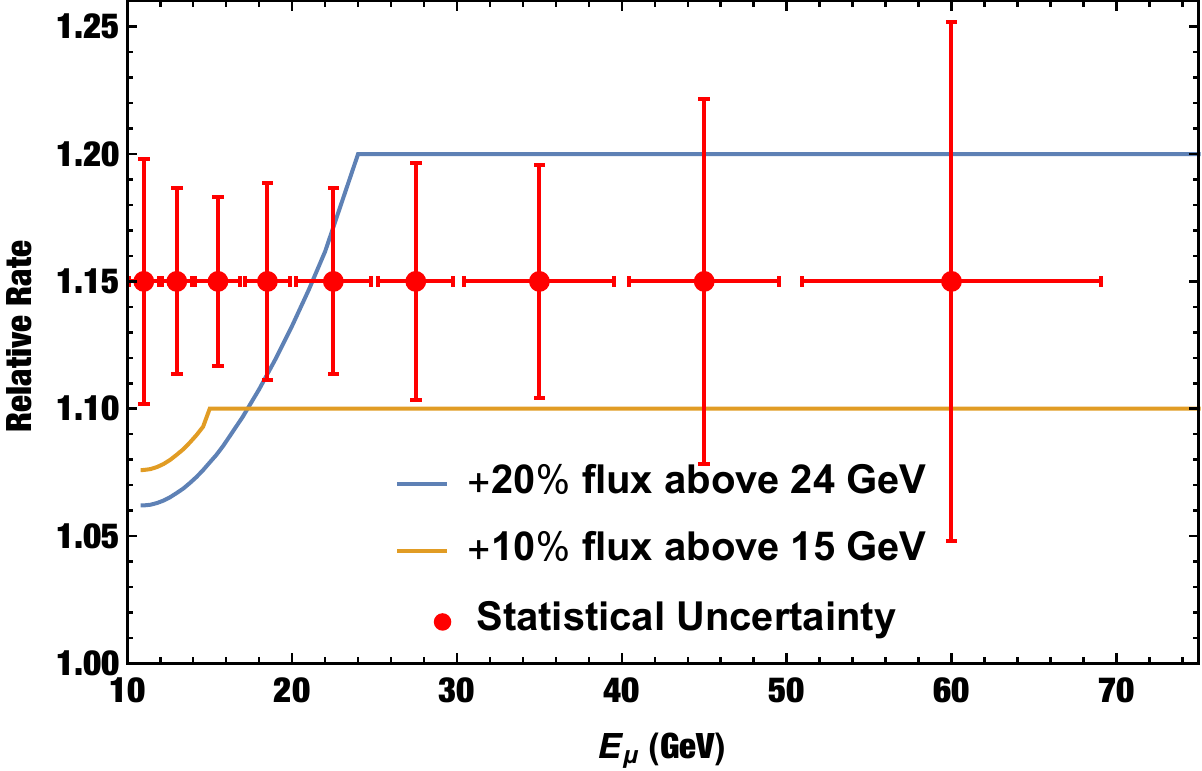}
  \fi
  \caption{The statistical uncertainty of a binned inverse muon decay sample shown against two possible spectral distortions, one of which increases the number of neutrinos above $15$~GeV by $10\%$, and one of which increases the number of neutrinos above $24$~GeV by $20\%$}
  \label{fig:IMD-sens}
\end{figure}

As previously noted in Section~\ref{sec:nue_scat_intro}, inverse muon decay (IMD), $\nu_\mu e^-\to\nu_e\mu^-$ has potential to constrain the high energy DUNE $\nu_\mu$ flux.  The rate of such events in a $30$~ton reference detector with a five year exposure is shown in Figure~\ref{fig:IMD-rate} as a function of both laboratory neutrino and muon energy.   There are of order $6\times 10^3$ events in such a sample produced with the neutrino mode beam, which would constrain the total rate of such high energy $\nu_\mu$.
The statistical sensitivity of such a sample to variations in this high energy $\nu_\mu$ spectrum is illustrated in Figure~\ref{fig:IMD-sens}.  The statistical uncertainty in such a sample would clearly allow either of two simulated neutrino spectral distortions, one of which increases the number of neutrinos above $15$~GeV by $10\%$, and one of which increases the number of neutrinos above $24$~GeV by $20\%$.  However, distinguishing between the two scenarios would be more difficult.

In addition to the limited use of such a probe of the high energy $\nu_\mu$ flux for DUNE's primarily neutrino oscillation mission, the design of the near detectors will likely not be optimized to measure these spectra.  Such high energy muons will not be contained in DUNE's near detectors, and measurement of the momentum by curvature in a magnetic field at such high muon energies will be difficult in a detector optimized for such measurements at significantly lower muon energies.

\section{Conclusions}
\label{sec:conclusions}

\blue{Because the neutrino-electron elastic scattering cross section is known, the flux can be extracted from the observed event rate. For realistic DUNE near detectors in the LBNF beam, it is possible to select a sample of many thousands of $\nu e^{-} \rightarrow \nu e^{-}$ events. In this work, we have investigated how well different potential DUNE near detector designs will be able to constrain the LBNF flux.}

\blue{We found that given realistic mass constraints, a 30t liquid argon detector is able to perform better than 5t low density trackers, even ones with significantly better tracking resolution. This is due to its higher statistics, and despite the superior angular resolution of lower density detectors. This was also found to be the case even for a 5t detector with perfect electron reconstruction and background rejection.}

\blue{With realistic systematic uncertainties, the uncertainty on the absolute neutrino flux in the 30t LAr detector is reduced from $\sim$8\% to $\sim$2\%. The uncertainty on the shape as a function of neutrino energy is also reduced by $\sim$20-30\%. This is partially due to the fact that the flux shape is better known {\em a priori} than the absolute normalization. The improved reconstruction performance of high resolution detectors has a stronger impact on the flux shape constraint, as expected, but for realistic detector sizes, a large liquid argon detector still outperforms a smaller detector with better resolution. It seems that detector mass is the most important factor for making a $\nu$--$e^{-}$ flux constraint, even for a 30t detector in the very intense LBNF beam. The intrinsic divergence of the beam is an important consideration which has the potential to limit the utility of a $\nu$--$e^{-}$ flux constraint, and as such was included in this study.}

\blue{As well as being able to reduce the neutrino flux uncertainties, we demonstrated that a $\nu$--$e^{-}$ sample with a large liquid argon detector is capable of identifying a large variation in the neutrino spectrum outside of predicted uncertainties, and is not biased to the input assumptions about the flux, despite the inability to directly distinguish neutrino flavor with a $\nu$--$e^{-}$ sample.}

\ifnum\sizecheck=0
\begin{acknowledgments}

CMM and KSM were supported by the Office of Science, Office of High Energy Physics, of the
U.S. Department of Energy under Contract Number DOE OHEP DE-AC02-05CH11231, and Award Number DE-SC0008475, respectively.
CW was supported by the Swiss National Science Foundation (SNSF) and Secretariat for Education, Research and Innovation (SERI). 

We are grateful to Steve Dennis who participated in early steps of the analysis that evolved into this work.
The VietNus 2017 workshop at the
International Center for Interdisciplinary Science and Education (ICISE), Qui Nhon, Vietnam, hosted our collaboration during these early steps.  We also thank the staff of the G-Life Coffee Shop, Qui Nhon, Vietnam, for their gracious hospitality, their dehumification devices that kept the air no worse than unpleasantly tepid, and their limitless supply of tasty iced beverages.   \blue{We also acknowledge the valuable feedback from many members of the DUNE collaboration as this work was presented in the context of the near detector design studies.}

\end{acknowledgments}
  \bibliography{DUNE-nue}
\fi

\clearpage

\appendix
\section{{\bf Electroweak radiative corrections to neutrino-electron scattering}}
\label{sec:radiative}
\newcommand{\nubar}{\overline{\nu}}

The cross section for tree-level neutrino-electron scattering is given in Eqn.~\ref{eqn:tree-xsec}, and this is the cross-section implemented in the GENIE 2.8 event generator~\cite{Andreopoulos:2009rq,Andreopoulos:2015wxa} which is used as the reference model in this study.  It is necessary to correct this model to use modern values of the electroweak couplings.  This is done by changing the chiral couplings, $C_{\mathrm{LL}}$ and $C_{\mathrm{LR}}$, to one-loop values predicted using global fits to electroweak data~\cite{Erler:2013xha}.  Table~\ref{tab:ewk-couplings} compares the values for these couplings GENIE to the values used in this analysis.

\begin{table}[b]
\begin{tabular}{c|ccc}
   & $C_{\mathrm{LL}}^{\nu_ee}$ & $C_{\mathrm{LL}}^{\nu_\mu e}$ & $C_{\mathrm{LR}}^{\nu e}$ \\ \hline
 \text{GENIE 2.8} & 0.7277 & -0.2723 & 0.2277 \\
 \text{One loop} & 0.7276 & -0.2730 & 0.2334 \\
\end{tabular}
\caption{Electroweak couplings in GENIE and in our one-loop calculation of $\nu e^-$ elastic scattering}
\label{tab:ewk-couplings}
\end{table}

We consider two possibilities for the one-loop electromagnetic radiative corrections, including the possibility of real photon emission.  Either the experiment measures truly the kinetic energy of the final state electron exclusive of any radiated photons and measures $y=T_e/E_\nu$, or the experiment measures the energy of radiated photons clustered together with emitted electrons and measures $y=(T_e+E_\gamma)/E_\nu$.  The first case would be relevant for low density trackers that measure the electron energy by curvature, and the second case would be relevant for calorimetric measurements of electron energy.

In the measurement of only electron energy, the corrections~\cite{Sarantakos:1982bp,Bahcall:1995mm} 
that modify the expressions for the $\nu_\mu e$, $\nubar_\mu e$, $\nu_e e$ and $\nubar_e e$ elastic scattering cross-sections in Eqn.~\ref{eqn:tree-xsec} as follows:
\begin{widetext}
\begin{eqnarray}
\label{eqn:nunloxsec}
\frac{d\sigma(\nu_\ell e^-\to \nu_\ell e^-)}{dy}=&\frac{G_{\mathrm{F}}^2 s}{\pi}&\left[
   \left( C_{\mathrm{LL}}^{\nu_\ell e}\right) ^2 (1+\frac{\alpha_{EM}}{\pi}X_1) +\left( C_{\mathrm{LR}}^{\nu e}\right) ^2 (1-y)^2 (1+\frac{\alpha_{EM}}{\pi}X_2) \right. \nonumber \\
&&\left. -\frac{C_{\mathrm{LL}}^{\nu_\ell e} C_{\mathrm{LR}}^{\nu_ e} m y}{E_\nu}(1+\frac{\alpha_{EM}}{\pi}X_3)\right]\\ 
\label{eqn:nubarnloxsec}
\frac{d\sigma(\nubar_\ell e^-\to \nubar_\ell e^-)}{dy}=&\frac{G_{\mathrm{F}}^2 s}{\pi}&\left[
   \left( C_{\mathrm{LR}}^{\nu e}\right) ^2 (1+\frac{\alpha_{EM}}{\pi}X_1) +\left( C_{\mathrm{LL}}^{\nu_\ell e}\right) ^2 (1-y)^2 (1+\frac{\alpha_{EM}}{\pi}X_2)  \right. \nonumber \\
&&\left. -\frac{C_{\mathrm{LL}}^{\nu_\ell e} C_{\mathrm{LR}}^{\nu e} m y}{E_\nu}(1+\frac{\alpha_{EM}}{\pi}X_3)\right]
\end{eqnarray}
\end{widetext}

\noindent  where $E_\nu$ is the neutrino energy, $s$ is the Mandelstam
invariant representing the square of the total energy in the center-of-mass frame, $m$ is the electron mass and $y=T_e/E_\nu$.  The $X_i$ correction terms are

\begin{widetext}
\begin{eqnarray}
X_1&=&\frac{1}{12} (6 y+12 \log (1-y)-6 \log (y)-5) \log \left(\frac{2 E_{\nu }}{m}\right) -\frac{\text{Li}_2(y)}{2}+\frac{y^2}{24}-\frac{11 y}{12} \nonumber \\
&&-\frac{1}{2} \log
   ^2\left(\frac{1}{y}-1\right)+y \log (y) -\frac{1}{12} (6 y+23) \log (1-y)+\frac{\pi ^2}{12}-\frac{47}{36},\\
\nonumber \\
X_2&=&\frac{\left(-4 y^2+\left(-6 y^2+6 y-3\right) \log (y)+11 y+6 (1-y)^2 \log (1-y)-7\right) \log \left(\frac{2 E_{\nu }}{m}\right)}{6
   (1-y)^2} \nonumber \\
&&+\frac{\left(-y^2+y-\frac{1}{2}\right) \left(\text{Li}_2(y)+\log ^2(y)-\frac{\pi ^2}{6}\right)}{(1-y)^2}+\frac{\left(4 y^2+2 y-3\right) \log (y)}{4
   (1-y)^2} \nonumber \\
&&-\frac{31-49 y}{72 (1-y)}+\frac{(10 y-7) \log (1-y)}{6 (1-y)}+\log (1-y) \left(\log (y)-\frac{1}{2} \log (1-y)\right) ,\\
\end{eqnarray}
\end{widetext}
\begin{eqnarray}
X_3&=\left(\frac{\left(m+y E_{\nu }\right) \log \left(\frac{\sqrt{y
   E_{\nu } \left(2 m+y E_{\nu }\right)}+m+y E_{\nu }}{m}\right)}{\sqrt{y E_{\nu } \left(2 m+y E_{\nu }\right)}}-1\right) \nonumber  \\
& \times\log\left(1-y-\frac{m}{\sqrt{y E_{\nu } \left(2 m+y E_{\nu }\right)}+m+y E_{\nu }}\right) ,
\end{eqnarray}

\noindent where Li$_2(z)$ is Spence's function, $\int_0^z\frac{-\log(1-u)}{u}du$.

In the second case, where $y\equiv (T_e+E_\gamma)/E_\nu$, the modifications to the cross-section
are more straightforward:
\begin{widetext}
\begin{eqnarray}
X_1&=&-\frac{2}{3} \log \left(\frac{2 y E_{\nu }}{m}\right)+\frac{y^2}{24}-\frac{5 y}{12}-\frac{\pi ^2}{6}+\frac{23}{72}\\
X_2&=&-\frac{2}{3} \log \left(\frac{2 y E_{\nu }}{m}\right)-\frac{y^2}{18(1-y)^2}-\frac{\pi ^2}{6}-\frac{2y}{9(1-y)^2}+\frac{23}{72(1-y)^2}
\end{eqnarray}
\end{widetext}
$X_3$ is not available in the calculation of Ref.~\cite{Bardin:1983yb}, \blue{although it has been recently calculated by the authors of Ref.~\cite{Tomalak:2019}}.  However, since $X_3$ enters into Eqns~\ref{eqn:nunloxsec} and \ref{eqn:nubarnloxsec} only in terms multiplied by $m_e/E_\nu$, it can be safely neglected.

The direction of the electron, however, in any detector under consideration is most likely to be measured as the electron's direction.  All of these calculations are done assuming collinear emission of photons along the lepton angle. Within that approximation, 
\begin{eqnarray}
E_\nu&=&\frac{2m(1-y)}{\theta_e^2} \nonumber\\
&\times&\left[ 1-\frac{(8-4y-y^2)}{12y(1-y)^2}\theta_e^2\right. \nonumber\\
&&\left. ~~-\frac{\epsilon}{4y(1-y)(1-\epsilon)}\theta_e^2+{\cal O}(\theta_e^4)\right] 
\end{eqnarray}
\noindent where $\epsilon\equiv E_\gamma/(T_e+E_\gamma)$.  Note that corrections from photon emission occurs only multiplied by the small $\theta_e^2$ and is thus negligible. 
This implies that there is no significant effect due to  real photon radiation on the reconstructed neutrino energy inferred from the electron angle and clustered electron plus photon energy, such as in a liquid argon TPC or other calorimetric detector.

\end{document}